\documentclass[aps,prl,reprint,superscriptaddress]{revtex4-2}

\usepackage{graphicx}
\usepackage{amsmath} 
\usepackage{color}
\usepackage{hyperref}
\usepackage[T1]{fontenc}  

\definecolor{ddgreen}{RGB}{0,150,50}
\definecolor{ddcyan}{RGB}{0,150,180}
\hypersetup{colorlinks=true,
    linkcolor=blue,
    filecolor=magenta,      
    urlcolor=ddcyan,
    citecolor=ddgreen}

\begin{document}


\title{Laser-probing the rotational cooling of molecular ions by electron
  collisions}


\author{\'Abel~K\'alosi} \email[email: ]{abel.kalosi@mpi-hd.mpg.de}
\affiliation{Max-Planck-Institut f\"ur Kernphysik, Saupfercheckweg 1, D-69117
  Heidelberg, Germany}
\affiliation{Columbia Astrophysics Laboratory, Columbia
  University, New York, NY 10027, USA}
\author{Manfred~Grieser}
\author{Robert~von~Hahn}
\author{Ulrich~Hechtfischer}
\altaffiliation{Now at: ASML Nederland B.V., Veldhoven, The Netherlands}
\author{Claude~Krantz} \altaffiliation{Present address: GSI Helmholtz Centre for
  Heavy Ion Research, 64291 Darmstadt} \author{Holger~Kreckel}
\author{Damian~M\"ull} \author{Daniel~Paul} \affiliation{Max-Planck-Institut
  f\"ur Kernphysik, Saupfercheckweg 1, D-69117 Heidelberg, Germany}
\author{Daniel~W.~Savin} \affiliation{Columbia Astrophysics Laboratory, Columbia
  University, New York, NY 10027, USA}
\author{Patrick Wilhelm}
\author{Andreas~Wolf}
\author{Old\v{r}ich~Novotn\'y}
\affiliation{Max-Planck-Institut f\"ur Kernphysik, Saupfercheckweg 1, D-69117
  Heidelberg, Germany}


\date{\today}%

\begin{abstract}
	We present state-selected measurements of rotational cooling and excitation rates of CH$^+$ molecular ions by inelastic electron collisions. The experiments are carried out at the Cryogenic Storage Ring, making use of a monoenergetic electron beam at matched velocity in combination with state-sensitive laser-dissociation of the CH$^+$ ions for simultaneous monitoring of the rotational level populations. Employing storage times of up to 600~s, we create conditions where electron-induced cooling to the $J = 0$ ground state dominates over radiative relaxation, allowing for the experimental determination of inelastic electron collision rates to benchmark state-of-the-art theoretical calculations. On a broader scale, our experiments pave the way to probe inelastic electron collisions for a variety of molecular ions relevant in various plasma environments.
\end{abstract}


\maketitle


Electron collisions with molecular ions are fundamental processes in partially ionized molecular gas, ranging from astrophysical environments and planetary atmospheres to low temperature plasma used in advanced technologies \cite{bartschat_electron_2016, bartschat_electron_2018}. Despite their fundamental relevance, detailed experimental data on the effect of electron collisions on the internal excitation of the molecular ions are essentially non-existent. Moreover, to enable computationally tractable calculations, most state-of-the-art theoretical methods require significant approximations \cite{dubernet_basecol2012_2013}, such as the Coulomb-Born (CB) \cite{boikova_rotational_1968, chu_rotational_1974,*chu_erratum_1975, bhattacharyya_rotational_1981, neufeld_electron-impact_1989} and R-matrix methods \cite{rabadan_calculated_1998, lim_electron-impact_1999, faure_electron-impact_2001,hamilton_electron-impact_2016, chakrabarti_r-matrix_2017, faure_state--state_2017, hamilton_electron-impact_2018, cooper_quantemol_2019}, which in this context have remained without experimental verification. For rotationally inelastic collisions between electrons and molecular ions, in particular, theoretical calculations have not yet converged \cite{boikova_rotational_1968, chu_rotational_1974,*chu_erratum_1975, bhattacharyya_rotational_1981, dickinson_electron_1981, neufeld_electron-impact_1989, rabadan_calculated_1998, flower_rotational_1999, lim_electron-impact_1999, faure_electron-impact_2001, liszt_rotational_2012, hamilton_electron-impact_2016, chakrabarti_r-matrix_2017, faure_state--state_2017} and we are unaware of any previous state-resolved experimental cross section data that can guide the theory.

For illustration, we consider the case of molecular ions in cold astrophysical environments \cite{black_molecules_1998,petrie_ions_2007}. They are mostly observed \cite{mcguire_2018_2018} by radio-astronomy via transitions between the lowest rotational states, requiring knowledge of the inelastic collision processes that contribute to rotational excitation and de-excitation. At the relevant temperatures of 10--100~K, the dominant partners for inelastic collisions are hydrogen atoms or molecules, and electrons \cite{liszt_rotational_2012,faure_state--state_2017}.  Electronic collisions are relevant even at the generally small electron density in the interstellar medium (typically $\sim$\,10$^{-4}$ of hydrogen) because of their higher rate coefficients ($>10^{-6}$ cm$^3$\,s$^{-1}$ at 10--100~K for dipolar molecular ions). For ions such as HCO$^+$, electron collisions are found to even dominate the inelastic rates \cite{liszt_rotational_2012}.

To represent, in particular, the astrophysical conditions, relevant laboratory
studies should ensure the low-density, binary-collision regime.  Hence, very
sensitive detection methods are required.  Moreover, the internal molecular
excitation must be probed for only a few populated rotational levels,
corresponding to the cold astrophysical temperatures.  We present such a
measurement using methylidyne (CH$^+$) cations circulating in a cryogenic
storage ring.  There, the rotational level populations are subject to radiative
cooling in the cryogenic environment and to controlled electron collisions in a
merged, velocity-matched beam of electrons.  We probe the evolution of the level
populations as a function of storage time by resonant photodissociation.  At
effective temperatures of $\sim$\,20~K for the radiation field and $\sim$\,26~K
for the electrons, we reach effective electron densities substantially above the
critical density needed for collisional transition rates to dominate over the
radiative ones.  Hence, rotational cooling by inelastic electron collisions
outperforms radiative cooling, leading to $J=0$ as the dominant rotational level
after a much shorter storage time than for radiative cooling alone.  Using laser
diagnostics, the electron-induced changes of the rotational populations are
measured and compared to theory.  This confirms recent predictions for the
low-temperature inelastic rate coefficients of CH$^+$ and demonstrates that, in
combination with suitable rotational diagnostics, cryogenic storage rings offer
a platform to measure rotationally inelastic electron-collision rates for a wide
range of molecular ions.

Merged-beams studies in room-temperature storage rings
\cite{shafir_rotational_2009} demonstrated the effect of electron collisions on
rotational excitation of HD$^+$ by measuring the fragmentation energies from
dissociative recombination (DR) events.  However, individual angular momenta $J$
could not be resolved and the effect of electron collisions on the rotational
populations was far from dominant.  Cryogenic storage rings for molecular ions
\cite{schmidt_electrostatic_2015,von_hahn_cryogenic_2016} now enable
state-selective studies for the lowest $J$-levels of elementary diatomic ions.
These studies addressed spontaneous radiative relaxation of rotational
excitation in a low-temperature blackbody field for CH$^+$
\cite{oconnor_photodissociation_2016} and OH$^-$
\cite{meyer_radiative_2017,schmidt_rotationally_2017,*schmidt_erratum_2018}.
Moreover, the rotational dependence of the low-energy DR rate was measured in
electron collision experiments on HeH$^+$
\cite{novotny_quantum-stateselective_2019}.  The present work extends these
studies by combining laser probing of the rotational level populations of CH$^+$
ions with the collisional rotational cooling of these ions by a merged electron
beam.

\begin{figure}
  \includegraphics[width=8.6cm]{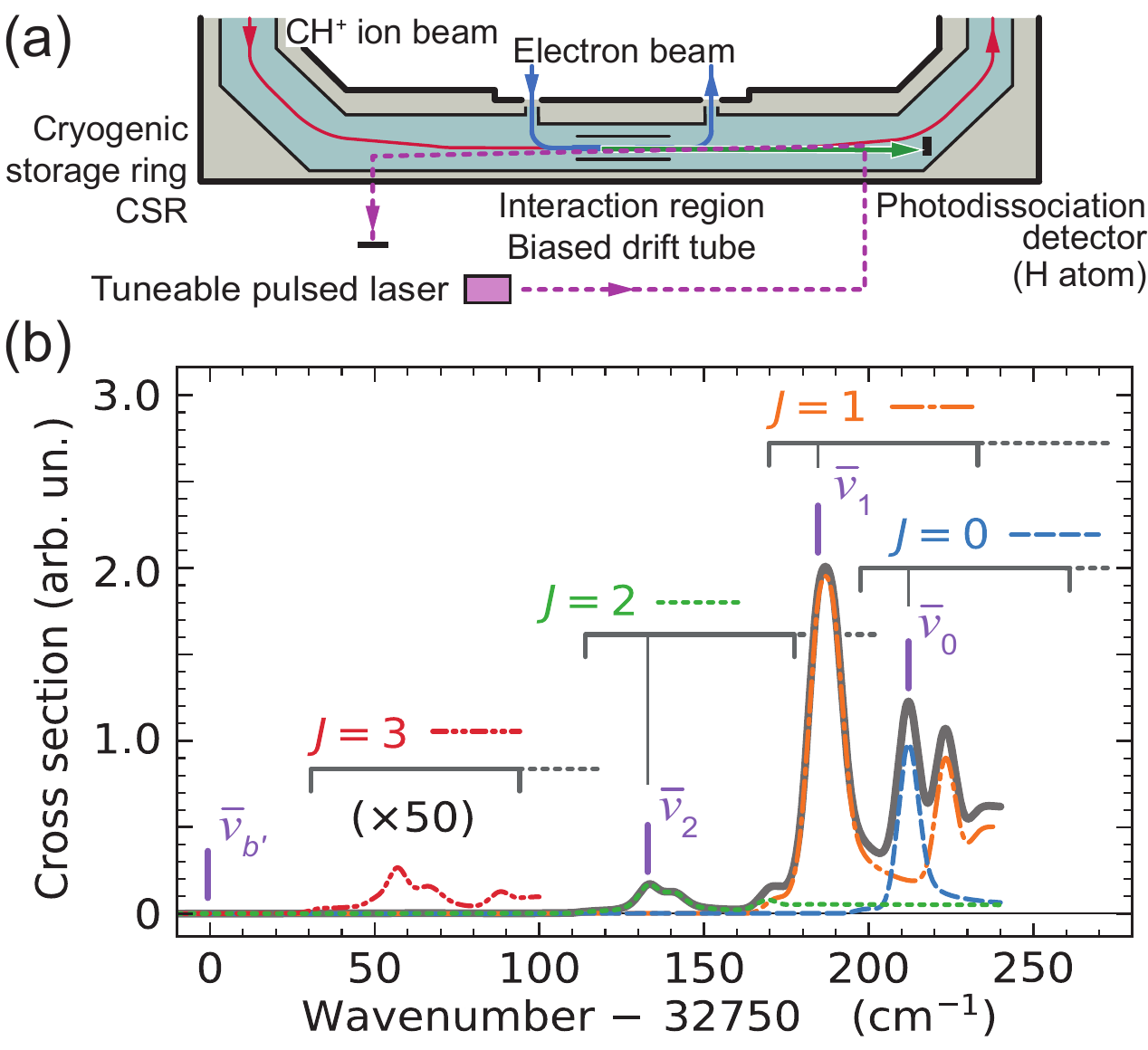}%
  \caption{\label{exp-method} (a) Schematic of the storage-ring section for
    laser probing and merged-beam electron--ion interaction. (b) Theoretical
    near-threshold photodissociation cross sections of CH$^+$ ions
    \cite{oconnor_photodissociation_2016} in $J\leq3$, weighted with a typical
    $J$ distribution during rotational relaxation in the CSR and convolved with
    the laser lineshape (FWHM 6.1\,cm$^{-1}$).  Horizontal brackets: wavenumber
    ranges between the thresholds for C$^+$($^2P_{1/2}$), lower, and
    C$^+$($^2P_{3/2}$), upper, for the indicated $J$.  Thick vertical lines mark
    the wavenumbers $\bar{\nu}_i$ ($i=0\ldots2$) and $\bar{\nu}_{b'}$ where
    laser probing signals were accumulated. The probing wavenumbers are Doppler shift corrected, as explained in the supplemental material \cite{sup}.  Individual $J$ contributions
    (broken curves) add up to the total probed signal (full curve).  The
    magnified contribution of $J=3$ is shown up to the onset of its continuum
    level only.}
\end{figure}

The experiments are performed in the cryogenic storage ring CSR
\cite{von_hahn_cryogenic_2016} at the Max Planck Institute for Nuclear Physics
in Heidelberg, Germany.  A 280-keV beam of CH$^+$ ions with a broad $J$
distribution is produced in a Penning ion source, electrostatically accelerated,
and injected into the CSR, where the ions then circulate with a mean storage
lifetime of $\sim$\,180 s.  For rotational diagnostics, they are illuminated in
a straight section of the ring with nanosecond laser pulses from an optical
parametric oscillator (OPO) with a tunable wavelength near 300 nm.  At certain
wavelengths, CH$^+$ ions in specific low-$J$ levels can be preferentially
addressed by resonant photodissociation, as already demonstrated at CSR
\cite{oconnor_photodissociation_2016}.  Neutral H atoms from the
photodissociation leave the closed ion orbit at a downstream electrostatic
deflector and are counted on a microchannel plate (MCP) detector
[Fig.~\ref{exp-method}(a)].

On each revolution, the ions interact with the velocity-matched, collinear
electron beam of the CSR electron cooler.  The electron cooler delivers a
continuous, nearly monoenergetic electron beam of $\sim$\,20~eV, guided by an
axial magnetic field (10~mT in the interaction region) to yield a 10-mm
diameter, nearly homogeneous current density profile in the collinear overlap
region with the ion beam.  The electron density is $7.0(6)\times10^5$ cm$^{-3}$
(here and below, the number in parentheses is the one-sigma uncertainty of the
final digit) and the relative velocities are mainly transverse to the beam
direction with an effective temperature $T_\perp \sim{}$26~K
[$k_BT_\perp=2.25(25)$~meV with $k_B$ being the Boltzmann constant; further
discussed below].  The overlap region is housed in a biased drift tube where the
electrons are decelerated to 11.8~eV, thus reaching the same velocity as the
stored CH$^+$ ions over an effective matched-velocity overlap region of
0.79(1)\,m.  The ion beam is centered within the electron beam and reaches an
effective full width at half-maximum (FWHM) diameter of $<$\,6~mm after 10~s of
phase-space cooling by the electrons \cite{sup}.  The DR rate and its dependence
on the electron--ion collision energy are measured using the MCP detector.  The
peak in the DR rate at the lowest achievable electron--ion collision energy
serves for matching the average ion and electron beam velocities.

CH$^+$ \cite{sup}%
\nocite{hechtfischer_photodissociation_2002, bernath_spectra_2005,
  hakalla_new_2006, hechtfischer_photodissociation_2007, kusunoki_triplet_1980,
  domenech_first_2018, cheng_dipole_2007, amitay_dissociative_1996,
  barinovs_ch_2004, orlov_cold_2003, pastuszka_electron_1996,
  budker_electron_1978, krantz_transverse_2021, van_hoof_accurate_2014, mcmc,
  Foreman_Mackey_2013, mezei_atom_2019, vazquez_rydbergch_2007, drbook} has an $X^1\Sigma^+$ ground state with a rotational
constant of $\sim$\,13.9~cm$^{-1}$ and a dissociation energy
$D_0=32\,946.7(11)$~cm$^{-1}$ \cite{hechtfischer_photodissociation_2002}.  A
metastable $a^3\Pi$ state with $\sim$\,1.2 eV excitation energy and measured
lifetime of $\sim$\,7~s \cite{amitay_dissociative_1996} is strongly populated in
the injected ion beam \cite{sup}. At a storage time $t\gtrsim20$~s, radiative
	relaxation has progressed along the vibrational cascades (all having radiative
	lifetimes $\tau_r < 0.1$~s) while the rotational cascade then has essentially depopulated the
	$J\geq4$ states [$\tau_r < 2$~s for $X^1\Sigma^+(v=0)$]. The population in the metastable $a^3\Pi$ levels was monitored in situ by storage-time-dependent
	molecular-fragment imaging making use of the unique signature of DR from these levels.
	Their fractional populations were found to be $< 0.1$ for $t > 25$~s and $< 0.05$ for $t > 40$~s \cite{sup}. For $t>20$~s, we probe
the populations of the remaining levels, $X^1\Sigma^+(v=0,J\leq3)$, via Feshbach
resonances in near-threshold photodissociation
\cite{oconnor_photodissociation_2016,
  hechtfischer_near-threshold_1998,hechtfischer_photodissociation_2002}.  In
this process, optical transitions excite the $A^1\Pi$ electronic state which
correlates to the atomic channel H $+$ C$^+$($^2P_{3/2}$) with a fine-structure
excitation of 63.42~cm$^{-1}$ \cite{NIST_ASD} and supports vibrational energy
levels above the H $+$ C$^+$($^2P_{1/2}$) channel.  These levels dissociate by
spin--orbit and rotational coupling while being long-lived enough to cause
distinct resonances \cite{hechtfischer_photodissociation_2002} in the
near-threshold photodissociation spectrum.

Considering the initial CH$^+(X^1\Sigma^+,v=0,J)$ levels, the photodissociation
yield as a function of the photon energy has characteristic resonances for each
$J$ [Fig.~\ref{exp-method}(b)].  In our laser probing scheme, we use three
photon wavenumbers $\bar{\nu}_i$ controlled by the OPO laser tuning, each
situated near a $J$ resonance peak for $J\leq2$.  Gated count rates $R_i$ were
accumulated in time windows synchronized with the arrival times of H atoms
produced by the $<$\,10-ns laser pulses (repetition rate 20~s$^{-1}$), where the arrival times were evaluated as the time difference between the trigger signal of a laser pulse and the detection time of a corresponding counted particle. The
background count rate $R_b$ on the MCP from CH$^+$ collisions with residual gas
molecules was recorded in the breaks between the synchronized gates.  The
wavenumbers were cycled and stayed $\sim$\,4~s at each $\bar{\nu}$.  The cycle
included a further wavenumber $\bar{\nu}_{b'}$ below the $J=3$ threshold to
probe a rate $R_{b'}$ for all backgrounds, including any laser-induced one.  The
resonant laser-induced rates were normalized to the measured laser pulse energy
$\epsilon_L$ and to $R_b$ (proportional to the stored ion number), yielding the
normalized signals $S_i=(R_i-R_{b'})/\epsilon_LR_b$.  Based on updated molecular
theory \cite{oconnor_photodissociation_2016} and on a precise study of the
resonance structure for our experimental parameters \cite{sup}, these normalized
signals can be expressed as $S_i= C \sum_J p_J\sigma_J(\bar{\nu}_i)$ with the
known cross sections $\sigma_J(\bar{\nu}_i)$ and an overall normalization $C$.
This relation is solved for the relative rotational populations $p_J$
\cite{sup}.  With electrons absent, we also performed finely spaced scans of
$\bar{\nu}$ to align the experimental peaks with the $J$-dependent theoretical
spectra.  Moreover, the spectra from laser scans at $t=14$--28\,s were fitted
with a linear combination of the calculated $\sigma_J(\bar{\nu})$ to derive
populations $\tilde{p}_J$ ($J\leq3$) \cite{sup} that we consider as the initial
populations at $t=21$\,s for the rotational levels in the $X^1\Sigma^+,v=0$
ground state.

\begin{figure}
  \includegraphics[width=8.5cm]{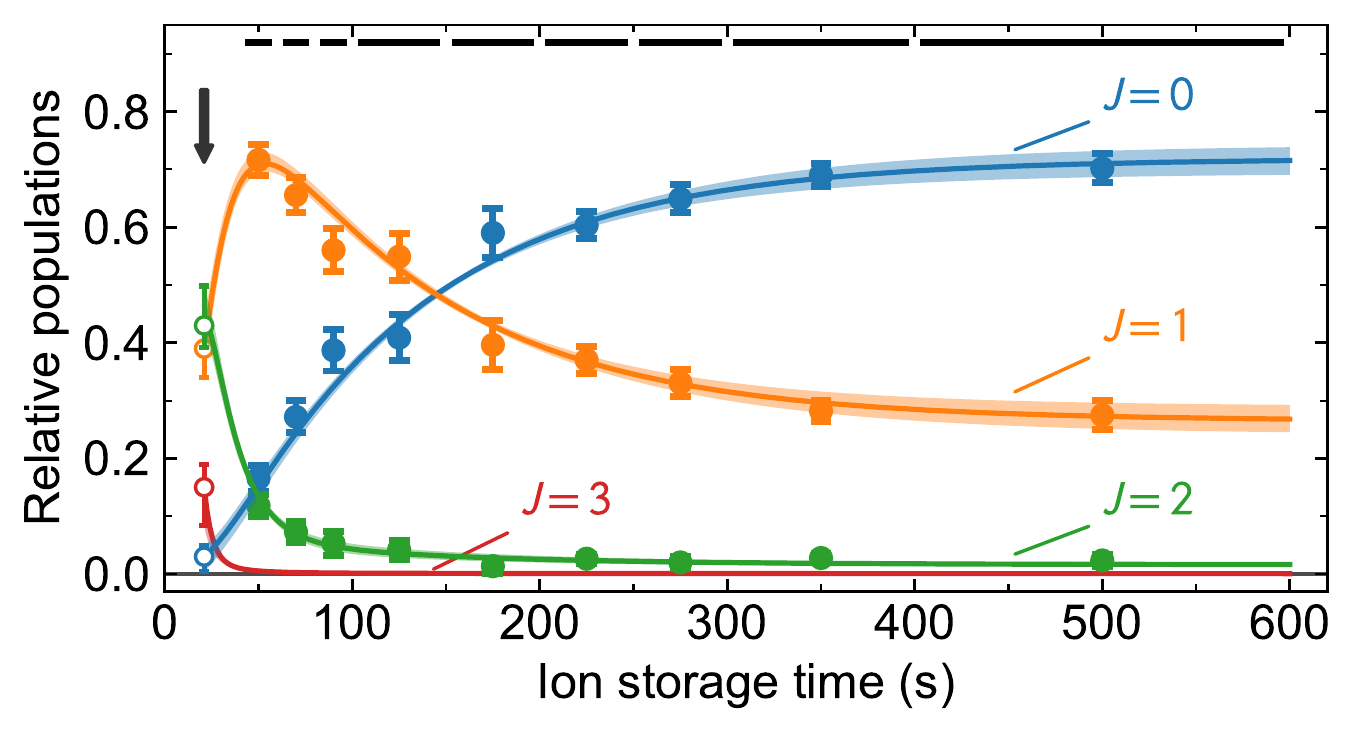}%
  \caption{\label{rad-results} Measured and modeled CH$^+$ rotational
    populations without electron interaction.  Data $p_J$ at $t>40$~s (filled
    symbols and one-sigma uncertainties) were determined from the signals $S_J$
    at the $J$-specific laser-probing wavelengths ($J\leq2$) averaged over the
    probing time windows indicated at the top (horizontal bars).  Specialized
    laser scans (see the text) yielded populations $\tilde{p}_J$ (unfilled
    symbols) at $t=21$~s (arrow) for $J\leq3$.  With these as initial
    conditions, the radiative model was fitted to the later $p_0$ and $p_2$
    ($p_1$ follows as $1-p_0-p_2$) varying $T_{\rm low}$ and $\epsilon$.  Light
    colored bands: one-sigma uncertainties of the fitted model populations
    (including the $\tilde{p}_J$ uncertainties).}
\end{figure}

\begin{figure}
  \includegraphics[width=8.6cm]{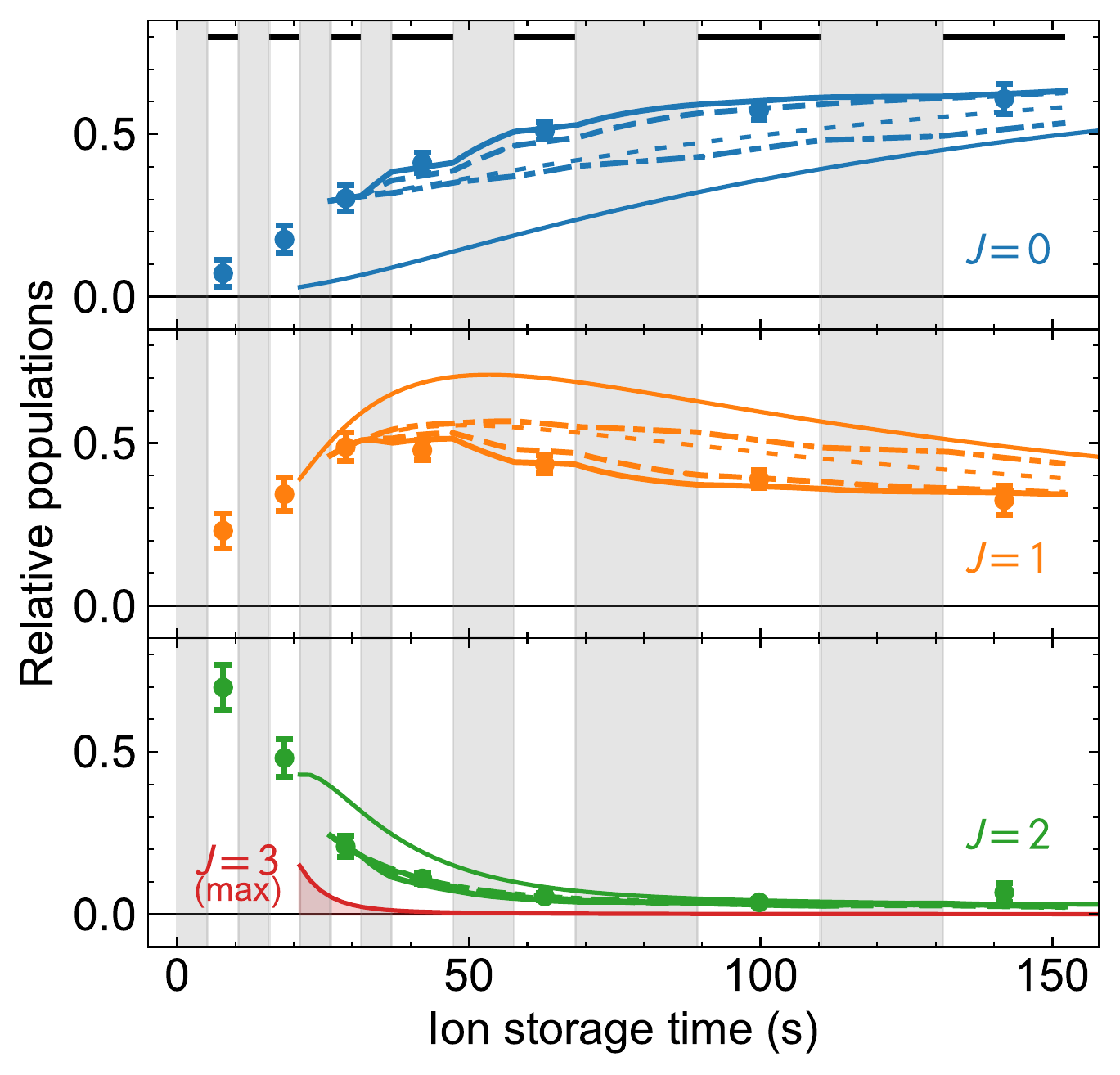}%
  \caption{\label{coll-results} Laser-probed CH$^+$ rotational populations $p_J$
    (symbols) with velocity-matched electrons applied in the shaded time
    windows.  Thin full line: fitted $p_J(t)$ for radiative relaxation only from
    Fig.\ \ref{rad-results}.  The other lines show modeled evolutions starting
    from the $p_J(t=28~{\rm s})$ measured here.  Short-dashed: radiation only;
    dash-dotted: depletion by $J$-dependent DR also included; long-dashed line:
    additionally including inelastic collisions using cross sections
    $\sigma_{J,J'}^{\rm CB}(E)$; thick full line: instead, using the $R$-matrix
    results \cite{hamilton_electron-impact_2016} for the inelastic collisions
    cross sections.  The red line and shaded area show the estimated maximum
    $J=3$ population.  Data points with one-sigma uncertainties.}
\end{figure}

As shown by the symbols in Fig.~\ref{rad-results}, relaxation by radiative
transitions alone was probed for ten windows of $t$ up to radiative equilibrium.
The deduced $p_J(t)$
were fitted by the corresponding $t$-window averages of $p_J$ from
rate-equations which included spontaneous radiative decay as well as the
transitions induced by a modeled blackbody radiation field of the CSR.  In this
radiative model \cite{sup}, the spectral intensity is composed of a cryogenic
component at a temperature $T_{\rm low}$ with a $1-\epsilon$ fraction of the
Planck blackbody spectrum plus a small fraction $\epsilon$ of a room-temperature
(300~K) blackbody spectrum from radiation leaks into CSR.  With the fitted
parameters $T_{\rm low}=14.6(21)$~K and $\epsilon=1.03(28)\times10^{-2}$, the
radiative model well reproduces the measured $p_J(t)$ and is used to predict the
$J$ populations in the stored CH$^+$ ions when collisional cooling by the
electrons is absent. Based on the in situ monitoring, the remaining small population in the metastable $a^3\Pi$ state is neglected.
Assuming a purely thermal radiation field, its effective temperature
$T_r^{\rm eff}$ can be derived directly from the asymptote of the population
ratio between the $J=0$ and 1 levels, $(p_1/p_0)_{\rm eq}=0.39(5)$, which yields
$T_r^{\rm eff}=19.6(11)$~K.

The effect of low-energy electron collisions was then measured
(Fig.~\ref{coll-results}) by merging the velocity-matched electron beam with the
CH$^+$ ions.  Phases with electrons on and off were alternated, with the
electron-off phases serving as laser-probing time windows.  This results in a
duty cycle for the inelastic electron interaction of $\sim$\,50\%.  The MCP
detector was activated for the laser-probing only and turned off when the
electrons were on, as the high DR event rate would saturate the MCP detector.

Compared to electrons off, the $J=0$ level now becomes populated much faster.
At only 30~s of storage, $p_0\approx0.3$ versus $p_0<0.1$ without electrons.
Moreover, the asymptote of the $J=1$ and 0 population ratio, representing the
equilibrium between rotational cooling and excitation, rises significantly to
$(p_1/p_0)_{\rm eq}=0.53(13)$.  Both observations reveal a strong collisional
effect and directly show that in the low-temperature regime the critical
electron density for the CH$^+$ $J=0\to1$ transition is similar to or lower than
our ring-averaged electron density $\bar{n}_e=2.06(21)\times10^4$ cm$^{-3}$.
The larger collisional value of $(p_1/p_0)_{\rm eq}$ indicates a somewhat higher
electron temperature as compared to $T_r^{\rm eff}$.

With the time dependence of the results in Fig.~\ref{coll-results}, we
quantitatively test the electron-collision rate coefficients.  We analyze the
evolution of the $p_J$ starting at $t=28$~s, when the $J=3$ relative population
has dropped to $\lesssim0.05$ and a three-level model ($J\le2$) is adequate.
Even disregarding inelastic collisions, already DR as a potentially $J$
dependent reactive process can influence the $p_J$.  Therefore, in addition to
the radiative rates, we must account for the variation of the $p_J$ due to
differences between the DR rate coefficients for the lowest $J$ levels.  These
were obtained in a separate run within the same experiment, measuring the time
dependence of the merged-beams $J$-averaged DR rate coefficient
$\bar{\alpha}_{\rm DR}^{\rm mb}(t)$ in the velocity-matched beams over 100~s of
storage.  These measurements required the electron energy to be rapidly
alternated from velocity-matching condition to higher values (0.2--4 eV
collision energy).  This led to modified $p_J(t)$ values that, during the times
of detuned electron velocity, were again laser-monitored.  Correlating the
changes of these $p_J(t)$ with those of $\bar{\alpha}_{\rm DR}^{\rm mb}(t)$,
differences between the $J$-specific DR rate coefficients could be derived
\cite{sup}.  The $p_J(t)$ modeled with radiative transitions plus $J$-selective
depletion by DR, but still excluding inelastic collisions, are separately shown
in Fig.\ \ref{coll-results}.

To include the inelastic collisions, cross sections $\sigma_{J,J'}(E)$ as
functions of the collision energy $E$ are used \cite{sup} to determine
merged-beams inelastic rate coefficients $\alpha_{J,J'}^{\rm mb}$, thus adding
the related $J \to J'$ transitions to the rate equations for $p_J(t)$.  We
perform this step for $\sigma_{J,J'}(E)$ calculated in the CB approximation
\cite{boikova_rotational_1968,chu_rotational_1974,*chu_erratum_1975,
  neufeld_electron-impact_1989,sup} as well as from the most recent $R$-matrix
calculations \cite{hamilton_electron-impact_2016}.  The latter are generally
higher than the CB results by a factor of $\sim$\,1.6.  The collision energy
distribution used in the determination of $\alpha_{J,J'}^{\rm mb}$ is modeled
for a set of electron temperatures $k_BT_\perp$ between 1.5 meV and 3.0 meV
\cite{sup}.  The modeled equilibrium population ratio $(p_1/p_0)_{\rm eq}$,
largely governed by $\alpha_{1,0}^{\rm mb}/\alpha_{0,1}^{\rm mb}$, depends
significantly on $k_BT_\perp$, but only weakly on the theoretical cross-section
size.  This size effectively cancels in the ratio
$\alpha_{1,0}^{\rm mb}/\alpha_{0,1}^{\rm mb}$ as the cooling and excitation
cross sections are related through detailed balance.  Comparing \cite{sup} the
model with the measured $p_J$ at $t=142$~s supports the experimental value of
$k_BT_\perp=2.25(25)$~meV [$T_\perp=26(3)$~K].

\begin{figure}
  \includegraphics[width=8.6cm]{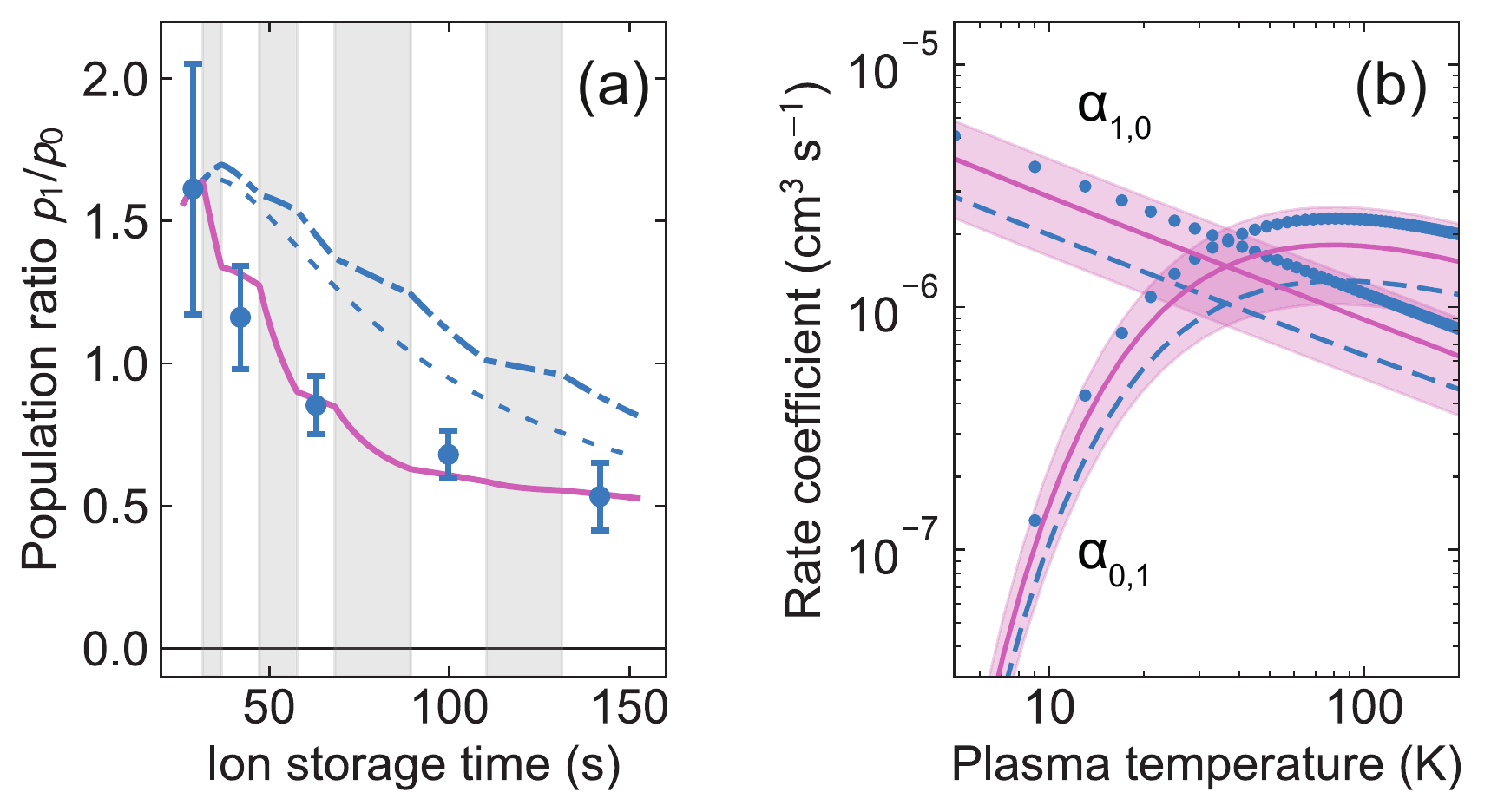}%
  \caption{\label{ratio-results} (a) Population ratios $p_1/p_0$ corresponding
    to Fig.~\ref{coll-results} (electrons applied in shaded time windows) with
    modeled evolutions starting from the measured populations at $t=28$~s.
    Short-dashed: radiation only; dash-dotted: depletion by $J$-dependent DR
    also included; full line: additionally including inelastic collisions with
    the simplified $\hat{\sigma}_{J,J'}(E)$ and fitting for
    $\hat{\sigma}^0_{0,1}$.  (b) Plasma rate coefficients $\alpha_{1,0}(T)$ and
    $\alpha_{0,1}(T)$ as constrained by the fitted $\hat{\sigma}^0_{0,1}$ (full
    lines and shaded one-sigma uncertainty), compared to $R$-matrix theory
    \cite{hamilton_electron-impact_2016} (coalescing dots) and to the CB
    calculation (dashed line).}
\end{figure}

The data measured at intermediate times for $p_{0}(t)$ and $p_{1}(t)$ can then
be compared to the model calculations employing the experimental $T_\perp$.
This probes the absolute values of the inelastic rate coefficients, dominantly
$\alpha_{1,0}^{\rm mb}$ and $\alpha_{0,1}^{\rm mb}$.  As seen in Fig.\
\ref{coll-results} the data are remarkably well reproduced when using the
$R$-matrix cross sections; slightly worse agreement is found for the CB
approximation.

Comparison with theory is also possible considering the thermal rate
coefficients implied by the experiment.  For the relevant collision energies,
the $J$-changing cross sections for an energy level pair spaced by
$\Delta E_{J,J'}$ are well approximated by a simplified, single-parameter
expression.  For excitation ($J'>J$), the approximation is
$\hat{\sigma}_{J,J'}(E) = \hat{\sigma}^0_{J,J'}\Delta E_{J,J'}/E$ (where
$E>\Delta E_{J,J'}$) with the threshold value $\hat{\sigma}^0_{J,J'}$ as only
parameter.  For de-excitation, detailed balance then yields, at all energies,
$\hat{\sigma}_{J',J}(E) = \hat{g}\hat{\sigma}^0_{J,J'}\Delta E_{J,J'}/E$ with
$\hat{g}=(2J+1)/(2J'+1)$.  An experimental value for $\hat{\sigma}^0_{0,1}$ can
be obtained from the time dependence of the ratio $p_1/p_0$ starting at $t=28$~s
[Fig.~\ref{ratio-results}(a)].  A fit of the parameter in the simplified model
yields
$\hat{\sigma}^0_{0,1}=1.07(23)_{\rm fit}(40)_{\rm syst}\times10^4$~\AA$^2$,
where the first uncertainty stems from the fitting and the second one from the
uncertainties in $T_{\perp}$, the initial condition at $t=28$~s, and
$\bar{n}_e$.  (Simplified $\hat{\sigma}_{J,J+1}(E)$ and
$\hat{\sigma}_{J+1,J}(E)$ for other level pairs were included in the model with
their $\hat{\sigma}^0_{J,J+1}$ matching Ref.\
\cite{hamilton_electron-impact_2016}, but reducing them to 0 changed the fit
result by only $<0.02\times10^4$~\AA$^2$.)  The simplified cross sections with
the experimental $\hat{\sigma}^0_{0,1}$ and the thermal electron energy
distribution at a kinetic temperature $T$ served to obtain the plasma rate
coefficients $\alpha_{1,0}(T)$ and $\alpha_{0,1}(T)$, shown in
Fig.~\ref{ratio-results}(b).  The corresponding $R$-matrix results
\cite{hamilton_electron-impact_2016} {and the CB approximation} lie well within
the experimental one-sigma uncertainty range of {$\pm$43\%}.

This demonstrates the use of a merged-beams electron target in a cryogenic
storage ring for quantitative probing of $J$-changing electron collisions
between the lowest rotational levels in small molecular ions.  The experimental
method for rotational state probing used here, employing $J$-dependent
resonances in single-step photodissociation, generally cannot be transferred to
other ion species.  However, effort is underway \cite{znotins_approach_2021} to
develop $J$-dependent photodissociation schemes for additional small ions in
cryogenic storage rings.  For most systems, such schemes will include
intermediate laser excitation steps similar to the resonant multiphoton
dissociation process earlier realized in ion traps
\cite{roth_rovibrational_2006}.  It will then be possible to extend the
measurements to other systems crucial for spectrocopic probing of the
interstellar medium, such as HCO$^+$.  The CSR merged-beams setup offers more
options for controlling and probing the electron collisions, beyond what has
been realized in the present survey.  Thus, future experiments may focus on
precisely probing the change of $p_{J}$ in a given electron interaction period
or apply detuned average velocities of the electron and ion beam. This will allow electron impact excitation and de-excitation cross sections of molecular cations to be investigated in more detail, over a wider range of electron collision energies relevant not only for astrophysical environments, and also for molecular plasma conditions in general.
\begin{acknowledgments}
  Financial support by the Max Planck Society is acknowledged.  A.K. and
  D.W.S. were supported in part by the U.S.\ National Science Foundation
  Division of Astronomical Sciences Astronomy and Astrophysics Grants program
  under AST-1907188.  We thank C.J. Williams for the NIST close coupling code
  used in the resonant photodissociation calculations and A. Faure for supplying
  the $R$-matrix cross sections underlying the rate coefficients published in
  Ref.\ \cite{hamilton_electron-impact_2016}.
\end{acknowledgments}


\begin{thebibliography}{25}%
\makeatletter
\providecommand \@ifxundefined [1]{%
 \@ifx{#1\undefined}
}%
\providecommand \@ifnum [1]{%
 \ifnum #1\expandafter \@firstoftwo
 \else \expandafter \@secondoftwo
 \fi
}%
\providecommand \@ifx [1]{%
 \ifx #1\expandafter \@firstoftwo
 \else \expandafter \@secondoftwo
 \fi
}%
\providecommand \natexlab [1]{#1}%
\providecommand \enquote  [1]{``#1''}%
\providecommand \bibnamefont  [1]{#1}%
\providecommand \bibfnamefont [1]{#1}%
\providecommand \citenamefont [1]{#1}%
\providecommand \href@noop [0]{\@secondoftwo}%
\providecommand \href [0]{\begingroup \@sanitize@url \@href}%
\providecommand \@href[1]{\@@startlink{#1}\@@href}%
\providecommand \@@href[1]{\endgroup#1\@@endlink}%
\providecommand \@sanitize@url [0]{\catcode `\\12\catcode `\$12\catcode
  `\&12\catcode `\#12\catcode `\^12\catcode `\_12\catcode `\%12\relax}%
\providecommand \@@startlink[1]{}%
\providecommand \@@endlink[0]{}%
\providecommand \url  [0]{\begingroup\@sanitize@url \@url }%
\providecommand \@url [1]{\endgroup\@href {#1}{\urlprefix }}%
\providecommand \urlprefix  [0]{URL }%
\providecommand \Eprint [0]{\href }%
\providecommand \doibase [0]{https://doi.org/}%
\providecommand \selectlanguage [0]{\@gobble}%
\providecommand \bibinfo  [0]{\@secondoftwo}%
\providecommand \bibfield  [0]{\@secondoftwo}%
\providecommand \translation [1]{[#1]}%
\providecommand \BibitemOpen [0]{}%
\providecommand \bibitemStop [0]{}%
\providecommand \bibitemNoStop [0]{.\EOS\space}%
\providecommand \EOS [0]{\spacefactor3000\relax}%
\providecommand \BibitemShut  [1]{\csname bibitem#1\endcsname}%
\let\auto@bib@innerbib\@empty
\bibitem [{\citenamefont {Hechtfischer}\ \emph {et~al.}(2002)\citenamefont
  {Hechtfischer}, \citenamefont {Williams}, \citenamefont {Lange},
  \citenamefont {Linkemann}, \citenamefont {Schwalm}, \citenamefont {Wester},
  \citenamefont {Wolf},\ and\ \citenamefont
  {Zajfman}}]{hechtfischer_photodissociation_2002}%
  \BibitemOpen
  \bibfield  {author} {\bibinfo {author} {\bibfnamefont {U.}~\bibnamefont
  {Hechtfischer}}, \bibinfo {author} {\bibfnamefont {C.~J.}\ \bibnamefont
  {Williams}}, \bibinfo {author} {\bibfnamefont {M.}~\bibnamefont {Lange}},
  \bibinfo {author} {\bibfnamefont {J.}~\bibnamefont {Linkemann}}, \bibinfo
  {author} {\bibfnamefont {D.}~\bibnamefont {Schwalm}}, \bibinfo {author}
  {\bibfnamefont {R.}~\bibnamefont {Wester}}, \bibinfo {author} {\bibfnamefont
  {A.}~\bibnamefont {Wolf}},\ and\ \bibinfo {author} {\bibfnamefont
  {D.}~\bibnamefont {Zajfman}},\ }\bibfield  {title} {\bibinfo {title}
  {Photodissociation spectroscopy of stored {CH$^+$} ions: {Detection},
  assignment, and close-coupled modeling of near-threshold {Feshbach}
  resonances},\ }\href {https://doi.org/10.1063/1.1513459} {\bibfield
  {journal} {\bibinfo  {journal} {J. Chem. Phys.}\ }\textbf {\bibinfo {volume}
  {117}},\ \bibinfo {pages} {8754} (\bibinfo {year} {2002})}\BibitemShut
  {NoStop}%
\bibitem [{\citenamefont {Bernath}(2005)}]{bernath_spectra_2005}%
  \BibitemOpen
  \bibfield  {author} {\bibinfo {author} {\bibfnamefont {P.~F.}\ \bibnamefont
  {Bernath}},\ }\href@noop {} {\emph {\bibinfo {title} {Spectra of {Atoms} and
  {Molecules}}}},\ \bibinfo {edition} {2nd}\ ed.\ (\bibinfo  {publisher}
  {Oxford University Press},\ \bibinfo {address} {New York},\ \bibinfo {year}
  {2005})\BibitemShut {NoStop}%
\bibitem [{\citenamefont {Hakalla}\ \emph {et~al.}(2006)\citenamefont
  {Hakalla}, \citenamefont {K\c{e}pa}, \citenamefont {Szajna},\ and\
  \citenamefont {Zachwieja}}]{hakalla_new_2006}%
  \BibitemOpen
  \bibfield  {author} {\bibinfo {author} {\bibfnamefont {R.}~\bibnamefont
  {Hakalla}}, \bibinfo {author} {\bibfnamefont {R.}~\bibnamefont {K\c{e}pa}},
  \bibinfo {author} {\bibfnamefont {W.}~\bibnamefont {Szajna}},\ and\ \bibinfo
  {author} {\bibfnamefont {M.}~\bibnamefont {Zachwieja}},\ }\bibfield  {title}
  {\bibinfo {title} {New analysis of the {Douglas}-{Herzberg} system
  {($A^1\Pi$-$X^1\Sigma^+$)} in the {CH$^+$} ion radical},\ }\href
  {https://doi.org/10.1140/epjd/e2006-00063-9} {\bibfield  {journal} {\bibinfo
  {journal} {Eur. Phys. J. D}\ }\textbf {\bibinfo {volume} {38}},\ \bibinfo
  {pages} {481} (\bibinfo {year} {2006})}\BibitemShut {NoStop}%
\bibitem [{\citenamefont {Hechtfischer}\ \emph {et~al.}(2007)\citenamefont
  {Hechtfischer}, \citenamefont {Rostas}, \citenamefont {Lange}, \citenamefont
  {Linkemann}, \citenamefont {Schwalm}, \citenamefont {Wester}, \citenamefont
  {Wolf},\ and\ \citenamefont {Zajfman}}]{hechtfischer_photodissociation_2007}%
  \BibitemOpen
  \bibfield  {author} {\bibinfo {author} {\bibfnamefont {U.}~\bibnamefont
  {Hechtfischer}}, \bibinfo {author} {\bibfnamefont {J.}~\bibnamefont
  {Rostas}}, \bibinfo {author} {\bibfnamefont {M.}~\bibnamefont {Lange}},
  \bibinfo {author} {\bibfnamefont {J.}~\bibnamefont {Linkemann}}, \bibinfo
  {author} {\bibfnamefont {D.}~\bibnamefont {Schwalm}}, \bibinfo {author}
  {\bibfnamefont {R.}~\bibnamefont {Wester}}, \bibinfo {author} {\bibfnamefont
  {A.}~\bibnamefont {Wolf}},\ and\ \bibinfo {author} {\bibfnamefont
  {D.}~\bibnamefont {Zajfman}},\ }\bibfield  {title} {\bibinfo {title}
  {Photodissociation spectroscopy of stored {CH$^+$} and {CD$^+$} ions:
  {Analysis} of the {$b^3\Sigma^-$--$a^3\Pi$} system},\ }\href
  {https://doi.org/10.1063/1.2800004} {\bibfield  {journal} {\bibinfo
  {journal} {J. Chem. Phys.}\ }\textbf {\bibinfo {volume} {127}},\ \bibinfo
  {pages} {204304} (\bibinfo {year} {2007})}\BibitemShut {NoStop}%
\bibitem [{\citenamefont {Kusunoki}\ and\ \citenamefont
  {Ottinger}(1980)}]{kusunoki_triplet_1980}%
  \BibitemOpen
  \bibfield  {author} {\bibinfo {author} {\bibfnamefont {I.}~\bibnamefont
  {Kusunoki}}\ and\ \bibinfo {author} {\bibfnamefont {C.}~\bibnamefont
  {Ottinger}},\ }\bibfield  {title} {\bibinfo {title} {Triplet {CH$^+$(CD$^+$)}
  emission from chemiluminescent ion--molecule reaction {C$^+$($^4P$)+
  H$_2$(D$_2$)}},\ }\href {https://doi.org/10.1063/1.440401} {\bibfield
  {journal} {\bibinfo  {journal} {J. Chem. Phys.}\ }\textbf {\bibinfo {volume}
  {73}},\ \bibinfo {pages} {2069} (\bibinfo {year} {1980})}\BibitemShut
  {NoStop}%
\bibitem [{\citenamefont {Dom\'{e}nech}\ \emph {et~al.}(2018)\citenamefont
  {Dom\'{e}nech}, \citenamefont {Jusko}, \citenamefont {Schlemmer},\ and\
  \citenamefont {Asvany}}]{domenech_first_2018}%
  \BibitemOpen
  \bibfield  {author} {\bibinfo {author} {\bibfnamefont {J.~L.}\ \bibnamefont
  {Dom\'{e}nech}}, \bibinfo {author} {\bibfnamefont {P.}~\bibnamefont {Jusko}},
  \bibinfo {author} {\bibfnamefont {S.}~\bibnamefont {Schlemmer}},\ and\
  \bibinfo {author} {\bibfnamefont {O.}~\bibnamefont {Asvany}},\ }\bibfield
  {title} {\bibinfo {title} {The first laboratory detection of
  vibration-rotation transitions of $^{\textrm{12}}${CH}$^{\textrm{+}}$ and
  $^{\textrm{13}}${CH}$^{\textrm{+}}$ and improved measurement of their
  rotational transition frequencies},\ }\href
  {https://doi.org/10.3847/1538-4357/aab36a} {\bibfield  {journal} {\bibinfo
  {journal} {Astrophys. J.}\ }\textbf {\bibinfo {volume} {857}},\ \bibinfo
  {pages} {61} (\bibinfo {year} {2018})}\BibitemShut {NoStop}%
\bibitem [{\citenamefont {Meyer}\ \emph {et~al.}(2017)\citenamefont {Meyer},
  \citenamefont {Becker}, \citenamefont {Blaum}, \citenamefont {Breitenfeldt},
  \citenamefont {George}, \citenamefont {G\"ock}, \citenamefont {Grieser},
  \citenamefont {Grussie}, \citenamefont {Guerin}, \citenamefont {von Hahn},
  \citenamefont {Herwig}, \citenamefont {Krantz}, \citenamefont {Kreckel},
  \citenamefont {Lion}, \citenamefont {Lohmann}, \citenamefont {Mishra},
  \citenamefont {Novotn\'y}, \citenamefont {O'Connor}, \citenamefont {Repnow},
  \citenamefont {Saurabh}, \citenamefont {Schwalm}, \citenamefont
  {Schweikhard}, \citenamefont {Spruck}, \citenamefont {Sunil~Kumar},
  \citenamefont {Vogel},\ and\ \citenamefont {Wolf}}]{meyer_radiative_2017}%
  \BibitemOpen
  \bibfield  {author} {\bibinfo {author} {\bibfnamefont {C.}~\bibnamefont
  {Meyer}}, \bibinfo {author} {\bibfnamefont {A.}~\bibnamefont {Becker}},
  \bibinfo {author} {\bibfnamefont {K.}~\bibnamefont {Blaum}}, \bibinfo
  {author} {\bibfnamefont {C.}~\bibnamefont {Breitenfeldt}}, \bibinfo {author}
  {\bibfnamefont {S.}~\bibnamefont {George}}, \bibinfo {author} {\bibfnamefont
  {J.}~\bibnamefont {G\"ock}}, \bibinfo {author} {\bibfnamefont
  {M.}~\bibnamefont {Grieser}}, \bibinfo {author} {\bibfnamefont
  {F.}~\bibnamefont {Grussie}}, \bibinfo {author} {\bibfnamefont
  {E.}~\bibnamefont {Guerin}}, \bibinfo {author} {\bibfnamefont
  {R.}~\bibnamefont {von Hahn}}, \bibinfo {author} {\bibfnamefont
  {P.}~\bibnamefont {Herwig}}, \bibinfo {author} {\bibfnamefont
  {C.}~\bibnamefont {Krantz}}, \bibinfo {author} {\bibfnamefont
  {H.}~\bibnamefont {Kreckel}}, \bibinfo {author} {\bibfnamefont
  {J.}~\bibnamefont {Lion}}, \bibinfo {author} {\bibfnamefont {S.}~\bibnamefont
  {Lohmann}}, \bibinfo {author} {\bibfnamefont {P.}~\bibnamefont {Mishra}},
  \bibinfo {author} {\bibfnamefont {O.}~\bibnamefont {Novotn\'y}}, \bibinfo
  {author} {\bibfnamefont {A.}~\bibnamefont {O'Connor}}, \bibinfo {author}
  {\bibfnamefont {R.}~\bibnamefont {Repnow}}, \bibinfo {author} {\bibfnamefont
  {S.}~\bibnamefont {Saurabh}}, \bibinfo {author} {\bibfnamefont
  {D.}~\bibnamefont {Schwalm}}, \bibinfo {author} {\bibfnamefont
  {L.}~\bibnamefont {Schweikhard}}, \bibinfo {author} {\bibfnamefont
  {K.}~\bibnamefont {Spruck}}, \bibinfo {author} {\bibfnamefont
  {S.}~\bibnamefont {Sunil~Kumar}}, \bibinfo {author} {\bibfnamefont
  {S.}~\bibnamefont {Vogel}},\ and\ \bibinfo {author} {\bibfnamefont
  {A.}~\bibnamefont {Wolf}},\ }\bibfield  {title} {\bibinfo {title} {Radiative
  rotational lifetimes and state-resolved relative detachment cross sections
  from photodetachment thermometry of molecular anions in a cryogenic storage
  ring},\ }\href {https://doi.org/10.1103/PhysRevLett.119.023202} {\bibfield
  {journal} {\bibinfo  {journal} {Phys. Rev. Lett.}\ }\textbf {\bibinfo
  {volume} {119}},\ \bibinfo {pages} {023202} (\bibinfo {year}
  {2017})}\BibitemShut {NoStop}%
\bibitem [{\citenamefont {Cheng}\ \emph {et~al.}(2007)\citenamefont {Cheng},
  \citenamefont {Brown}, \citenamefont {Rosmus}, \citenamefont {Linguerri},
  \citenamefont {Komiha},\ and\ \citenamefont {Myers}}]{cheng_dipole_2007}%
  \BibitemOpen
  \bibfield  {author} {\bibinfo {author} {\bibfnamefont {M.}~\bibnamefont
  {Cheng}}, \bibinfo {author} {\bibfnamefont {J.~M.}\ \bibnamefont {Brown}},
  \bibinfo {author} {\bibfnamefont {P.}~\bibnamefont {Rosmus}}, \bibinfo
  {author} {\bibfnamefont {R.}~\bibnamefont {Linguerri}}, \bibinfo {author}
  {\bibfnamefont {N.}~\bibnamefont {Komiha}},\ and\ \bibinfo {author}
  {\bibfnamefont {E.~G.}\ \bibnamefont {Myers}},\ }\bibfield  {title} {\bibinfo
  {title} {Dipole moments and orientation polarizabilities of diatomic
  molecular ions for precision atomic mass measurement},\ }\href
  {https://doi.org/10.1103/PhysRevA.75.012502} {\bibfield  {journal} {\bibinfo
  {journal} {Phys. Rev. A}\ }\textbf {\bibinfo {volume} {75}},\ \bibinfo
  {pages} {012502} (\bibinfo {year} {2007})}\BibitemShut {NoStop}%
\bibitem [{\citenamefont {Amitay}\ \emph {et~al.}(1996)\citenamefont {Amitay},
  \citenamefont {Zajfman}, \citenamefont {Forck}, \citenamefont {Hechtfischer},
  \citenamefont {Seidel}, \citenamefont {Grieser}, \citenamefont {Habs},
  \citenamefont {Repnow}, \citenamefont {Schwalm},\ and\ \citenamefont
  {Wolf}}]{amitay_dissociative_1996}%
  \BibitemOpen
  \bibfield  {author} {\bibinfo {author} {\bibfnamefont {Z.}~\bibnamefont
  {Amitay}}, \bibinfo {author} {\bibfnamefont {D.}~\bibnamefont {Zajfman}},
  \bibinfo {author} {\bibfnamefont {P.}~\bibnamefont {Forck}}, \bibinfo
  {author} {\bibfnamefont {U.}~\bibnamefont {Hechtfischer}}, \bibinfo {author}
  {\bibfnamefont {B.}~\bibnamefont {Seidel}}, \bibinfo {author} {\bibfnamefont
  {M.}~\bibnamefont {Grieser}}, \bibinfo {author} {\bibfnamefont
  {D.}~\bibnamefont {Habs}}, \bibinfo {author} {\bibfnamefont {R.}~\bibnamefont
  {Repnow}}, \bibinfo {author} {\bibfnamefont {D.}~\bibnamefont {Schwalm}},\
  and\ \bibinfo {author} {\bibfnamefont {A.}~\bibnamefont {Wolf}},\ }\bibfield
  {title} {\bibinfo {title} {Dissociative recombination of {CH$^+$}: {Cross}
  section and final states},\ }\href {https://doi.org/10.1103/PhysRevA.54.4032}
  {\bibfield  {journal} {\bibinfo  {journal} {Phys. Rev. A}\ }\textbf {\bibinfo
  {volume} {54}},\ \bibinfo {pages} {4032} (\bibinfo {year}
  {1996})}\BibitemShut {NoStop}%
\bibitem [{\citenamefont {O'Connor}\ \emph {et~al.}(2016)\citenamefont
  {O'Connor}, \citenamefont {Becker}, \citenamefont {Blaum}, \citenamefont
  {Breitenfeldt}, \citenamefont {George}, \citenamefont {G\"ock}, \citenamefont
  {Grieser}, \citenamefont {Grussie}, \citenamefont {Guerin}, \citenamefont
  {von Hahn}, \citenamefont {Hechtfischer}, \citenamefont {Herwig},
  \citenamefont {Karthein}, \citenamefont {Krantz}, \citenamefont {Kreckel},
  \citenamefont {Lohmann}, \citenamefont {Meyer}, \citenamefont {Mishra},
  \citenamefont {Novotn\'y}, \citenamefont {Repnow}, \citenamefont {Saurabh},
  \citenamefont {Schwalm}, \citenamefont {Spruck}, \citenamefont {Sunil~Kumar},
  \citenamefont {Vogel},\ and\ \citenamefont
  {Wolf}}]{oconnor_photodissociation_2016}%
  \BibitemOpen
  \bibfield  {author} {\bibinfo {author} {\bibfnamefont {A.}~\bibnamefont
  {O'Connor}}, \bibinfo {author} {\bibfnamefont {A.}~\bibnamefont {Becker}},
  \bibinfo {author} {\bibfnamefont {K.}~\bibnamefont {Blaum}}, \bibinfo
  {author} {\bibfnamefont {C.}~\bibnamefont {Breitenfeldt}}, \bibinfo {author}
  {\bibfnamefont {S.}~\bibnamefont {George}}, \bibinfo {author} {\bibfnamefont
  {J.}~\bibnamefont {G\"ock}}, \bibinfo {author} {\bibfnamefont
  {M.}~\bibnamefont {Grieser}}, \bibinfo {author} {\bibfnamefont
  {F.}~\bibnamefont {Grussie}}, \bibinfo {author} {\bibfnamefont
  {E.}~\bibnamefont {Guerin}}, \bibinfo {author} {\bibfnamefont
  {R.}~\bibnamefont {von Hahn}}, \bibinfo {author} {\bibfnamefont
  {U.}~\bibnamefont {Hechtfischer}}, \bibinfo {author} {\bibfnamefont
  {P.}~\bibnamefont {Herwig}}, \bibinfo {author} {\bibfnamefont
  {J.}~\bibnamefont {Karthein}}, \bibinfo {author} {\bibfnamefont
  {C.}~\bibnamefont {Krantz}}, \bibinfo {author} {\bibfnamefont
  {H.}~\bibnamefont {Kreckel}}, \bibinfo {author} {\bibfnamefont
  {S.}~\bibnamefont {Lohmann}}, \bibinfo {author} {\bibfnamefont
  {C.}~\bibnamefont {Meyer}}, \bibinfo {author} {\bibfnamefont
  {P.}~\bibnamefont {Mishra}}, \bibinfo {author} {\bibfnamefont
  {O.}~\bibnamefont {Novotn\'y}}, \bibinfo {author} {\bibfnamefont
  {R.}~\bibnamefont {Repnow}}, \bibinfo {author} {\bibfnamefont
  {S.}~\bibnamefont {Saurabh}}, \bibinfo {author} {\bibfnamefont
  {D.}~\bibnamefont {Schwalm}}, \bibinfo {author} {\bibfnamefont
  {K.}~\bibnamefont {Spruck}}, \bibinfo {author} {\bibfnamefont
  {S.}~\bibnamefont {Sunil~Kumar}}, \bibinfo {author} {\bibfnamefont
  {S.}~\bibnamefont {Vogel}},\ and\ \bibinfo {author} {\bibfnamefont
  {A.}~\bibnamefont {Wolf}},\ }\bibfield  {title} {\bibinfo {title}
  {Photodissociation of an internally cold beam of {CH$^+$} ions in a cryogenic
  storage ring},\ }\href {https://doi.org/10.1103/PhysRevLett.116.113002}
  {\bibfield  {journal} {\bibinfo  {journal} {Phys. Rev. Lett.}\ }\textbf
  {\bibinfo {volume} {116}},\ \bibinfo {pages} {113002} (\bibinfo {year}
  {2016})}\BibitemShut {NoStop}%
\bibitem [{\citenamefont {Barinovs}\ and\ \citenamefont {van
  Hemert}(2004)}]{barinovs_ch_2004}%
  \BibitemOpen
  \bibfield  {author} {\bibinfo {author} {\bibfnamefont {{\u{G}}.}~\bibnamefont
  {Barinovs}}\ and\ \bibinfo {author} {\bibfnamefont {M.~C.}\ \bibnamefont {van
  Hemert}},\ }\bibfield  {title} {\bibinfo {title} {{CH$^+$} potential energy
  curves and photodissociation cross-section},\ }\href
  {https://doi.org/10.1016/j.cplett.2004.10.035} {\bibfield  {journal}
  {\bibinfo  {journal} {Chem. Phys. Lett.}\ }\textbf {\bibinfo {volume}
  {399}},\ \bibinfo {pages} {406} (\bibinfo {year} {2004})}\BibitemShut
  {NoStop}%
\bibitem [{\citenamefont {Orlov}\ \emph {et~al.}(2003)\citenamefont {Orlov},
  \citenamefont {Weigel}, \citenamefont {Hoppe}, \citenamefont {Schwalm},
  \citenamefont {Jaroshevich}, \citenamefont {Terekhov},\ and\ \citenamefont
  {Wolf}}]{orlov_cold_2003}%
  \BibitemOpen
  \bibfield  {author} {\bibinfo {author} {\bibfnamefont {D.~A.}\ \bibnamefont
  {Orlov}}, \bibinfo {author} {\bibfnamefont {U.}~\bibnamefont {Weigel}},
  \bibinfo {author} {\bibfnamefont {M.}~\bibnamefont {Hoppe}}, \bibinfo
  {author} {\bibfnamefont {D.}~\bibnamefont {Schwalm}}, \bibinfo {author}
  {\bibfnamefont {A.~S.}\ \bibnamefont {Jaroshevich}}, \bibinfo {author}
  {\bibfnamefont {A.~S.}\ \bibnamefont {Terekhov}},\ and\ \bibinfo {author}
  {\bibfnamefont {A.}~\bibnamefont {Wolf}},\ }\bibfield  {title} {\bibinfo
  {title} {Cold electrons from cryogenic {GaAs} photocathodes: {Energetic} and
  angular distributions},\ }\href
  {https://doi.org/10.1023/B:HYPE.0000004226.23869.2b} {\bibfield  {journal}
  {\bibinfo  {journal} {Hyperfine Interact.}\ }\textbf {\bibinfo {volume}
  {146/147}},\ \bibinfo {pages} {215} (\bibinfo {year} {2003})}\BibitemShut
  {NoStop}%
\bibitem [{\citenamefont {Pastuszka}\ \emph {et~al.}(1996)\citenamefont
  {Pastuszka}, \citenamefont {Schramm}, \citenamefont {Grieser}, \citenamefont
  {Broude}, \citenamefont {Grimm}, \citenamefont {Habs}, \citenamefont
  {Kenntner}, \citenamefont {Miesner}, \citenamefont {Sch\"{u}\ss{}ler},
  \citenamefont {Schwalm},\ and\ \citenamefont
  {Wolf}}]{pastuszka_electron_1996}%
  \BibitemOpen
  \bibfield  {author} {\bibinfo {author} {\bibfnamefont {S.}~\bibnamefont
  {Pastuszka}}, \bibinfo {author} {\bibfnamefont {U.}~\bibnamefont {Schramm}},
  \bibinfo {author} {\bibfnamefont {M.}~\bibnamefont {Grieser}}, \bibinfo
  {author} {\bibfnamefont {C.}~\bibnamefont {Broude}}, \bibinfo {author}
  {\bibfnamefont {R.}~\bibnamefont {Grimm}}, \bibinfo {author} {\bibfnamefont
  {D.}~\bibnamefont {Habs}}, \bibinfo {author} {\bibfnamefont {J.}~\bibnamefont
  {Kenntner}}, \bibinfo {author} {\bibfnamefont {H.-J.}\ \bibnamefont
  {Miesner}}, \bibinfo {author} {\bibfnamefont {T.}~\bibnamefont
  {Sch\"{u}\ss{}ler}}, \bibinfo {author} {\bibfnamefont {D.}~\bibnamefont
  {Schwalm}},\ and\ \bibinfo {author} {\bibfnamefont {A.}~\bibnamefont
  {Wolf}},\ }\bibfield  {title} {\bibinfo {title} {Electron cooling and
  recombination experiments with an adiabatically expanded electron beam},\
  }\href {https://doi.org/10.1016/0168-9002(95)00786-5} {\bibfield  {journal}
  {\bibinfo  {journal} {Nucl. Instrum. Methods Phys. Res. Sect. A -- Accel.
  Spectrom. Dect. Assoc. Equip.}\ }\textbf {\bibinfo {volume} {369}},\ \bibinfo
  {pages} {11} (\bibinfo {year} {1996})}\BibitemShut {NoStop}%
\bibitem [{bud()}]{budker_electron_1978}%
  \BibitemOpen
  \href@noop {} {}\bibinfo {note} {G. I. Budker and A. N. Skrinski\u{\i},
  Electron cooling and new possibilities in elementary particle physics, Usp.
  Fiz. Nauk {\bf 124}, 561--595 (1978)
  \href{https://iopscience.iop.org/article/10.1070/PU1978v021n04ABEH005537/meta}
  {[Sov. Phys. Usp. {\bf 21}, 277--296 (1986)]}}\BibitemShut {NoStop}%
\bibitem [{\citenamefont {Krantz}\ \emph {et~al.}(2021)\citenamefont {Krantz},
  \citenamefont {Buhr}, \citenamefont {Grieser}, \citenamefont {Lestinsky},
  \citenamefont {Novotn\'y}, \citenamefont {Novotny}, \citenamefont {Orlov},
  \citenamefont {Repnow}, \citenamefont {Terekhov}, \citenamefont {Wilhelm},\
  and\ \citenamefont {Wolf}}]{krantz_transverse_2021}%
  \BibitemOpen
  \bibfield  {author} {\bibinfo {author} {\bibfnamefont {C.}~\bibnamefont
  {Krantz}}, \bibinfo {author} {\bibfnamefont {H.}~\bibnamefont {Buhr}},
  \bibinfo {author} {\bibfnamefont {M.}~\bibnamefont {Grieser}}, \bibinfo
  {author} {\bibfnamefont {M.}~\bibnamefont {Lestinsky}}, \bibinfo {author}
  {\bibfnamefont {O.}~\bibnamefont {Novotn\'y}}, \bibinfo {author}
  {\bibfnamefont {S.}~\bibnamefont {Novotny}}, \bibinfo {author} {\bibfnamefont
  {D.}~\bibnamefont {Orlov}}, \bibinfo {author} {\bibfnamefont
  {R.}~\bibnamefont {Repnow}}, \bibinfo {author} {\bibfnamefont
  {A.}~\bibnamefont {Terekhov}}, \bibinfo {author} {\bibfnamefont
  {P.}~\bibnamefont {Wilhelm}},\ and\ \bibinfo {author} {\bibfnamefont
  {A.}~\bibnamefont {Wolf}},\ }\bibfield  {title} {\bibinfo {title} {Transverse
  electron cooling of heavy molecular ions},\ }\href
  {https://doi.org/10.1103/PhysRevAccelBeams.24.050101} {\bibfield  {journal}
  {\bibinfo  {journal} {Phys. Rev. Accel. Beams}\ }\textbf {\bibinfo {volume}
  {24}},\ \bibinfo {pages} {050101} (\bibinfo {year} {2021})}\BibitemShut
  {NoStop}%
\bibitem [{\citenamefont {von Hahn}\ \emph {et~al.}(2016)\citenamefont {von
  Hahn}, \citenamefont {Becker}, \citenamefont {Berg}, \citenamefont {Blaum},
  \citenamefont {Breitenfeldt}, \citenamefont {Fadil}, \citenamefont
  {Fellenberger}, \citenamefont {Froese}, \citenamefont {George}, \citenamefont
  {G\"ock}, \citenamefont {Grieser}, \citenamefont {Grussie}, \citenamefont
  {Guerin}, \citenamefont {Heber}, \citenamefont {Herwig}, \citenamefont
  {Karthein}, \citenamefont {Krantz}, \citenamefont {Kreckel}, \citenamefont
  {Lange}, \citenamefont {Laux}, \citenamefont {Lohmann}, \citenamefont {Menk},
  \citenamefont {Meyer}, \citenamefont {Mishra}, \citenamefont {Novotn\'y},
  \citenamefont {O'Connor}, \citenamefont {Orlov}, \citenamefont {Rappaport},
  \citenamefont {Repnow}, \citenamefont {Saurabh}, \citenamefont {Schippers},
  \citenamefont {Schr\"oter}, \citenamefont {Schwalm}, \citenamefont
  {Schweikhard}, \citenamefont {Sieber}, \citenamefont {Shornikov},
  \citenamefont {Spruck}, \citenamefont {Sunil~Kumar}, \citenamefont {Ullrich},
  \citenamefont {Urbain}, \citenamefont {Vogel}, \citenamefont {Wilhelm},
  \citenamefont {Wolf},\ and\ \citenamefont
  {Zajfman}}]{von_hahn_cryogenic_2016}%
  \BibitemOpen
  \bibfield  {author} {\bibinfo {author} {\bibfnamefont {R.}~\bibnamefont {von
  Hahn}}, \bibinfo {author} {\bibfnamefont {A.}~\bibnamefont {Becker}},
  \bibinfo {author} {\bibfnamefont {F.}~\bibnamefont {Berg}}, \bibinfo {author}
  {\bibfnamefont {K.}~\bibnamefont {Blaum}}, \bibinfo {author} {\bibfnamefont
  {C.}~\bibnamefont {Breitenfeldt}}, \bibinfo {author} {\bibfnamefont
  {H.}~\bibnamefont {Fadil}}, \bibinfo {author} {\bibfnamefont
  {F.}~\bibnamefont {Fellenberger}}, \bibinfo {author} {\bibfnamefont
  {M.}~\bibnamefont {Froese}}, \bibinfo {author} {\bibfnamefont
  {S.}~\bibnamefont {George}}, \bibinfo {author} {\bibfnamefont
  {J.}~\bibnamefont {G\"ock}}, \bibinfo {author} {\bibfnamefont
  {M.}~\bibnamefont {Grieser}}, \bibinfo {author} {\bibfnamefont
  {F.}~\bibnamefont {Grussie}}, \bibinfo {author} {\bibfnamefont {E.~A.}\
  \bibnamefont {Guerin}}, \bibinfo {author} {\bibfnamefont {O.}~\bibnamefont
  {Heber}}, \bibinfo {author} {\bibfnamefont {P.}~\bibnamefont {Herwig}},
  \bibinfo {author} {\bibfnamefont {J.}~\bibnamefont {Karthein}}, \bibinfo
  {author} {\bibfnamefont {C.}~\bibnamefont {Krantz}}, \bibinfo {author}
  {\bibfnamefont {H.}~\bibnamefont {Kreckel}}, \bibinfo {author} {\bibfnamefont
  {M.}~\bibnamefont {Lange}}, \bibinfo {author} {\bibfnamefont
  {F.}~\bibnamefont {Laux}}, \bibinfo {author} {\bibfnamefont {S.}~\bibnamefont
  {Lohmann}}, \bibinfo {author} {\bibfnamefont {S.}~\bibnamefont {Menk}},
  \bibinfo {author} {\bibfnamefont {C.}~\bibnamefont {Meyer}}, \bibinfo
  {author} {\bibfnamefont {P.~M.}\ \bibnamefont {Mishra}}, \bibinfo {author}
  {\bibfnamefont {O.}~\bibnamefont {Novotn\'y}}, \bibinfo {author}
  {\bibfnamefont {A.~P.}\ \bibnamefont {O'Connor}}, \bibinfo {author}
  {\bibfnamefont {D.~A.}\ \bibnamefont {Orlov}}, \bibinfo {author}
  {\bibfnamefont {M.~L.}\ \bibnamefont {Rappaport}}, \bibinfo {author}
  {\bibfnamefont {R.}~\bibnamefont {Repnow}}, \bibinfo {author} {\bibfnamefont
  {S.}~\bibnamefont {Saurabh}}, \bibinfo {author} {\bibfnamefont
  {S.}~\bibnamefont {Schippers}}, \bibinfo {author} {\bibfnamefont {C.~D.}\
  \bibnamefont {Schr\"oter}}, \bibinfo {author} {\bibfnamefont
  {D.}~\bibnamefont {Schwalm}}, \bibinfo {author} {\bibfnamefont
  {L.}~\bibnamefont {Schweikhard}}, \bibinfo {author} {\bibfnamefont
  {T.}~\bibnamefont {Sieber}}, \bibinfo {author} {\bibfnamefont
  {A.}~\bibnamefont {Shornikov}}, \bibinfo {author} {\bibfnamefont
  {K.}~\bibnamefont {Spruck}}, \bibinfo {author} {\bibfnamefont
  {S.}~\bibnamefont {Sunil~Kumar}}, \bibinfo {author} {\bibfnamefont
  {J.}~\bibnamefont {Ullrich}}, \bibinfo {author} {\bibfnamefont
  {X.}~\bibnamefont {Urbain}}, \bibinfo {author} {\bibfnamefont
  {S.}~\bibnamefont {Vogel}}, \bibinfo {author} {\bibfnamefont
  {P.}~\bibnamefont {Wilhelm}}, \bibinfo {author} {\bibfnamefont
  {A.}~\bibnamefont {Wolf}},\ and\ \bibinfo {author} {\bibfnamefont
  {D.}~\bibnamefont {Zajfman}},\ }\bibfield  {title} {\bibinfo {title} {The
  cryogenic storage ring {CSR}},\ }\href {https://doi.org/10.1063/1.4953888}
  {\bibfield  {journal} {\bibinfo  {journal} {Rev. Sci. Instrum.}\ }\textbf
  {\bibinfo {volume} {87}},\ \bibinfo {pages} {063115} (\bibinfo {year}
  {2016})}\BibitemShut {NoStop}%
\bibitem [{boi()}]{boikova_rotational_1968}%
  \BibitemOpen
  \href@noop {} {}\bibinfo {note} {R. F. Bo\u{\i}kova and V. D. Ob'{}'edkov,
  Rotational and vibrational excitation of molecular ions by electrons, Zh.
  Eksp. Teor. Fiz. {\bf 54}, 1439 (1968)
  \href{http://jetp.ras.ru/cgi-bin/e/index/e/27/5/p772?a=list} {[Sov. Phys.
  JETP {\bf 27}, 772--774 (1968)]}}\BibitemShut {NoStop}%
\bibitem [{\citenamefont {Neufeld}\ and\ \citenamefont
  {Dalgarno}(1989)}]{neufeld_electron-impact_1989}%
  \BibitemOpen
  \bibfield  {author} {\bibinfo {author} {\bibfnamefont {D.~A.}\ \bibnamefont
  {Neufeld}}\ and\ \bibinfo {author} {\bibfnamefont {A.}~\bibnamefont
  {Dalgarno}},\ }\bibfield  {title} {\bibinfo {title} {Electron-impact
  excitation of molecular ions},\ }\href
  {https://doi.org/10.1103/PhysRevA.40.633} {\bibfield  {journal} {\bibinfo
  {journal} {Phys. Rev. A}\ }\textbf {\bibinfo {volume} {40}},\ \bibinfo
  {pages} {633} (\bibinfo {year} {1989})}\BibitemShut {NoStop}%
\bibitem [{\citenamefont {Hamilton}\ \emph {et~al.}(2016)\citenamefont
  {Hamilton}, \citenamefont {Faure},\ and\ \citenamefont
  {Tennyson}}]{hamilton_electron-impact_2016}%
  \BibitemOpen
  \bibfield  {author} {\bibinfo {author} {\bibfnamefont {J.~R.}\ \bibnamefont
  {Hamilton}}, \bibinfo {author} {\bibfnamefont {A.}~\bibnamefont {Faure}},\
  and\ \bibinfo {author} {\bibfnamefont {J.}~\bibnamefont {Tennyson}},\
  }\bibfield  {title} {\bibinfo {title} {Electron-impact excitation of diatomic
  hydride cations -- {I.} {HeH$^+$}, {CH$^+$}, {ArH$^+$}},\ }\href
  {https://doi.org/10.1093/mnras/stv2429} {\bibfield  {journal} {\bibinfo
  {journal} {Mon. Not. R. Astron. Soc.}\ }\textbf {\bibinfo {volume} {455}},\
  \bibinfo {pages} {3281} (\bibinfo {year} {2016})}\BibitemShut {NoStop}%
\bibitem [{\citenamefont {van Hoof}\ \emph {et~al.}(2014)\citenamefont {van
  Hoof}, \citenamefont {Williams}, \citenamefont {Volk}, \citenamefont
  {Chatzikos}, \citenamefont {Ferland}, \citenamefont {Lykins}, \citenamefont
  {Porter},\ and\ \citenamefont {Wang}}]{van_hoof_accurate_2014}%
  \BibitemOpen
  \bibfield  {author} {\bibinfo {author} {\bibfnamefont {P.~A.~M.}\
  \bibnamefont {van Hoof}}, \bibinfo {author} {\bibfnamefont {R.~J.~R.}\
  \bibnamefont {Williams}}, \bibinfo {author} {\bibfnamefont {K.}~\bibnamefont
  {Volk}}, \bibinfo {author} {\bibfnamefont {M.}~\bibnamefont {Chatzikos}},
  \bibinfo {author} {\bibfnamefont {G.~J.}\ \bibnamefont {Ferland}}, \bibinfo
  {author} {\bibfnamefont {M.}~\bibnamefont {Lykins}}, \bibinfo {author}
  {\bibfnamefont {R.~L.}\ \bibnamefont {Porter}},\ and\ \bibinfo {author}
  {\bibfnamefont {Y.}~\bibnamefont {Wang}},\ }\bibfield  {title} {\bibinfo
  {title} {Accurate determination of the free--free {Gaunt} factor -- {I}.
  {Non}-relativistic {Gaunt} factors},\ }\href
  {https://doi.org/10.1093/mnras/stu1438} {\bibfield  {journal} {\bibinfo
  {journal} {Mon. Not. R. Astron. Soc.}\ }\textbf {\bibinfo {volume} {444}},\
  \bibinfo {pages} {420} (\bibinfo {year} {2014})}\BibitemShut {NoStop}%
\bibitem [{\citenamefont {Mezei}\ \emph {et~al.}(2019)\citenamefont {Mezei},
  \citenamefont {Ep{\'e}e~Ep{\'e}e}, \citenamefont {Motapon},\ and\
  \citenamefont {Schneider}}]{mezei_atom_2019}%
  \BibitemOpen
  \bibfield  {author} {\bibinfo {author} {\bibfnamefont {Z.~J.}\ \bibnamefont
  {Mezei}}, \bibinfo {author} {\bibfnamefont {M.~D.}\ \bibnamefont
  {Ep{\'e}e~Ep{\'e}e}}, \bibinfo {author} {\bibfnamefont {O.}~\bibnamefont
  {Motapon}},\ and\ \bibinfo {author} {\bibfnamefont {I.~F.}\ \bibnamefont
  {Schneider}},\ }\bibfield  {title} {\bibinfo {title} {Dissociative
  recombination of {CH}$^+$ molecular ion induced by very low energy
  electrons},\ }\href {https://doi.org/10.3390/atoms7030082} {\bibfield
  {journal} {\bibinfo  {journal} {Atoms}\ }\textbf {\bibinfo {volume} {7}},\
  \bibinfo {pages} {82} (\bibinfo {year} {2019})}\BibitemShut {NoStop}%
\bibitem [{\citenamefont {V{\' a}zquez}\ \emph {et~al.}(2007)\citenamefont
  {V{\' a}zquez}, \citenamefont {Amero}, \citenamefont {Liebermann},
  \citenamefont {Buenker},\ and\ \citenamefont
  {Lefebvre-Brion}}]{vazquez_rydbergch_2007}%
  \BibitemOpen
  \bibfield  {author} {\bibinfo {author} {\bibfnamefont {G.~J.}\ \bibnamefont
  {V{\' a}zquez}}, \bibinfo {author} {\bibfnamefont {J.~M.}\ \bibnamefont
  {Amero}}, \bibinfo {author} {\bibfnamefont {H.~P.}\ \bibnamefont
  {Liebermann}}, \bibinfo {author} {\bibfnamefont {R.~J.}\ \bibnamefont
  {Buenker}},\ and\ \bibinfo {author} {\bibfnamefont {H.}~\bibnamefont
  {Lefebvre-Brion}},\ }\bibfield  {title} {\bibinfo {title} {Insight into the
  {Rydberg} states of {CH}},\ }\href {https://doi.org/10.1063/1.2721535}
  {\bibfield  {journal} {\bibinfo  {journal} {J. Chem. Phys.}\ }\textbf
  {\bibinfo {volume} {126}},\ \bibinfo {pages} {164302} (\bibinfo {year}
  {2007})}\BibitemShut {NoStop}%
\bibitem [{\citenamefont {Larsson}\ and\ \citenamefont {Orel}(2008)}]{drbook}%
  \BibitemOpen
  \bibfield  {author} {\bibinfo {author} {\bibfnamefont {M.}~\bibnamefont
  {Larsson}}\ and\ \bibinfo {author} {\bibfnamefont {A.~E.}\ \bibnamefont
  {Orel}},\ }\href@noop {} {\emph {\bibinfo {title} {Dissociative Recombination
  of Molecular Ions}}}\ (\bibinfo  {publisher} {Cambridge University Press,
  Cambridge},\ \bibinfo {year} {2008})\BibitemShut {NoStop}%
\bibitem [{\citenamefont {Hogg}\ and\ \citenamefont
  {Foreman-Mackey}(2018)}]{mcmc}%
  \BibitemOpen
  \bibfield  {author} {\bibinfo {author} {\bibfnamefont {D.~W.}\ \bibnamefont
  {Hogg}}\ and\ \bibinfo {author} {\bibfnamefont {D.}~\bibnamefont
  {Foreman-Mackey}},\ }\bibfield  {title} {\bibinfo {title} {Data analysis
  recipes: {Using} {Markov} {Chain} {Monte} {Carlo}},\ }\href
  {https://doi.org/10.3847/1538-4365/aab76e} {\bibfield  {journal} {\bibinfo
  {journal} {Astrophs. J. Suppl. Ser.}\ }\textbf {\bibinfo {volume} {236}},\
  \bibinfo {pages} {11} (\bibinfo {year} {2018})}\BibitemShut {NoStop}%
\bibitem [{\citenamefont {Foreman-Mackey}\ \emph {et~al.}(2013)\citenamefont
  {Foreman-Mackey}, \citenamefont {Hogg}, \citenamefont {Lang},\ and\
  \citenamefont {Goodman}}]{Foreman_Mackey_2013}%
  \BibitemOpen
  \bibfield  {author} {\bibinfo {author} {\bibfnamefont {D.}~\bibnamefont
  {Foreman-Mackey}}, \bibinfo {author} {\bibfnamefont {D.~W.}\ \bibnamefont
  {Hogg}}, \bibinfo {author} {\bibfnamefont {D.}~\bibnamefont {Lang}},\ and\
  \bibinfo {author} {\bibfnamefont {J.}~\bibnamefont {Goodman}},\ }\bibfield
  {title} {\bibinfo {title} {{emcee}: The {MCMC} hammer},\ }\href
  {https://doi.org/10.1086/670067} {\bibfield  {journal} {\bibinfo  {journal}
  {Publ. Astron. Soc. Pac.}\ }\textbf {\bibinfo {volume} {125}},\ \bibinfo
  {pages} {306} (\bibinfo {year} {2013})}\BibitemShut {NoStop}%
\end{thebibliography}%


\begin{thebibliography}{54}%
	\makeatletter
	\providecommand \@ifxundefined [1]{%
		\@ifx{#1\undefined}
	}%
	\providecommand \@ifnum [1]{%
		\ifnum #1\expandafter \@firstoftwo
		\else \expandafter \@secondoftwo
		\fi
	}%
	\providecommand \@ifx [1]{%
		\ifx #1\expandafter \@firstoftwo
		\else \expandafter \@secondoftwo
		\fi
	}%
	\providecommand \natexlab [1]{#1}%
	\providecommand \enquote  [1]{``#1''}%
	\providecommand \bibnamefont  [1]{#1}%
	\providecommand \bibfnamefont [1]{#1}%
	\providecommand \citenamefont [1]{#1}%
	\providecommand \href@noop [0]{\@secondoftwo}%
	\providecommand \href [0]{\begingroup \@sanitize@url \@href}%
	\providecommand \@href[1]{\@@startlink{#1}\@@href}%
	\providecommand \@@href[1]{\endgroup#1\@@endlink}%
	\providecommand \@sanitize@url [0]{\catcode `\\12\catcode `\$12\catcode
		`\&12\catcode `\#12\catcode `\^12\catcode `\_12\catcode `\%12\relax}%
	\providecommand \@@startlink[1]{}%
	\providecommand \@@endlink[0]{}%
	\providecommand \url  [0]{\begingroup\@sanitize@url \@url }%
	\providecommand \@url [1]{\endgroup\@href {#1}{\urlprefix }}%
	\providecommand \urlprefix  [0]{URL }%
	\providecommand \Eprint [0]{\href }%
	\providecommand \doibase [0]{https://doi.org/}%
	\providecommand \selectlanguage [0]{\@gobble}%
	\providecommand \bibinfo  [0]{\@secondoftwo}%
	\providecommand \bibfield  [0]{\@secondoftwo}%
	\providecommand \translation [1]{[#1]}%
	\providecommand \BibitemOpen [0]{}%
	\providecommand \bibitemStop [0]{}%
	\providecommand \bibitemNoStop [0]{.\EOS\space}%
	\providecommand \EOS [0]{\spacefactor3000\relax}%
	\providecommand \BibitemShut  [1]{\csname bibitem#1\endcsname}%
	\let\auto@bib@innerbib\@empty
	\bibitem [{\citenamefont {Bartschat}\ and\ \citenamefont
		{Kushner}(2016)}]{bartschat_electron_2016}%
	\BibitemOpen
	\bibfield  {author} {\bibinfo {author} {\bibfnamefont {K.}~\bibnamefont
			{Bartschat}}\ and\ \bibinfo {author} {\bibfnamefont {M.~J.}\ \bibnamefont
			{Kushner}},\ }\bibfield  {title} {\bibinfo {title} {Electron collisions with
			atoms, ions, molecules, and surfaces: {Fundamental} science empowering
			advances in technology},\ }\href {https://doi.org/10.1073/pnas.1606132113}
	{\bibfield  {journal} {\bibinfo  {journal} {Proc. Natl. Acad. Sci. USA}\
		}\textbf {\bibinfo {volume} {113}},\ \bibinfo {pages} {7026} (\bibinfo {year}
		{2016})}\BibitemShut {NoStop}%
	\bibitem [{\citenamefont {Bartschat}(2018)}]{bartschat_electron_2018}%
	\BibitemOpen
	\bibfield  {author} {\bibinfo {author} {\bibfnamefont {K.}~\bibnamefont
			{Bartschat}},\ }\bibfield  {title} {\bibinfo {title} {Electron
			collisions—experiment, theory, and applications},\ }\href
	{https://doi.org/10.1088/1361-6455/aac5aa} {\bibfield  {journal} {\bibinfo
			{journal} {J. Phys. B: At. Mol. Opt. Phys.}\ }\textbf {\bibinfo {volume}
			{51}},\ \bibinfo {pages} {132001} (\bibinfo {year} {2018})}\BibitemShut
	{NoStop}%
	\bibitem [{\citenamefont {Dubernet}\ \emph {et~al.}(2013)\citenamefont
		{Dubernet}, \citenamefont {Alexander}, \citenamefont {Ba}, \citenamefont
		{Balakrishnan}, \citenamefont {Balan\c{c}a}, \citenamefont {Ceccarelli},
		\citenamefont {Cernicharo}, \citenamefont {Daniel}, \citenamefont {Dayou},
		\citenamefont {Doronin}, \citenamefont {Dumouchel}, \citenamefont {Faure},
		\citenamefont {Feautrier}, \citenamefont {Flower}, \citenamefont {Grosjean},
		\citenamefont {Halvick}, \citenamefont {K\l{}os}, \citenamefont {Lique},
		\citenamefont {McBane}, \citenamefont {Marinakis}, \citenamefont {Moreau},
		\citenamefont {Moszynski}, \citenamefont {Neufeld}, \citenamefont {Roueff},
		\citenamefont {Schilke}, \citenamefont {Spielfiedel}, \citenamefont
		{Stancil}, \citenamefont {Stoecklin}, \citenamefont {Tennyson}, \citenamefont
		{Yang}, \citenamefont {Vasserot},\ and\ \citenamefont
		{Wiesenfeld}}]{dubernet_basecol2012_2013}%
	\BibitemOpen
	\bibfield  {author} {\bibinfo {author} {\bibfnamefont {M.-L.}\ \bibnamefont
			{Dubernet}}, \bibinfo {author} {\bibfnamefont {M.~H.}\ \bibnamefont
			{Alexander}}, \bibinfo {author} {\bibfnamefont {Y.~A.}\ \bibnamefont {Ba}},
		\bibinfo {author} {\bibfnamefont {N.}~\bibnamefont {Balakrishnan}}, \bibinfo
		{author} {\bibfnamefont {C.}~\bibnamefont {Balan\c{c}a}}, \bibinfo {author}
		{\bibfnamefont {C.}~\bibnamefont {Ceccarelli}}, \bibinfo {author}
		{\bibfnamefont {J.}~\bibnamefont {Cernicharo}}, \bibinfo {author}
		{\bibfnamefont {F.}~\bibnamefont {Daniel}}, \bibinfo {author} {\bibfnamefont
			{F.}~\bibnamefont {Dayou}}, \bibinfo {author} {\bibfnamefont
			{M.}~\bibnamefont {Doronin}}, \bibinfo {author} {\bibfnamefont
			{F.}~\bibnamefont {Dumouchel}}, \bibinfo {author} {\bibfnamefont
			{A.}~\bibnamefont {Faure}}, \bibinfo {author} {\bibfnamefont
			{N.}~\bibnamefont {Feautrier}}, \bibinfo {author} {\bibfnamefont {D.~R.}\
			\bibnamefont {Flower}}, \bibinfo {author} {\bibfnamefont {A.}~\bibnamefont
			{Grosjean}}, \bibinfo {author} {\bibfnamefont {P.}~\bibnamefont {Halvick}},
		\bibinfo {author} {\bibfnamefont {J.}~\bibnamefont {K\l{}os}}, \bibinfo
		{author} {\bibfnamefont {F.}~\bibnamefont {Lique}}, \bibinfo {author}
		{\bibfnamefont {G.~C.}\ \bibnamefont {McBane}}, \bibinfo {author}
		{\bibfnamefont {S.}~\bibnamefont {Marinakis}}, \bibinfo {author}
		{\bibfnamefont {N.}~\bibnamefont {Moreau}}, \bibinfo {author} {\bibfnamefont
			{R.}~\bibnamefont {Moszynski}}, \bibinfo {author} {\bibfnamefont {D.~A.}\
			\bibnamefont {Neufeld}}, \bibinfo {author} {\bibfnamefont {E.}~\bibnamefont
			{Roueff}}, \bibinfo {author} {\bibfnamefont {P.}~\bibnamefont {Schilke}},
		\bibinfo {author} {\bibfnamefont {A.}~\bibnamefont {Spielfiedel}}, \bibinfo
		{author} {\bibfnamefont {P.~C.}\ \bibnamefont {Stancil}}, \bibinfo {author}
		{\bibfnamefont {T.}~\bibnamefont {Stoecklin}}, \bibinfo {author}
		{\bibfnamefont {J.}~\bibnamefont {Tennyson}}, \bibinfo {author}
		{\bibfnamefont {B.}~\bibnamefont {Yang}}, \bibinfo {author} {\bibfnamefont
			{A.-M.}\ \bibnamefont {Vasserot}},\ and\ \bibinfo {author} {\bibfnamefont
			{L.}~\bibnamefont {Wiesenfeld}},\ }\bibfield  {title} {\bibinfo {title}
		{{BASECOL2012}: {A} collisional database repository and web service within
			the {Virtual} {Atomic} and {Molecular} {Data} {Centre} ({VAMDC})},\ }\href
	{https://doi.org/10.1051/0004-6361/201220630} {\bibfield  {journal} {\bibinfo
			{journal} {Astron. Astrophys.}\ }\textbf {\bibinfo {volume} {553}},\
		\bibinfo {pages} {A50} (\bibinfo {year} {2013})}\BibitemShut {NoStop}%
	\bibitem [{boi()}]{boikova_rotational_1968}%
	\BibitemOpen
	\href@noop {} {}\bibinfo {note} {R. F. Bo\u{\i}kova and V. D. Ob'{}'edkov,
		Rotational and vibrational excitation of molecular ions by electrons, Zh.
		Eksp. Teor. Fiz. {\bf 54}, 1439 (1968)
		\href{http://jetp.ras.ru/cgi-bin/e/index/e/27/5/p772?a=list} {[Sov. Phys.
			JETP {\bf 27}, 772--774 (1968)]}}\BibitemShut {NoStop}%
	\bibitem [{\citenamefont {Chu}\ and\ \citenamefont
		{Dalgarno}(1974)}]{chu_rotational_1974}%
	\BibitemOpen
	\bibfield  {author} {\bibinfo {author} {\bibfnamefont {S.-I.}\ \bibnamefont
			{Chu}}\ and\ \bibinfo {author} {\bibfnamefont {A.}~\bibnamefont {Dalgarno}},\
	}\bibfield  {title} {\bibinfo {title} {Rotational excitation of {CH$^+$} by
			electron impact},\ }\href {https://doi.org/10.1103/PhysRevA.10.788}
	{\bibfield  {journal} {\bibinfo  {journal} {Phys. Rev. A}\ }\textbf {\bibinfo
			{volume} {10}},\ \bibinfo {pages} {788} (\bibinfo {year} {1974})}\BibitemShut
	{NoStop}%
	\bibitem [{\citenamefont {Chu}\ and\ \citenamefont
		{Dalgarno}(1975)}]{chu_erratum_1975}%
	\BibitemOpen
	\bibfield  {author} {\bibinfo {author} {\bibfnamefont {S.-I.}\ \bibnamefont
			{Chu}}\ and\ \bibinfo {author} {\bibfnamefont {A.}~\bibnamefont {Dalgarno}},\
	}\href {https://doi.org/10.1103/PhysRevA.12.725} {\bibfield  {journal}
		{\bibinfo  {journal} {Phys. Rev. A}\ }\textbf {\bibinfo {volume} {12}},\
		\bibinfo {pages} {725} (\bibinfo {year} {1975})}\BibitemShut {NoStop}%
	\bibitem [{\citenamefont {Bhattacharyya}\ \emph {et~al.}(1981)\citenamefont
		{Bhattacharyya}, \citenamefont {Bhattacharyya},\ and\ \citenamefont
		{Narayan}}]{bhattacharyya_rotational_1981}%
	\BibitemOpen
	\bibfield  {author} {\bibinfo {author} {\bibfnamefont {S.~S.}\ \bibnamefont
			{Bhattacharyya}}, \bibinfo {author} {\bibfnamefont {B.}~\bibnamefont
			{Bhattacharyya}},\ and\ \bibinfo {author} {\bibfnamefont {M.~V.}\
			\bibnamefont {Narayan}},\ }\bibfield  {title} {\bibinfo {title} {Rotational
			excitation of molecular ions by electron impact under interstellar
			conditions},\ }\href {https://doi.org/10.1086/159102} {\bibfield  {journal}
		{\bibinfo  {journal} {Astrophys. J.}\ }\textbf {\bibinfo {volume} {247}},\
		\bibinfo {pages} {936} (\bibinfo {year} {1981})}\BibitemShut {NoStop}%
	\bibitem [{\citenamefont {Neufeld}\ and\ \citenamefont
		{Dalgarno}(1989)}]{neufeld_electron-impact_1989}%
	\BibitemOpen
	\bibfield  {author} {\bibinfo {author} {\bibfnamefont {D.~A.}\ \bibnamefont
			{Neufeld}}\ and\ \bibinfo {author} {\bibfnamefont {A.}~\bibnamefont
			{Dalgarno}},\ }\bibfield  {title} {\bibinfo {title} {Electron-impact
			excitation of molecular ions},\ }\href
	{https://doi.org/10.1103/PhysRevA.40.633} {\bibfield  {journal} {\bibinfo
			{journal} {Phys. Rev. A}\ }\textbf {\bibinfo {volume} {40}},\ \bibinfo
		{pages} {633} (\bibinfo {year} {1989})}\BibitemShut {NoStop}%
	\bibitem [{\citenamefont {Rabad\'{a}n}\ \emph {et~al.}(1998)\citenamefont
		{Rabad\'{a}n}, \citenamefont {Sarpal},\ and\ \citenamefont
		{Tennyson}}]{rabadan_calculated_1998}%
	\BibitemOpen
	\bibfield  {author} {\bibinfo {author} {\bibfnamefont {I.}~\bibnamefont
			{Rabad\'{a}n}}, \bibinfo {author} {\bibfnamefont {B.~K.}\ \bibnamefont
			{Sarpal}},\ and\ \bibinfo {author} {\bibfnamefont {J.}~\bibnamefont
			{Tennyson}},\ }\bibfield  {title} {\bibinfo {title} {Calculated rotational
			and vibrational excitation rates for electron--{HeH}$^+$ collisions},\ }\href
	{https://doi.org/10.1046/j.1365-8711.1998.01741.x} {\bibfield  {journal}
		{\bibinfo  {journal} {Mon. Not. R. Astron. Soc.}\ }\textbf {\bibinfo {volume}
			{299}},\ \bibinfo {pages} {171} (\bibinfo {year} {1998})}\BibitemShut
	{NoStop}%
	\bibitem [{\citenamefont {Lim}\ \emph {et~al.}(1999)\citenamefont {Lim},
		\citenamefont {Rabad\'an},\ and\ \citenamefont
		{Tennyson}}]{lim_electron-impact_1999}%
	\BibitemOpen
	\bibfield  {author} {\bibinfo {author} {\bibfnamefont {A.~J.}\ \bibnamefont
			{Lim}}, \bibinfo {author} {\bibfnamefont {I.}~\bibnamefont {Rabad\'an}},\
		and\ \bibinfo {author} {\bibfnamefont {J.}~\bibnamefont {Tennyson}},\
	}\bibfield  {title} {\bibinfo {title} {Electron-impact rotational excitation
			of {CH}$^+$},\ }\href {https://doi.org/10.1046/j.1365-8711.1999.02533.x}
	{\bibfield  {journal} {\bibinfo  {journal} {Mon. Not. R. Astron. Soc.}\
		}\textbf {\bibinfo {volume} {306}},\ \bibinfo {pages} {473} (\bibinfo {year}
		{1999})}\BibitemShut {NoStop}%
	\bibitem [{\citenamefont {Faure}\ and\ \citenamefont
		{Tennyson}(2001)}]{faure_electron-impact_2001}%
	\BibitemOpen
	\bibfield  {author} {\bibinfo {author} {\bibfnamefont {A.}~\bibnamefont
			{Faure}}\ and\ \bibinfo {author} {\bibfnamefont {J.}~\bibnamefont
			{Tennyson}},\ }\bibfield  {title} {\bibinfo {title} {Electron-impact
			rotational excitation of linear molecular ions},\ }\href
	{https://doi.org/10.1046/j.1365-8711.2001.04480.x} {\bibfield  {journal}
		{\bibinfo  {journal} {Mon. Not. R. Astron. Soc.}\ }\textbf {\bibinfo {volume}
			{325}},\ \bibinfo {pages} {443} (\bibinfo {year} {2001})}\BibitemShut
	{NoStop}%
	\bibitem [{\citenamefont {Hamilton}\ \emph {et~al.}(2016)\citenamefont
		{Hamilton}, \citenamefont {Faure},\ and\ \citenamefont
		{Tennyson}}]{hamilton_electron-impact_2016}%
	\BibitemOpen
	\bibfield  {author} {\bibinfo {author} {\bibfnamefont {J.~R.}\ \bibnamefont
			{Hamilton}}, \bibinfo {author} {\bibfnamefont {A.}~\bibnamefont {Faure}},\
		and\ \bibinfo {author} {\bibfnamefont {J.}~\bibnamefont {Tennyson}},\
	}\bibfield  {title} {\bibinfo {title} {Electron-impact excitation of diatomic
			hydride cations -- {I.} {HeH$^+$}, {CH$^+$}, {ArH$^+$}},\ }\href
	{https://doi.org/10.1093/mnras/stv2429} {\bibfield  {journal} {\bibinfo
			{journal} {Mon. Not. R. Astron. Soc.}\ }\textbf {\bibinfo {volume} {455}},\
		\bibinfo {pages} {3281} (\bibinfo {year} {2016})}\BibitemShut {NoStop}%
	\bibitem [{\citenamefont {Chakrabarti}\ \emph {et~al.}(2017)\citenamefont
		{Chakrabarti}, \citenamefont {Dora}, \citenamefont {Ghosh}, \citenamefont
		{Choudhury},\ and\ \citenamefont {Tennyson}}]{chakrabarti_r-matrix_2017}%
	\BibitemOpen
	\bibfield  {author} {\bibinfo {author} {\bibfnamefont {K.}~\bibnamefont
			{Chakrabarti}}, \bibinfo {author} {\bibfnamefont {A.}~\bibnamefont {Dora}},
		\bibinfo {author} {\bibfnamefont {R.}~\bibnamefont {Ghosh}}, \bibinfo
		{author} {\bibfnamefont {B.~S.}\ \bibnamefont {Choudhury}},\ and\ \bibinfo
		{author} {\bibfnamefont {J.}~\bibnamefont {Tennyson}},\ }\bibfield  {title}
	{\bibinfo {title} {{$R$}-matrix study of electron impact excitation and
			dissociation of {CH}$^+$ ions},\ }\href
	{https://doi.org/10.1088/1361-6455/aa8377} {\bibfield  {journal} {\bibinfo
			{journal} {J. Phys. B: At. Mol. Opt. Phys.}\ }\textbf {\bibinfo {volume}
			{50}},\ \bibinfo {pages} {175202} (\bibinfo {year} {2017})}\BibitemShut
	{NoStop}%
	\bibitem [{\citenamefont {Faure}\ \emph {et~al.}(2017)\citenamefont {Faure},
		\citenamefont {Halvick}, \citenamefont {Stoecklin}, \citenamefont {Honvault},
		\citenamefont {Ep\'{e}e~Ep\'{e}e}, \citenamefont {Mezei}, \citenamefont
		{Motapon}, \citenamefont {Schneider}, \citenamefont {Tennyson}, \citenamefont
		{Roncero}, \citenamefont {Bulut},\ and\ \citenamefont
		{Zanchet}}]{faure_state--state_2017}%
	\BibitemOpen
	\bibfield  {author} {\bibinfo {author} {\bibfnamefont {A.}~\bibnamefont
			{Faure}}, \bibinfo {author} {\bibfnamefont {P.}~\bibnamefont {Halvick}},
		\bibinfo {author} {\bibfnamefont {T.}~\bibnamefont {Stoecklin}}, \bibinfo
		{author} {\bibfnamefont {P.}~\bibnamefont {Honvault}}, \bibinfo {author}
		{\bibfnamefont {M.~D.}\ \bibnamefont {Ep\'{e}e~Ep\'{e}e}}, \bibinfo {author}
		{\bibfnamefont {J.~Z.}\ \bibnamefont {Mezei}}, \bibinfo {author}
		{\bibfnamefont {O.}~\bibnamefont {Motapon}}, \bibinfo {author} {\bibfnamefont
			{I.~F.}\ \bibnamefont {Schneider}}, \bibinfo {author} {\bibfnamefont
			{J.}~\bibnamefont {Tennyson}}, \bibinfo {author} {\bibfnamefont
			{O.}~\bibnamefont {Roncero}}, \bibinfo {author} {\bibfnamefont
			{N.}~\bibnamefont {Bulut}},\ and\ \bibinfo {author} {\bibfnamefont
			{A.}~\bibnamefont {Zanchet}},\ }\bibfield  {title} {\bibinfo {title}
		{State-to-state chemistry and rotational excitation of {CH}$^+$ in
			photon-dominated regions},\ }\href {https://doi.org/10.1093/mnras/stx892}
	{\bibfield  {journal} {\bibinfo  {journal} {Mon. Not. R. Astron. Soc.}\
		}\textbf {\bibinfo {volume} {469}},\ \bibinfo {pages} {612} (\bibinfo {year}
		{2017})}\BibitemShut {NoStop}%
	\bibitem [{\citenamefont {Hamilton}\ \emph {et~al.}(2018)\citenamefont
		{Hamilton}, \citenamefont {Faure},\ and\ \citenamefont
		{Tennyson}}]{hamilton_electron-impact_2018}%
	\BibitemOpen
	\bibfield  {author} {\bibinfo {author} {\bibfnamefont {J.~R.}\ \bibnamefont
			{Hamilton}}, \bibinfo {author} {\bibfnamefont {A.}~\bibnamefont {Faure}},\
		and\ \bibinfo {author} {\bibfnamefont {J.}~\bibnamefont {Tennyson}},\
	}\bibfield  {title} {\bibinfo {title} {Electron-impact excitation of diatomic
			hydride cations {II}: {OH$^+$} and {SH$^+$}},\ }\href
	{https://doi.org/10.1093/mnras/sty437} {\bibfield  {journal} {\bibinfo
			{journal} {Mon. Not. R. Astron. Soc.}\ }\textbf {\bibinfo {volume} {476}},\
		\bibinfo {pages} {2931} (\bibinfo {year} {2018})}\BibitemShut {NoStop}%
	\bibitem [{\citenamefont {Cooper}\ \emph {et~al.}(2019)\citenamefont {Cooper},
		\citenamefont {Tudorovskaya}, \citenamefont {Mohr}, \citenamefont {O’Hare},
		\citenamefont {Hanicinec}, \citenamefont {Dzarasova}, \citenamefont
		{Gorfinkiel}, \citenamefont {Benda}, \citenamefont {Ma\u{s}\'i{}n},
		\citenamefont {Al-Refaie}, \citenamefont {Knowles},\ and\ \citenamefont
		{Tennyson}}]{cooper_quantemol_2019}%
	\BibitemOpen
	\bibfield  {author} {\bibinfo {author} {\bibfnamefont {B.}~\bibnamefont
			{Cooper}}, \bibinfo {author} {\bibfnamefont {M.}~\bibnamefont
			{Tudorovskaya}}, \bibinfo {author} {\bibfnamefont {S.}~\bibnamefont {Mohr}},
		\bibinfo {author} {\bibfnamefont {A.}~\bibnamefont {O’Hare}}, \bibinfo
		{author} {\bibfnamefont {M.}~\bibnamefont {Hanicinec}}, \bibinfo {author}
		{\bibfnamefont {A.}~\bibnamefont {Dzarasova}}, \bibinfo {author}
		{\bibfnamefont {J.}~\bibnamefont {Gorfinkiel}}, \bibinfo {author}
		{\bibfnamefont {J.}~\bibnamefont {Benda}}, \bibinfo {author} {\bibfnamefont
			{Z.}~\bibnamefont {Ma\u{s}\'i{}n}}, \bibinfo {author} {\bibfnamefont
			{A.}~\bibnamefont {Al-Refaie}}, \bibinfo {author} {\bibfnamefont
			{P.}~\bibnamefont {Knowles}},\ and\ \bibinfo {author} {\bibfnamefont
			{J.}~\bibnamefont {Tennyson}},\ }\bibfield  {title} {\bibinfo {title}
		{Quantemol electron collisions ({QEC}): {An} enhanced expert system for
			performing electron molecule collision calculations using the {R}-matrix
			method},\ }\href {https://doi.org/10.3390/atoms7040097} {\bibfield  {journal}
		{\bibinfo  {journal} {Atoms}\ }\textbf {\bibinfo {volume} {7}},\ \bibinfo
		{pages} {97} (\bibinfo {year} {2019})}\BibitemShut {NoStop}%
	\bibitem [{\citenamefont {Dickinson}\ and\ \citenamefont
		{Flower}(1981)}]{dickinson_electron_1981}%
	\BibitemOpen
	\bibfield  {author} {\bibinfo {author} {\bibfnamefont {A.~S.}\ \bibnamefont
			{Dickinson}}\ and\ \bibinfo {author} {\bibfnamefont {D.~R.}\ \bibnamefont
			{Flower}},\ }\bibfield  {title} {\bibinfo {title} {Electron collisional
			excitation of interstellar molecular ions},\ }\href
	{https://doi.org/10.1093/mnras/196.2.297} {\bibfield  {journal} {\bibinfo
			{journal} {Mon. Not. R. Astron. Soc.}\ }\textbf {\bibinfo {volume} {196}},\
		\bibinfo {pages} {297} (\bibinfo {year} {1981})}\BibitemShut {NoStop}%
	\bibitem [{\citenamefont {Flower}(1999)}]{flower_rotational_1999}%
	\BibitemOpen
	\bibfield  {author} {\bibinfo {author} {\bibfnamefont {D.~R.}\ \bibnamefont
			{Flower}},\ }\bibfield  {title} {\bibinfo {title} {Rotational excitation of
			{HCO}$^+$ by {H$_2$}},\ }\href
	{https://doi.org/10.1046/j.1365-8711.1999.02451.x} {\bibfield  {journal}
		{\bibinfo  {journal} {Mon. Not. R. Astron. Soc.}\ }\textbf {\bibinfo {volume}
			{305}},\ \bibinfo {pages} {651} (\bibinfo {year} {1999})}\BibitemShut
	{NoStop}%
	\bibitem [{\citenamefont {Liszt}(2012)}]{liszt_rotational_2012}%
	\BibitemOpen
	\bibfield  {author} {\bibinfo {author} {\bibfnamefont {H.~S.}\ \bibnamefont
			{Liszt}},\ }\bibfield  {title} {\bibinfo {title} {Rotational excitation of
			simple polar molecules by {H$_2$} and electrons in diffuse clouds},\ }\href
	{https://doi.org/10.1051/0004-6361/201117882} {\bibfield  {journal} {\bibinfo
			{journal} {Astron. Astrophys.}\ }\textbf {\bibinfo {volume} {538}},\
		\bibinfo {pages} {A27} (\bibinfo {year} {2012})}\BibitemShut {NoStop}%
	\bibitem [{\citenamefont {Black}(1998)}]{black_molecules_1998}%
	\BibitemOpen
	\bibfield  {author} {\bibinfo {author} {\bibfnamefont {J.~H.}\ \bibnamefont
			{Black}},\ }\bibfield  {title} {\bibinfo {title} {Molecules in harsh
			environments},\ }\href {https://doi.org/10.1039/a800591e} {\bibfield
		{journal} {\bibinfo  {journal} {Faraday Disc.}\ }\textbf {\bibinfo {volume}
			{109}},\ \bibinfo {pages} {257} (\bibinfo {year} {1998})}\BibitemShut
	{NoStop}%
	\bibitem [{\citenamefont {Petrie}\ and\ \citenamefont
		{Bohme}(2007)}]{petrie_ions_2007}%
	\BibitemOpen
	\bibfield  {author} {\bibinfo {author} {\bibfnamefont {S.}~\bibnamefont
			{Petrie}}\ and\ \bibinfo {author} {\bibfnamefont {D.~K.}\ \bibnamefont
			{Bohme}},\ }\bibfield  {title} {\bibinfo {title} {Ions in space},\ }\href
	{https://doi.org/10.1002/mas.20114} {\bibfield  {journal} {\bibinfo
			{journal} {Mass Spectrometry Reviews}\ }\textbf {\bibinfo {volume} {26}},\
		\bibinfo {pages} {258} (\bibinfo {year} {2007})}\BibitemShut {NoStop}%
	\bibitem [{\citenamefont {McGuire}(2018)}]{mcguire_2018_2018}%
	\BibitemOpen
	\bibfield  {author} {\bibinfo {author} {\bibfnamefont {B.~A.}\ \bibnamefont
			{McGuire}},\ }\bibfield  {title} {\bibinfo {title} {2018 census of
			interstellar, circumstellar, extragalactic, protoplanetary disk, and
			exoplanetary molecules},\ }\href {https://doi.org/10.3847/1538-4365/aae5d2}
	{\bibfield  {journal} {\bibinfo  {journal} {Astrophys. J. Suppl. Ser.}\
		}\textbf {\bibinfo {volume} {239}},\ \bibinfo {pages} {17} (\bibinfo {year}
		{2018})}\BibitemShut {NoStop}%
	\bibitem [{\citenamefont {Shafir}\ \emph {et~al.}(2009)\citenamefont {Shafir},
		\citenamefont {Novotny}, \citenamefont {Buhr}, \citenamefont {Altevogt},
		\citenamefont {Faure}, \citenamefont {Grieser}, \citenamefont {Harvey},
		\citenamefont {Heber}, \citenamefont {Hoffmann}, \citenamefont {Kreckel},
		\citenamefont {Lammich}, \citenamefont {Nevo}, \citenamefont {Pedersen},
		\citenamefont {Rubinstein}, \citenamefont {Schneider}, \citenamefont
		{Schwalm}, \citenamefont {Tennyson}, \citenamefont {Wolf},\ and\
		\citenamefont {Zajfman}}]{shafir_rotational_2009}%
	\BibitemOpen
	\bibfield  {author} {\bibinfo {author} {\bibfnamefont {D.}~\bibnamefont
			{Shafir}}, \bibinfo {author} {\bibfnamefont {S.}~\bibnamefont {Novotny}},
		\bibinfo {author} {\bibfnamefont {H.}~\bibnamefont {Buhr}}, \bibinfo {author}
		{\bibfnamefont {S.}~\bibnamefont {Altevogt}}, \bibinfo {author}
		{\bibfnamefont {A.}~\bibnamefont {Faure}}, \bibinfo {author} {\bibfnamefont
			{M.}~\bibnamefont {Grieser}}, \bibinfo {author} {\bibfnamefont {A.~G.}\
			\bibnamefont {Harvey}}, \bibinfo {author} {\bibfnamefont {O.}~\bibnamefont
			{Heber}}, \bibinfo {author} {\bibfnamefont {J.}~\bibnamefont {Hoffmann}},
		\bibinfo {author} {\bibfnamefont {H.}~\bibnamefont {Kreckel}}, \bibinfo
		{author} {\bibfnamefont {L.}~\bibnamefont {Lammich}}, \bibinfo {author}
		{\bibfnamefont {I.}~\bibnamefont {Nevo}}, \bibinfo {author} {\bibfnamefont
			{H.~B.}\ \bibnamefont {Pedersen}}, \bibinfo {author} {\bibfnamefont
			{H.}~\bibnamefont {Rubinstein}}, \bibinfo {author} {\bibfnamefont {I.~F.}\
			\bibnamefont {Schneider}}, \bibinfo {author} {\bibfnamefont {D.}~\bibnamefont
			{Schwalm}}, \bibinfo {author} {\bibfnamefont {J.}~\bibnamefont {Tennyson}},
		\bibinfo {author} {\bibfnamefont {A.}~\bibnamefont {Wolf}},\ and\ \bibinfo
		{author} {\bibfnamefont {D.}~\bibnamefont {Zajfman}},\ }\bibfield  {title}
	{\bibinfo {title} {Rotational cooling of {HD${^+}$} molecular ions by
			superelastic collisions with electrons},\ }\href
	{https://doi.org/10.1103/PhysRevLett.102.223202} {\bibfield  {journal}
		{\bibinfo  {journal} {Phys. Rev. Lett.}\ }\textbf {\bibinfo {volume} {102}},\
		\bibinfo {pages} {223202} (\bibinfo {year} {2009})}\BibitemShut {NoStop}%
	\bibitem [{\citenamefont {Schmidt}(2015)}]{schmidt_electrostatic_2015}%
	\BibitemOpen
	\bibfield  {author} {\bibinfo {author} {\bibfnamefont {H.~T.}\ \bibnamefont
			{Schmidt}},\ }\bibfield  {title} {\bibinfo {title} {Electrostatic storage
			rings for atomic and molecular physics},\ }\href
	{https://doi.org/10.1088/0031-8949/2015/T166/014063} {\bibfield  {journal}
		{\bibinfo  {journal} {Phys. Scripta}\ }\textbf {\bibinfo {volume} {T166}},\
		\bibinfo {pages} {014063} (\bibinfo {year} {2015})}\BibitemShut {NoStop}%
	\bibitem [{\citenamefont {von Hahn}\ \emph {et~al.}(2016)\citenamefont {von
			Hahn}, \citenamefont {Becker}, \citenamefont {Berg}, \citenamefont {Blaum},
		\citenamefont {Breitenfeldt}, \citenamefont {Fadil}, \citenamefont
		{Fellenberger}, \citenamefont {Froese}, \citenamefont {George}, \citenamefont
		{G\"ock}, \citenamefont {Grieser}, \citenamefont {Grussie}, \citenamefont
		{Guerin}, \citenamefont {Heber}, \citenamefont {Herwig}, \citenamefont
		{Karthein}, \citenamefont {Krantz}, \citenamefont {Kreckel}, \citenamefont
		{Lange}, \citenamefont {Laux}, \citenamefont {Lohmann}, \citenamefont {Menk},
		\citenamefont {Meyer}, \citenamefont {Mishra}, \citenamefont {Novotn\'y},
		\citenamefont {O'Connor}, \citenamefont {Orlov}, \citenamefont {Rappaport},
		\citenamefont {Repnow}, \citenamefont {Saurabh}, \citenamefont {Schippers},
		\citenamefont {Schr\"oter}, \citenamefont {Schwalm}, \citenamefont
		{Schweikhard}, \citenamefont {Sieber}, \citenamefont {Shornikov},
		\citenamefont {Spruck}, \citenamefont {Sunil~Kumar}, \citenamefont {Ullrich},
		\citenamefont {Urbain}, \citenamefont {Vogel}, \citenamefont {Wilhelm},
		\citenamefont {Wolf},\ and\ \citenamefont
		{Zajfman}}]{von_hahn_cryogenic_2016}%
	\BibitemOpen
	\bibfield  {author} {\bibinfo {author} {\bibfnamefont {R.}~\bibnamefont {von
				Hahn}}, \bibinfo {author} {\bibfnamefont {A.}~\bibnamefont {Becker}},
		\bibinfo {author} {\bibfnamefont {F.}~\bibnamefont {Berg}}, \bibinfo {author}
		{\bibfnamefont {K.}~\bibnamefont {Blaum}}, \bibinfo {author} {\bibfnamefont
			{C.}~\bibnamefont {Breitenfeldt}}, \bibinfo {author} {\bibfnamefont
			{H.}~\bibnamefont {Fadil}}, \bibinfo {author} {\bibfnamefont
			{F.}~\bibnamefont {Fellenberger}}, \bibinfo {author} {\bibfnamefont
			{M.}~\bibnamefont {Froese}}, \bibinfo {author} {\bibfnamefont
			{S.}~\bibnamefont {George}}, \bibinfo {author} {\bibfnamefont
			{J.}~\bibnamefont {G\"ock}}, \bibinfo {author} {\bibfnamefont
			{M.}~\bibnamefont {Grieser}}, \bibinfo {author} {\bibfnamefont
			{F.}~\bibnamefont {Grussie}}, \bibinfo {author} {\bibfnamefont {E.~A.}\
			\bibnamefont {Guerin}}, \bibinfo {author} {\bibfnamefont {O.}~\bibnamefont
			{Heber}}, \bibinfo {author} {\bibfnamefont {P.}~\bibnamefont {Herwig}},
		\bibinfo {author} {\bibfnamefont {J.}~\bibnamefont {Karthein}}, \bibinfo
		{author} {\bibfnamefont {C.}~\bibnamefont {Krantz}}, \bibinfo {author}
		{\bibfnamefont {H.}~\bibnamefont {Kreckel}}, \bibinfo {author} {\bibfnamefont
			{M.}~\bibnamefont {Lange}}, \bibinfo {author} {\bibfnamefont
			{F.}~\bibnamefont {Laux}}, \bibinfo {author} {\bibfnamefont {S.}~\bibnamefont
			{Lohmann}}, \bibinfo {author} {\bibfnamefont {S.}~\bibnamefont {Menk}},
		\bibinfo {author} {\bibfnamefont {C.}~\bibnamefont {Meyer}}, \bibinfo
		{author} {\bibfnamefont {P.~M.}\ \bibnamefont {Mishra}}, \bibinfo {author}
		{\bibfnamefont {O.}~\bibnamefont {Novotn\'y}}, \bibinfo {author}
		{\bibfnamefont {A.~P.}\ \bibnamefont {O'Connor}}, \bibinfo {author}
		{\bibfnamefont {D.~A.}\ \bibnamefont {Orlov}}, \bibinfo {author}
		{\bibfnamefont {M.~L.}\ \bibnamefont {Rappaport}}, \bibinfo {author}
		{\bibfnamefont {R.}~\bibnamefont {Repnow}}, \bibinfo {author} {\bibfnamefont
			{S.}~\bibnamefont {Saurabh}}, \bibinfo {author} {\bibfnamefont
			{S.}~\bibnamefont {Schippers}}, \bibinfo {author} {\bibfnamefont {C.~D.}\
			\bibnamefont {Schr\"oter}}, \bibinfo {author} {\bibfnamefont
			{D.}~\bibnamefont {Schwalm}}, \bibinfo {author} {\bibfnamefont
			{L.}~\bibnamefont {Schweikhard}}, \bibinfo {author} {\bibfnamefont
			{T.}~\bibnamefont {Sieber}}, \bibinfo {author} {\bibfnamefont
			{A.}~\bibnamefont {Shornikov}}, \bibinfo {author} {\bibfnamefont
			{K.}~\bibnamefont {Spruck}}, \bibinfo {author} {\bibfnamefont
			{S.}~\bibnamefont {Sunil~Kumar}}, \bibinfo {author} {\bibfnamefont
			{J.}~\bibnamefont {Ullrich}}, \bibinfo {author} {\bibfnamefont
			{X.}~\bibnamefont {Urbain}}, \bibinfo {author} {\bibfnamefont
			{S.}~\bibnamefont {Vogel}}, \bibinfo {author} {\bibfnamefont
			{P.}~\bibnamefont {Wilhelm}}, \bibinfo {author} {\bibfnamefont
			{A.}~\bibnamefont {Wolf}},\ and\ \bibinfo {author} {\bibfnamefont
			{D.}~\bibnamefont {Zajfman}},\ }\bibfield  {title} {\bibinfo {title} {The
			cryogenic storage ring {CSR}},\ }\href {https://doi.org/10.1063/1.4953888}
	{\bibfield  {journal} {\bibinfo  {journal} {Rev. Sci. Instrum.}\ }\textbf
		{\bibinfo {volume} {87}},\ \bibinfo {pages} {063115} (\bibinfo {year}
		{2016})}\BibitemShut {NoStop}%
	\bibitem [{\citenamefont {O'Connor}\ \emph {et~al.}(2016)\citenamefont
		{O'Connor}, \citenamefont {Becker}, \citenamefont {Blaum}, \citenamefont
		{Breitenfeldt}, \citenamefont {George}, \citenamefont {G\"ock}, \citenamefont
		{Grieser}, \citenamefont {Grussie}, \citenamefont {Guerin}, \citenamefont
		{von Hahn}, \citenamefont {Hechtfischer}, \citenamefont {Herwig},
		\citenamefont {Karthein}, \citenamefont {Krantz}, \citenamefont {Kreckel},
		\citenamefont {Lohmann}, \citenamefont {Meyer}, \citenamefont {Mishra},
		\citenamefont {Novotn\'y}, \citenamefont {Repnow}, \citenamefont {Saurabh},
		\citenamefont {Schwalm}, \citenamefont {Spruck}, \citenamefont {Sunil~Kumar},
		\citenamefont {Vogel},\ and\ \citenamefont
		{Wolf}}]{oconnor_photodissociation_2016}%
	\BibitemOpen
	\bibfield  {author} {\bibinfo {author} {\bibfnamefont {A.}~\bibnamefont
			{O'Connor}}, \bibinfo {author} {\bibfnamefont {A.}~\bibnamefont {Becker}},
		\bibinfo {author} {\bibfnamefont {K.}~\bibnamefont {Blaum}}, \bibinfo
		{author} {\bibfnamefont {C.}~\bibnamefont {Breitenfeldt}}, \bibinfo {author}
		{\bibfnamefont {S.}~\bibnamefont {George}}, \bibinfo {author} {\bibfnamefont
			{J.}~\bibnamefont {G\"ock}}, \bibinfo {author} {\bibfnamefont
			{M.}~\bibnamefont {Grieser}}, \bibinfo {author} {\bibfnamefont
			{F.}~\bibnamefont {Grussie}}, \bibinfo {author} {\bibfnamefont
			{E.}~\bibnamefont {Guerin}}, \bibinfo {author} {\bibfnamefont
			{R.}~\bibnamefont {von Hahn}}, \bibinfo {author} {\bibfnamefont
			{U.}~\bibnamefont {Hechtfischer}}, \bibinfo {author} {\bibfnamefont
			{P.}~\bibnamefont {Herwig}}, \bibinfo {author} {\bibfnamefont
			{J.}~\bibnamefont {Karthein}}, \bibinfo {author} {\bibfnamefont
			{C.}~\bibnamefont {Krantz}}, \bibinfo {author} {\bibfnamefont
			{H.}~\bibnamefont {Kreckel}}, \bibinfo {author} {\bibfnamefont
			{S.}~\bibnamefont {Lohmann}}, \bibinfo {author} {\bibfnamefont
			{C.}~\bibnamefont {Meyer}}, \bibinfo {author} {\bibfnamefont
			{P.}~\bibnamefont {Mishra}}, \bibinfo {author} {\bibfnamefont
			{O.}~\bibnamefont {Novotn\'y}}, \bibinfo {author} {\bibfnamefont
			{R.}~\bibnamefont {Repnow}}, \bibinfo {author} {\bibfnamefont
			{S.}~\bibnamefont {Saurabh}}, \bibinfo {author} {\bibfnamefont
			{D.}~\bibnamefont {Schwalm}}, \bibinfo {author} {\bibfnamefont
			{K.}~\bibnamefont {Spruck}}, \bibinfo {author} {\bibfnamefont
			{S.}~\bibnamefont {Sunil~Kumar}}, \bibinfo {author} {\bibfnamefont
			{S.}~\bibnamefont {Vogel}},\ and\ \bibinfo {author} {\bibfnamefont
			{A.}~\bibnamefont {Wolf}},\ }\bibfield  {title} {\bibinfo {title}
		{Photodissociation of an internally cold beam of {CH$^+$} ions in a cryogenic
			storage ring},\ }\href {https://doi.org/10.1103/PhysRevLett.116.113002}
	{\bibfield  {journal} {\bibinfo  {journal} {Phys. Rev. Lett.}\ }\textbf
		{\bibinfo {volume} {116}},\ \bibinfo {pages} {113002} (\bibinfo {year}
		{2016})}\BibitemShut {NoStop}%
	\bibitem [{\citenamefont {Meyer}\ \emph {et~al.}(2017)\citenamefont {Meyer},
		\citenamefont {Becker}, \citenamefont {Blaum}, \citenamefont {Breitenfeldt},
		\citenamefont {George}, \citenamefont {G\"ock}, \citenamefont {Grieser},
		\citenamefont {Grussie}, \citenamefont {Guerin}, \citenamefont {von Hahn},
		\citenamefont {Herwig}, \citenamefont {Krantz}, \citenamefont {Kreckel},
		\citenamefont {Lion}, \citenamefont {Lohmann}, \citenamefont {Mishra},
		\citenamefont {Novotn\'y}, \citenamefont {O'Connor}, \citenamefont {Repnow},
		\citenamefont {Saurabh}, \citenamefont {Schwalm}, \citenamefont
		{Schweikhard}, \citenamefont {Spruck}, \citenamefont {Sunil~Kumar},
		\citenamefont {Vogel},\ and\ \citenamefont {Wolf}}]{meyer_radiative_2017}%
	\BibitemOpen
	\bibfield  {author} {\bibinfo {author} {\bibfnamefont {C.}~\bibnamefont
			{Meyer}}, \bibinfo {author} {\bibfnamefont {A.}~\bibnamefont {Becker}},
		\bibinfo {author} {\bibfnamefont {K.}~\bibnamefont {Blaum}}, \bibinfo
		{author} {\bibfnamefont {C.}~\bibnamefont {Breitenfeldt}}, \bibinfo {author}
		{\bibfnamefont {S.}~\bibnamefont {George}}, \bibinfo {author} {\bibfnamefont
			{J.}~\bibnamefont {G\"ock}}, \bibinfo {author} {\bibfnamefont
			{M.}~\bibnamefont {Grieser}}, \bibinfo {author} {\bibfnamefont
			{F.}~\bibnamefont {Grussie}}, \bibinfo {author} {\bibfnamefont
			{E.}~\bibnamefont {Guerin}}, \bibinfo {author} {\bibfnamefont
			{R.}~\bibnamefont {von Hahn}}, \bibinfo {author} {\bibfnamefont
			{P.}~\bibnamefont {Herwig}}, \bibinfo {author} {\bibfnamefont
			{C.}~\bibnamefont {Krantz}}, \bibinfo {author} {\bibfnamefont
			{H.}~\bibnamefont {Kreckel}}, \bibinfo {author} {\bibfnamefont
			{J.}~\bibnamefont {Lion}}, \bibinfo {author} {\bibfnamefont {S.}~\bibnamefont
			{Lohmann}}, \bibinfo {author} {\bibfnamefont {P.}~\bibnamefont {Mishra}},
		\bibinfo {author} {\bibfnamefont {O.}~\bibnamefont {Novotn\'y}}, \bibinfo
		{author} {\bibfnamefont {A.}~\bibnamefont {O'Connor}}, \bibinfo {author}
		{\bibfnamefont {R.}~\bibnamefont {Repnow}}, \bibinfo {author} {\bibfnamefont
			{S.}~\bibnamefont {Saurabh}}, \bibinfo {author} {\bibfnamefont
			{D.}~\bibnamefont {Schwalm}}, \bibinfo {author} {\bibfnamefont
			{L.}~\bibnamefont {Schweikhard}}, \bibinfo {author} {\bibfnamefont
			{K.}~\bibnamefont {Spruck}}, \bibinfo {author} {\bibfnamefont
			{S.}~\bibnamefont {Sunil~Kumar}}, \bibinfo {author} {\bibfnamefont
			{S.}~\bibnamefont {Vogel}},\ and\ \bibinfo {author} {\bibfnamefont
			{A.}~\bibnamefont {Wolf}},\ }\bibfield  {title} {\bibinfo {title} {Radiative
			rotational lifetimes and state-resolved relative detachment cross sections
			from photodetachment thermometry of molecular anions in a cryogenic storage
			ring},\ }\href {https://doi.org/10.1103/PhysRevLett.119.023202} {\bibfield
		{journal} {\bibinfo  {journal} {Phys. Rev. Lett.}\ }\textbf {\bibinfo
			{volume} {119}},\ \bibinfo {pages} {023202} (\bibinfo {year}
		{2017})}\BibitemShut {NoStop}%
	\bibitem [{\citenamefont {Schmidt}\ \emph {et~al.}(2017)\citenamefont
		{Schmidt}, \citenamefont {Eklund}, \citenamefont {Chartkunchand},
		\citenamefont {Anderson}, \citenamefont {Kami\'nska}, \citenamefont
		{de~Ruette}, \citenamefont {Thomas}, \citenamefont {Kristiansson},
		\citenamefont {Gatchell}, \citenamefont {Reinhed}, \citenamefont {Ros\'en},
		\citenamefont {Simonsson}, \citenamefont {K\"a{}llberg}, \citenamefont
		{L\"o{}fgren}, \citenamefont {Mannervik}, \citenamefont {Zettergren},\ and\
		\citenamefont {Cederquist}}]{schmidt_rotationally_2017}%
	\BibitemOpen
	\bibfield  {author} {\bibinfo {author} {\bibfnamefont {H.}~\bibnamefont
			{Schmidt}}, \bibinfo {author} {\bibfnamefont {G.}~\bibnamefont {Eklund}},
		\bibinfo {author} {\bibfnamefont {K.}~\bibnamefont {Chartkunchand}}, \bibinfo
		{author} {\bibfnamefont {E.}~\bibnamefont {Anderson}}, \bibinfo {author}
		{\bibfnamefont {M.}~\bibnamefont {Kami\'nska}}, \bibinfo {author}
		{\bibfnamefont {N.}~\bibnamefont {de~Ruette}}, \bibinfo {author}
		{\bibfnamefont {R.}~\bibnamefont {Thomas}}, \bibinfo {author} {\bibfnamefont
			{M.}~\bibnamefont {Kristiansson}}, \bibinfo {author} {\bibfnamefont
			{M.}~\bibnamefont {Gatchell}}, \bibinfo {author} {\bibfnamefont
			{P.}~\bibnamefont {Reinhed}}, \bibinfo {author} {\bibfnamefont
			{S.}~\bibnamefont {Ros\'en}}, \bibinfo {author} {\bibfnamefont
			{A.}~\bibnamefont {Simonsson}}, \bibinfo {author} {\bibfnamefont
			{A.}~\bibnamefont {K\"a{}llberg}}, \bibinfo {author} {\bibfnamefont
			{P.}~\bibnamefont {L\"o{}fgren}}, \bibinfo {author} {\bibfnamefont
			{S.}~\bibnamefont {Mannervik}}, \bibinfo {author} {\bibfnamefont
			{H.}~\bibnamefont {Zettergren}},\ and\ \bibinfo {author} {\bibfnamefont
			{H.}~\bibnamefont {Cederquist}},\ }\bibfield  {title} {\bibinfo {title}
		{Rotationally cold {OH$^-$} ions in the cryogenic electrostatic ion-beam
			storage ring {DESIREE}},\ }\href
	{https://doi.org/10.1103/PhysRevLett.119.073001} {\bibfield  {journal}
		{\bibinfo  {journal} {Phys. Rev. Lett.}\ }\textbf {\bibinfo {volume} {119}},\
		\bibinfo {pages} {073001} (\bibinfo {year} {2017})}\BibitemShut {NoStop}%
	\bibitem [{\citenamefont {Schmidt}\ \emph {et~al.}(2018)\citenamefont
		{Schmidt}, \citenamefont {Eklund}, \citenamefont {Chartkunchand},
		\citenamefont {Anderson}, \citenamefont {Kami\'nska}, \citenamefont
		{de~Ruette}, \citenamefont {Thomas}, \citenamefont {Kristiansson},
		\citenamefont {Gatchell}, \citenamefont {Reinhed}, \citenamefont {Ros\'en},
		\citenamefont {Simonsson}, \citenamefont {K\"a{}llberg}, \citenamefont
		{L\"o{}fgren}, \citenamefont {Mannervik}, \citenamefont {Zettergren},\ and\
		\citenamefont {Cederquist}}]{schmidt_erratum_2018}%
	\BibitemOpen
	\bibfield  {author} {\bibinfo {author} {\bibfnamefont {H.}~\bibnamefont
			{Schmidt}}, \bibinfo {author} {\bibfnamefont {G.}~\bibnamefont {Eklund}},
		\bibinfo {author} {\bibfnamefont {K.}~\bibnamefont {Chartkunchand}}, \bibinfo
		{author} {\bibfnamefont {E.}~\bibnamefont {Anderson}}, \bibinfo {author}
		{\bibfnamefont {M.}~\bibnamefont {Kami\'nska}}, \bibinfo {author}
		{\bibfnamefont {N.}~\bibnamefont {de~Ruette}}, \bibinfo {author}
		{\bibfnamefont {R.}~\bibnamefont {Thomas}}, \bibinfo {author} {\bibfnamefont
			{M.}~\bibnamefont {Kristiansson}}, \bibinfo {author} {\bibfnamefont
			{M.}~\bibnamefont {Gatchell}}, \bibinfo {author} {\bibfnamefont
			{P.}~\bibnamefont {Reinhed}}, \bibinfo {author} {\bibfnamefont
			{S.}~\bibnamefont {Ros\'en}}, \bibinfo {author} {\bibfnamefont
			{A.}~\bibnamefont {Simonsson}}, \bibinfo {author} {\bibfnamefont
			{A.}~\bibnamefont {K\"a{}llberg}}, \bibinfo {author} {\bibfnamefont
			{P.}~\bibnamefont {L\"o{}fgren}}, \bibinfo {author} {\bibfnamefont
			{S.}~\bibnamefont {Mannervik}}, \bibinfo {author} {\bibfnamefont
			{H.}~\bibnamefont {Zettergren}},\ and\ \bibinfo {author} {\bibfnamefont
			{H.}~\bibnamefont {Cederquist}},\ }\href
	{https://doi.org/10.1103/PhysRevLett.121.079901} {\bibfield  {journal}
		{\bibinfo  {journal} {Phys. Rev. Lett.}\ }\textbf {\bibinfo {volume} {121}},\
		\bibinfo {pages} {079901} (\bibinfo {year} {2018})}\BibitemShut {NoStop}%
	\bibitem [{\citenamefont {Novotn\'y}\ \emph {et~al.}(2019)\citenamefont
		{Novotn\'y}, \citenamefont {Wilhelm}, \citenamefont {Paul}, \citenamefont
		{K\'alosi}, \citenamefont {Saurabh}, \citenamefont {Becker}, \citenamefont
		{Blaum}, \citenamefont {George}, \citenamefont {G\"ock}, \citenamefont
		{Grieser}, \citenamefont {Grussie}, \citenamefont {von Hahn}, \citenamefont
		{Krantz}, \citenamefont {Kreckel}, \citenamefont {Meyer}, \citenamefont
		{Mishra}, \citenamefont {Muell}, \citenamefont {Nuesslein}, \citenamefont
		{Orlov}, \citenamefont {Rimmler}, \citenamefont {Schmidt}, \citenamefont
		{Shornikov}, \citenamefont {Terekhov}, \citenamefont {Vogel}, \citenamefont
		{Zajfman},\ and\ \citenamefont {Wolf}}]{novotny_quantum-stateselective_2019}%
	\BibitemOpen
	\bibfield  {author} {\bibinfo {author} {\bibfnamefont {O.}~\bibnamefont
			{Novotn\'y}}, \bibinfo {author} {\bibfnamefont {P.}~\bibnamefont {Wilhelm}},
		\bibinfo {author} {\bibfnamefont {D.}~\bibnamefont {Paul}}, \bibinfo {author}
		{\bibfnamefont {A.}~\bibnamefont {K\'alosi}}, \bibinfo {author}
		{\bibfnamefont {S.}~\bibnamefont {Saurabh}}, \bibinfo {author} {\bibfnamefont
			{A.}~\bibnamefont {Becker}}, \bibinfo {author} {\bibfnamefont
			{K.}~\bibnamefont {Blaum}}, \bibinfo {author} {\bibfnamefont
			{S.}~\bibnamefont {George}}, \bibinfo {author} {\bibfnamefont
			{J.}~\bibnamefont {G\"ock}}, \bibinfo {author} {\bibfnamefont
			{M.}~\bibnamefont {Grieser}}, \bibinfo {author} {\bibfnamefont
			{F.}~\bibnamefont {Grussie}}, \bibinfo {author} {\bibfnamefont
			{R.}~\bibnamefont {von Hahn}}, \bibinfo {author} {\bibfnamefont
			{C.}~\bibnamefont {Krantz}}, \bibinfo {author} {\bibfnamefont
			{H.}~\bibnamefont {Kreckel}}, \bibinfo {author} {\bibfnamefont
			{C.}~\bibnamefont {Meyer}}, \bibinfo {author} {\bibfnamefont {P.~M.}\
			\bibnamefont {Mishra}}, \bibinfo {author} {\bibfnamefont {D.}~\bibnamefont
			{Muell}}, \bibinfo {author} {\bibfnamefont {F.}~\bibnamefont {Nuesslein}},
		\bibinfo {author} {\bibfnamefont {D.~A.}\ \bibnamefont {Orlov}}, \bibinfo
		{author} {\bibfnamefont {M.}~\bibnamefont {Rimmler}}, \bibinfo {author}
		{\bibfnamefont {V.~C.}\ \bibnamefont {Schmidt}}, \bibinfo {author}
		{\bibfnamefont {A.}~\bibnamefont {Shornikov}}, \bibinfo {author}
		{\bibfnamefont {A.~S.}\ \bibnamefont {Terekhov}}, \bibinfo {author}
		{\bibfnamefont {S.}~\bibnamefont {Vogel}}, \bibinfo {author} {\bibfnamefont
			{D.}~\bibnamefont {Zajfman}},\ and\ \bibinfo {author} {\bibfnamefont
			{A.}~\bibnamefont {Wolf}},\ }\bibfield  {title} {\bibinfo {title}
		{Quantum-state-selective electron recombination studies suggest enhanced
			abundance of primordial {HeH}$^{+}$},\ }\href
	{https://doi.org/10.1126/science.aax5921} {\bibfield  {journal} {\bibinfo
			{journal} {Science}\ }\textbf {\bibinfo {volume} {365}},\ \bibinfo {pages}
		{676} (\bibinfo {year} {2019})}\BibitemShut {NoStop}%
	\bibitem [{sup()}]{sup}%
	\BibitemOpen
	\href@noop {} {}\bibinfo {note} {Supplemental material on the molecular
		parameters of CH$^+$, the experimental method, and the modelling of the
		inelastic rates, which contains Refs.\ [30--49]}\BibitemShut {NoStop}%
	\bibitem [{\citenamefont {Hechtfischer}\ \emph {et~al.}(2002)\citenamefont
		{Hechtfischer}, \citenamefont {Williams}, \citenamefont {Lange},
		\citenamefont {Linkemann}, \citenamefont {Schwalm}, \citenamefont {Wester},
		\citenamefont {Wolf},\ and\ \citenamefont
		{Zajfman}}]{hechtfischer_photodissociation_2002}%
	\BibitemOpen
	\bibfield  {author} {\bibinfo {author} {\bibfnamefont {U.}~\bibnamefont
			{Hechtfischer}}, \bibinfo {author} {\bibfnamefont {C.~J.}\ \bibnamefont
			{Williams}}, \bibinfo {author} {\bibfnamefont {M.}~\bibnamefont {Lange}},
		\bibinfo {author} {\bibfnamefont {J.}~\bibnamefont {Linkemann}}, \bibinfo
		{author} {\bibfnamefont {D.}~\bibnamefont {Schwalm}}, \bibinfo {author}
		{\bibfnamefont {R.}~\bibnamefont {Wester}}, \bibinfo {author} {\bibfnamefont
			{A.}~\bibnamefont {Wolf}},\ and\ \bibinfo {author} {\bibfnamefont
			{D.}~\bibnamefont {Zajfman}},\ }\bibfield  {title} {\bibinfo {title}
		{Photodissociation spectroscopy of stored {CH$^+$} ions: {Detection},
			assignment, and close-coupled modeling of near-threshold {Feshbach}
			resonances},\ }\href {https://doi.org/10.1063/1.1513459} {\bibfield
		{journal} {\bibinfo  {journal} {J. Chem. Phys.}\ }\textbf {\bibinfo {volume}
			{117}},\ \bibinfo {pages} {8754} (\bibinfo {year} {2002})}\BibitemShut
	{NoStop}%
	\bibitem [{\citenamefont {Bernath}(2005)}]{bernath_spectra_2005}%
	\BibitemOpen
	\bibfield  {author} {\bibinfo {author} {\bibfnamefont {P.~F.}\ \bibnamefont
			{Bernath}},\ }\href@noop {} {\emph {\bibinfo {title} {Spectra of {Atoms} and
				{Molecules}}}},\ \bibinfo {edition} {2nd}\ ed.\ (\bibinfo  {publisher}
	{Oxford University Press},\ \bibinfo {address} {New York},\ \bibinfo {year}
	{2005})\BibitemShut {NoStop}%
	\bibitem [{\citenamefont {Hakalla}\ \emph {et~al.}(2006)\citenamefont
		{Hakalla}, \citenamefont {K\c{e}pa}, \citenamefont {Szajna},\ and\
		\citenamefont {Zachwieja}}]{hakalla_new_2006}%
	\BibitemOpen
	\bibfield  {author} {\bibinfo {author} {\bibfnamefont {R.}~\bibnamefont
			{Hakalla}}, \bibinfo {author} {\bibfnamefont {R.}~\bibnamefont {K\c{e}pa}},
		\bibinfo {author} {\bibfnamefont {W.}~\bibnamefont {Szajna}},\ and\ \bibinfo
		{author} {\bibfnamefont {M.}~\bibnamefont {Zachwieja}},\ }\bibfield  {title}
	{\bibinfo {title} {New analysis of the {Douglas}-{Herzberg} system
			{($A^1\Pi$-$X^1\Sigma^+$)} in the {CH$^+$} ion radical},\ }\href
	{https://doi.org/10.1140/epjd/e2006-00063-9} {\bibfield  {journal} {\bibinfo
			{journal} {Eur. Phys. J. D}\ }\textbf {\bibinfo {volume} {38}},\ \bibinfo
		{pages} {481} (\bibinfo {year} {2006})}\BibitemShut {NoStop}%
	\bibitem [{\citenamefont {Hechtfischer}\ \emph {et~al.}(2007)\citenamefont
		{Hechtfischer}, \citenamefont {Rostas}, \citenamefont {Lange}, \citenamefont
		{Linkemann}, \citenamefont {Schwalm}, \citenamefont {Wester}, \citenamefont
		{Wolf},\ and\ \citenamefont {Zajfman}}]{hechtfischer_photodissociation_2007}%
	\BibitemOpen
	\bibfield  {author} {\bibinfo {author} {\bibfnamefont {U.}~\bibnamefont
			{Hechtfischer}}, \bibinfo {author} {\bibfnamefont {J.}~\bibnamefont
			{Rostas}}, \bibinfo {author} {\bibfnamefont {M.}~\bibnamefont {Lange}},
		\bibinfo {author} {\bibfnamefont {J.}~\bibnamefont {Linkemann}}, \bibinfo
		{author} {\bibfnamefont {D.}~\bibnamefont {Schwalm}}, \bibinfo {author}
		{\bibfnamefont {R.}~\bibnamefont {Wester}}, \bibinfo {author} {\bibfnamefont
			{A.}~\bibnamefont {Wolf}},\ and\ \bibinfo {author} {\bibfnamefont
			{D.}~\bibnamefont {Zajfman}},\ }\bibfield  {title} {\bibinfo {title}
		{Photodissociation spectroscopy of stored {CH$^+$} and {CD$^+$} ions:
			{Analysis} of the {$b^3\Sigma^-$--$a^3\Pi$} system},\ }\href
	{https://doi.org/10.1063/1.2800004} {\bibfield  {journal} {\bibinfo
			{journal} {J. Chem. Phys.}\ }\textbf {\bibinfo {volume} {127}},\ \bibinfo
		{pages} {204304} (\bibinfo {year} {2007})}\BibitemShut {NoStop}%
	\bibitem [{\citenamefont {Kusunoki}\ and\ \citenamefont
		{Ottinger}(1980)}]{kusunoki_triplet_1980}%
	\BibitemOpen
	\bibfield  {author} {\bibinfo {author} {\bibfnamefont {I.}~\bibnamefont
			{Kusunoki}}\ and\ \bibinfo {author} {\bibfnamefont {C.}~\bibnamefont
			{Ottinger}},\ }\bibfield  {title} {\bibinfo {title} {Triplet {CH$^+$(CD$^+$)}
			emission from chemiluminescent ion--molecule reaction {C$^+$($^4P$)+
				H$_2$(D$_2$)}},\ }\href {https://doi.org/10.1063/1.440401} {\bibfield
		{journal} {\bibinfo  {journal} {J. Chem. Phys.}\ }\textbf {\bibinfo {volume}
			{73}},\ \bibinfo {pages} {2069} (\bibinfo {year} {1980})}\BibitemShut
	{NoStop}%
	\bibitem [{\citenamefont {Dom\'{e}nech}\ \emph {et~al.}(2018)\citenamefont
		{Dom\'{e}nech}, \citenamefont {Jusko}, \citenamefont {Schlemmer},\ and\
		\citenamefont {Asvany}}]{domenech_first_2018}%
	\BibitemOpen
	\bibfield  {author} {\bibinfo {author} {\bibfnamefont {J.~L.}\ \bibnamefont
			{Dom\'{e}nech}}, \bibinfo {author} {\bibfnamefont {P.}~\bibnamefont {Jusko}},
		\bibinfo {author} {\bibfnamefont {S.}~\bibnamefont {Schlemmer}},\ and\
		\bibinfo {author} {\bibfnamefont {O.}~\bibnamefont {Asvany}},\ }\bibfield
	{title} {\bibinfo {title} {The first laboratory detection of
			vibration-rotation transitions of $^{\textrm{12}}${CH}$^{\textrm{+}}$ and
			$^{\textrm{13}}${CH}$^{\textrm{+}}$ and improved measurement of their
			rotational transition frequencies},\ }\href
	{https://doi.org/10.3847/1538-4357/aab36a} {\bibfield  {journal} {\bibinfo
			{journal} {Astrophys. J.}\ }\textbf {\bibinfo {volume} {857}},\ \bibinfo
		{pages} {61} (\bibinfo {year} {2018})}\BibitemShut {NoStop}%
	\bibitem [{\citenamefont {Cheng}\ \emph {et~al.}(2007)\citenamefont {Cheng},
		\citenamefont {Brown}, \citenamefont {Rosmus}, \citenamefont {Linguerri},
		\citenamefont {Komiha},\ and\ \citenamefont {Myers}}]{cheng_dipole_2007}%
	\BibitemOpen
	\bibfield  {author} {\bibinfo {author} {\bibfnamefont {M.}~\bibnamefont
			{Cheng}}, \bibinfo {author} {\bibfnamefont {J.~M.}\ \bibnamefont {Brown}},
		\bibinfo {author} {\bibfnamefont {P.}~\bibnamefont {Rosmus}}, \bibinfo
		{author} {\bibfnamefont {R.}~\bibnamefont {Linguerri}}, \bibinfo {author}
		{\bibfnamefont {N.}~\bibnamefont {Komiha}},\ and\ \bibinfo {author}
		{\bibfnamefont {E.~G.}\ \bibnamefont {Myers}},\ }\bibfield  {title} {\bibinfo
		{title} {Dipole moments and orientation polarizabilities of diatomic
			molecular ions for precision atomic mass measurement},\ }\href
	{https://doi.org/10.1103/PhysRevA.75.012502} {\bibfield  {journal} {\bibinfo
			{journal} {Phys. Rev. A}\ }\textbf {\bibinfo {volume} {75}},\ \bibinfo
		{pages} {012502} (\bibinfo {year} {2007})}\BibitemShut {NoStop}%
	\bibitem [{\citenamefont {Amitay}\ \emph {et~al.}(1996)\citenamefont {Amitay},
		\citenamefont {Zajfman}, \citenamefont {Forck}, \citenamefont {Hechtfischer},
		\citenamefont {Seidel}, \citenamefont {Grieser}, \citenamefont {Habs},
		\citenamefont {Repnow}, \citenamefont {Schwalm},\ and\ \citenamefont
		{Wolf}}]{amitay_dissociative_1996}%
	\BibitemOpen
	\bibfield  {author} {\bibinfo {author} {\bibfnamefont {Z.}~\bibnamefont
			{Amitay}}, \bibinfo {author} {\bibfnamefont {D.}~\bibnamefont {Zajfman}},
		\bibinfo {author} {\bibfnamefont {P.}~\bibnamefont {Forck}}, \bibinfo
		{author} {\bibfnamefont {U.}~\bibnamefont {Hechtfischer}}, \bibinfo {author}
		{\bibfnamefont {B.}~\bibnamefont {Seidel}}, \bibinfo {author} {\bibfnamefont
			{M.}~\bibnamefont {Grieser}}, \bibinfo {author} {\bibfnamefont
			{D.}~\bibnamefont {Habs}}, \bibinfo {author} {\bibfnamefont {R.}~\bibnamefont
			{Repnow}}, \bibinfo {author} {\bibfnamefont {D.}~\bibnamefont {Schwalm}},\
		and\ \bibinfo {author} {\bibfnamefont {A.}~\bibnamefont {Wolf}},\ }\bibfield
	{title} {\bibinfo {title} {Dissociative recombination of {CH$^+$}: {Cross}
			section and final states},\ }\href {https://doi.org/10.1103/PhysRevA.54.4032}
	{\bibfield  {journal} {\bibinfo  {journal} {Phys. Rev. A}\ }\textbf {\bibinfo
			{volume} {54}},\ \bibinfo {pages} {4032} (\bibinfo {year}
		{1996})}\BibitemShut {NoStop}%
	\bibitem [{\citenamefont {Barinovs}\ and\ \citenamefont {van
			Hemert}(2004)}]{barinovs_ch_2004}%
	\BibitemOpen
	\bibfield  {author} {\bibinfo {author} {\bibfnamefont {{\u{G}}.}~\bibnamefont
			{Barinovs}}\ and\ \bibinfo {author} {\bibfnamefont {M.~C.}\ \bibnamefont {van
				Hemert}},\ }\bibfield  {title} {\bibinfo {title} {{CH$^+$} potential energy
			curves and photodissociation cross-section},\ }\href
	{https://doi.org/10.1016/j.cplett.2004.10.035} {\bibfield  {journal}
		{\bibinfo  {journal} {Chem. Phys. Lett.}\ }\textbf {\bibinfo {volume}
			{399}},\ \bibinfo {pages} {406} (\bibinfo {year} {2004})}\BibitemShut
	{NoStop}%
	\bibitem [{\citenamefont {Orlov}\ \emph {et~al.}(2003)\citenamefont {Orlov},
		\citenamefont {Weigel}, \citenamefont {Hoppe}, \citenamefont {Schwalm},
		\citenamefont {Jaroshevich}, \citenamefont {Terekhov},\ and\ \citenamefont
		{Wolf}}]{orlov_cold_2003}%
	\BibitemOpen
	\bibfield  {author} {\bibinfo {author} {\bibfnamefont {D.~A.}\ \bibnamefont
			{Orlov}}, \bibinfo {author} {\bibfnamefont {U.}~\bibnamefont {Weigel}},
		\bibinfo {author} {\bibfnamefont {M.}~\bibnamefont {Hoppe}}, \bibinfo
		{author} {\bibfnamefont {D.}~\bibnamefont {Schwalm}}, \bibinfo {author}
		{\bibfnamefont {A.~S.}\ \bibnamefont {Jaroshevich}}, \bibinfo {author}
		{\bibfnamefont {A.~S.}\ \bibnamefont {Terekhov}},\ and\ \bibinfo {author}
		{\bibfnamefont {A.}~\bibnamefont {Wolf}},\ }\bibfield  {title} {\bibinfo
		{title} {Cold electrons from cryogenic {GaAs} photocathodes: {Energetic} and
			angular distributions},\ }\href
	{https://doi.org/10.1023/B:HYPE.0000004226.23869.2b} {\bibfield  {journal}
		{\bibinfo  {journal} {Hyperfine Interact.}\ }\textbf {\bibinfo {volume}
			{146/147}},\ \bibinfo {pages} {215} (\bibinfo {year} {2003})}\BibitemShut
	{NoStop}%
	\bibitem [{\citenamefont {Pastuszka}\ \emph {et~al.}(1996)\citenamefont
		{Pastuszka}, \citenamefont {Schramm}, \citenamefont {Grieser}, \citenamefont
		{Broude}, \citenamefont {Grimm}, \citenamefont {Habs}, \citenamefont
		{Kenntner}, \citenamefont {Miesner}, \citenamefont {Sch\"{u}\ss{}ler},
		\citenamefont {Schwalm},\ and\ \citenamefont
		{Wolf}}]{pastuszka_electron_1996}%
	\BibitemOpen
	\bibfield  {author} {\bibinfo {author} {\bibfnamefont {S.}~\bibnamefont
			{Pastuszka}}, \bibinfo {author} {\bibfnamefont {U.}~\bibnamefont {Schramm}},
		\bibinfo {author} {\bibfnamefont {M.}~\bibnamefont {Grieser}}, \bibinfo
		{author} {\bibfnamefont {C.}~\bibnamefont {Broude}}, \bibinfo {author}
		{\bibfnamefont {R.}~\bibnamefont {Grimm}}, \bibinfo {author} {\bibfnamefont
			{D.}~\bibnamefont {Habs}}, \bibinfo {author} {\bibfnamefont {J.}~\bibnamefont
			{Kenntner}}, \bibinfo {author} {\bibfnamefont {H.-J.}\ \bibnamefont
			{Miesner}}, \bibinfo {author} {\bibfnamefont {T.}~\bibnamefont
			{Sch\"{u}\ss{}ler}}, \bibinfo {author} {\bibfnamefont {D.}~\bibnamefont
			{Schwalm}},\ and\ \bibinfo {author} {\bibfnamefont {A.}~\bibnamefont
			{Wolf}},\ }\bibfield  {title} {\bibinfo {title} {Electron cooling and
			recombination experiments with an adiabatically expanded electron beam},\
	}\href {https://doi.org/10.1016/0168-9002(95)00786-5} {\bibfield  {journal}
		{\bibinfo  {journal} {Nucl. Instrum. Methods Phys. Res. Sect. A -- Accel.
				Spectrom. Dect. Assoc. Equip.}\ }\textbf {\bibinfo {volume} {369}},\ \bibinfo
		{pages} {11} (\bibinfo {year} {1996})}\BibitemShut {NoStop}%
	\bibitem [{bud()}]{budker_electron_1978}%
	\BibitemOpen
	\href@noop {} {}\bibinfo {note} {G. I. Budker and A. N. Skrinski\u{\i},
		Electron cooling and new possibilities in elementary particle physics, Usp.
		Fiz. Nauk {\bf 124}, 561--595 (1978)
		\href{https://iopscience.iop.org/article/10.1070/PU1978v021n04ABEH005537/meta}
		{[Sov. Phys. Usp. {\bf 21}, 277--296 (1986)]}}\BibitemShut {NoStop}%
	\bibitem [{\citenamefont {Krantz}\ \emph {et~al.}(2021)\citenamefont {Krantz},
		\citenamefont {Buhr}, \citenamefont {Grieser}, \citenamefont {Lestinsky},
		\citenamefont {Novotn\'y}, \citenamefont {Novotny}, \citenamefont {Orlov},
		\citenamefont {Repnow}, \citenamefont {Terekhov}, \citenamefont {Wilhelm},\
		and\ \citenamefont {Wolf}}]{krantz_transverse_2021}%
	\BibitemOpen
	\bibfield  {author} {\bibinfo {author} {\bibfnamefont {C.}~\bibnamefont
			{Krantz}}, \bibinfo {author} {\bibfnamefont {H.}~\bibnamefont {Buhr}},
		\bibinfo {author} {\bibfnamefont {M.}~\bibnamefont {Grieser}}, \bibinfo
		{author} {\bibfnamefont {M.}~\bibnamefont {Lestinsky}}, \bibinfo {author}
		{\bibfnamefont {O.}~\bibnamefont {Novotn\'y}}, \bibinfo {author}
		{\bibfnamefont {S.}~\bibnamefont {Novotny}}, \bibinfo {author} {\bibfnamefont
			{D.}~\bibnamefont {Orlov}}, \bibinfo {author} {\bibfnamefont
			{R.}~\bibnamefont {Repnow}}, \bibinfo {author} {\bibfnamefont
			{A.}~\bibnamefont {Terekhov}}, \bibinfo {author} {\bibfnamefont
			{P.}~\bibnamefont {Wilhelm}},\ and\ \bibinfo {author} {\bibfnamefont
			{A.}~\bibnamefont {Wolf}},\ }\bibfield  {title} {\bibinfo {title} {Transverse
			electron cooling of heavy molecular ions},\ }\href
	{https://doi.org/10.1103/PhysRevAccelBeams.24.050101} {\bibfield  {journal}
		{\bibinfo  {journal} {Phys. Rev. Accel. Beams}\ }\textbf {\bibinfo {volume}
			{24}},\ \bibinfo {pages} {050101} (\bibinfo {year} {2021})}\BibitemShut
	{NoStop}%
	\bibitem [{\citenamefont {van Hoof}\ \emph {et~al.}(2014)\citenamefont {van
			Hoof}, \citenamefont {Williams}, \citenamefont {Volk}, \citenamefont
		{Chatzikos}, \citenamefont {Ferland}, \citenamefont {Lykins}, \citenamefont
		{Porter},\ and\ \citenamefont {Wang}}]{van_hoof_accurate_2014}%
	\BibitemOpen
	\bibfield  {author} {\bibinfo {author} {\bibfnamefont {P.~A.~M.}\
			\bibnamefont {van Hoof}}, \bibinfo {author} {\bibfnamefont {R.~J.~R.}\
			\bibnamefont {Williams}}, \bibinfo {author} {\bibfnamefont {K.}~\bibnamefont
			{Volk}}, \bibinfo {author} {\bibfnamefont {M.}~\bibnamefont {Chatzikos}},
		\bibinfo {author} {\bibfnamefont {G.~J.}\ \bibnamefont {Ferland}}, \bibinfo
		{author} {\bibfnamefont {M.}~\bibnamefont {Lykins}}, \bibinfo {author}
		{\bibfnamefont {R.~L.}\ \bibnamefont {Porter}},\ and\ \bibinfo {author}
		{\bibfnamefont {Y.}~\bibnamefont {Wang}},\ }\bibfield  {title} {\bibinfo
		{title} {Accurate determination of the free--free {Gaunt} factor -- {I}.
			{Non}-relativistic {Gaunt} factors},\ }\href
	{https://doi.org/10.1093/mnras/stu1438} {\bibfield  {journal} {\bibinfo
			{journal} {Mon. Not. R. Astron. Soc.}\ }\textbf {\bibinfo {volume} {444}},\
		\bibinfo {pages} {420} (\bibinfo {year} {2014})}\BibitemShut {NoStop}%
	\bibitem [{\citenamefont {Hogg}\ and\ \citenamefont
		{Foreman-Mackey}(2018)}]{mcmc}%
	\BibitemOpen
	\bibfield  {author} {\bibinfo {author} {\bibfnamefont {D.~W.}\ \bibnamefont
			{Hogg}}\ and\ \bibinfo {author} {\bibfnamefont {D.}~\bibnamefont
			{Foreman-Mackey}},\ }\bibfield  {title} {\bibinfo {title} {Data analysis
			recipes: {Using} {Markov} {Chain} {Monte} {Carlo}},\ }\href
	{https://doi.org/10.3847/1538-4365/aab76e} {\bibfield  {journal} {\bibinfo
			{journal} {Astrophs. J. Suppl. Ser.}\ }\textbf {\bibinfo {volume} {236}},\
		\bibinfo {pages} {11} (\bibinfo {year} {2018})}\BibitemShut {NoStop}%
	\bibitem [{\citenamefont {Foreman-Mackey}\ \emph {et~al.}(2013)\citenamefont
		{Foreman-Mackey}, \citenamefont {Hogg}, \citenamefont {Lang},\ and\
		\citenamefont {Goodman}}]{Foreman_Mackey_2013}%
	\BibitemOpen
	\bibfield  {author} {\bibinfo {author} {\bibfnamefont {D.}~\bibnamefont
			{Foreman-Mackey}}, \bibinfo {author} {\bibfnamefont {D.~W.}\ \bibnamefont
			{Hogg}}, \bibinfo {author} {\bibfnamefont {D.}~\bibnamefont {Lang}},\ and\
		\bibinfo {author} {\bibfnamefont {J.}~\bibnamefont {Goodman}},\ }\bibfield
	{title} {\bibinfo {title} {{emcee}: The {MCMC} hammer},\ }\href
	{https://doi.org/10.1086/670067} {\bibfield  {journal} {\bibinfo  {journal}
			{Publ. Astron. Soc. Pac.}\ }\textbf {\bibinfo {volume} {125}},\ \bibinfo
		{pages} {306} (\bibinfo {year} {2013})}\BibitemShut {NoStop}%
	\bibitem [{\citenamefont {Mezei}\ \emph {et~al.}(2019)\citenamefont {Mezei},
		\citenamefont {Ep{\'e}e~Ep{\'e}e}, \citenamefont {Motapon},\ and\
		\citenamefont {Schneider}}]{mezei_atom_2019}%
	\BibitemOpen
	\bibfield  {author} {\bibinfo {author} {\bibfnamefont {Z.~J.}\ \bibnamefont
			{Mezei}}, \bibinfo {author} {\bibfnamefont {M.~D.}\ \bibnamefont
			{Ep{\'e}e~Ep{\'e}e}}, \bibinfo {author} {\bibfnamefont {O.}~\bibnamefont
			{Motapon}},\ and\ \bibinfo {author} {\bibfnamefont {I.~F.}\ \bibnamefont
			{Schneider}},\ }\bibfield  {title} {\bibinfo {title} {Dissociative
			recombination of {CH}$^+$ molecular ion induced by very low energy
			electrons},\ }\href {https://doi.org/10.3390/atoms7030082} {\bibfield
		{journal} {\bibinfo  {journal} {Atoms}\ }\textbf {\bibinfo {volume} {7}},\
		\bibinfo {pages} {82} (\bibinfo {year} {2019})}\BibitemShut {NoStop}%
	\bibitem [{\citenamefont {V{\' a}zquez}\ \emph {et~al.}(2007)\citenamefont
		{V{\' a}zquez}, \citenamefont {Amero}, \citenamefont {Liebermann},
		\citenamefont {Buenker},\ and\ \citenamefont
		{Lefebvre-Brion}}]{vazquez_rydbergch_2007}%
	\BibitemOpen
	\bibfield  {author} {\bibinfo {author} {\bibfnamefont {G.~J.}\ \bibnamefont
			{V{\' a}zquez}}, \bibinfo {author} {\bibfnamefont {J.~M.}\ \bibnamefont
			{Amero}}, \bibinfo {author} {\bibfnamefont {H.~P.}\ \bibnamefont
			{Liebermann}}, \bibinfo {author} {\bibfnamefont {R.~J.}\ \bibnamefont
			{Buenker}},\ and\ \bibinfo {author} {\bibfnamefont {H.}~\bibnamefont
			{Lefebvre-Brion}},\ }\bibfield  {title} {\bibinfo {title} {Insight into the
			{Rydberg} states of {CH}},\ }\href {https://doi.org/10.1063/1.2721535}
	{\bibfield  {journal} {\bibinfo  {journal} {J. Chem. Phys.}\ }\textbf
		{\bibinfo {volume} {126}},\ \bibinfo {pages} {164302} (\bibinfo {year}
		{2007})}\BibitemShut {NoStop}%
	\bibitem [{\citenamefont {Larsson}\ and\ \citenamefont {Orel}(2008)}]{drbook}%
	\BibitemOpen
	\bibfield  {author} {\bibinfo {author} {\bibfnamefont {M.}~\bibnamefont
			{Larsson}}\ and\ \bibinfo {author} {\bibfnamefont {A.~E.}\ \bibnamefont
			{Orel}},\ }\href@noop {} {\emph {\bibinfo {title} {Dissociative Recombination
				of Molecular Ions}}}\ (\bibinfo  {publisher} {Cambridge University Press,
		Cambridge},\ \bibinfo {year} {2008})\BibitemShut {NoStop}%
	\bibitem [{\citenamefont {Hechtfischer}\ \emph {et~al.}(1998)\citenamefont
		{Hechtfischer}, \citenamefont {Amitay}, \citenamefont {Forck}, \citenamefont
		{Lange}, \citenamefont {Linkemann}, \citenamefont {Schmitt}, \citenamefont
		{Schramm}, \citenamefont {Schwalm}, \citenamefont {Wester}, \citenamefont
		{Zajfman},\ and\ \citenamefont {Wolf}}]{hechtfischer_near-threshold_1998}%
	\BibitemOpen
	\bibfield  {author} {\bibinfo {author} {\bibfnamefont {U.}~\bibnamefont
			{Hechtfischer}}, \bibinfo {author} {\bibfnamefont {Z.}~\bibnamefont
			{Amitay}}, \bibinfo {author} {\bibfnamefont {P.}~\bibnamefont {Forck}},
		\bibinfo {author} {\bibfnamefont {M.}~\bibnamefont {Lange}}, \bibinfo
		{author} {\bibfnamefont {J.}~\bibnamefont {Linkemann}}, \bibinfo {author}
		{\bibfnamefont {M.}~\bibnamefont {Schmitt}}, \bibinfo {author} {\bibfnamefont
			{U.}~\bibnamefont {Schramm}}, \bibinfo {author} {\bibfnamefont
			{D.}~\bibnamefont {Schwalm}}, \bibinfo {author} {\bibfnamefont
			{R.}~\bibnamefont {Wester}}, \bibinfo {author} {\bibfnamefont
			{D.}~\bibnamefont {Zajfman}},\ and\ \bibinfo {author} {\bibfnamefont
			{A.}~\bibnamefont {Wolf}},\ }\bibfield  {title} {\bibinfo {title}
		{Near-threshold photodissociation of cold {CH$^+$} in a storage ring},\
	}\href {https://doi.org/10.1103/PhysRevLett.80.2809} {\bibfield  {journal}
		{\bibinfo  {journal} {Phys. Rev. Lett.}\ }\textbf {\bibinfo {volume} {80}},\
		\bibinfo {pages} {2809} (\bibinfo {year} {1998})}\BibitemShut {NoStop}%
	\bibitem [{\citenamefont {Kramida}\ \emph {et~al.}(2020)\citenamefont
		{Kramida}, \citenamefont {{Yu.~Ralchenko}}, \citenamefont {Reader},\ and\
		\citenamefont {{NIST ASD Team}}}]{NIST_ASD}%
	\BibitemOpen
	\bibfield  {author} {\bibinfo {author} {\bibfnamefont {A.}~\bibnamefont
			{Kramida}}, \bibinfo {author} {\bibnamefont {{Yu.~Ralchenko}}}, \bibinfo
		{author} {\bibfnamefont {J.}~\bibnamefont {Reader}},\ and\ \bibinfo {author}
		{\bibnamefont {{NIST ASD Team}}},\ }\href@noop {} {}\bibinfo {howpublished}
	{{NIST Atomic Spectra Database (ver. 5.8), [Online]. Available:
			{\url{https://physics.nist.gov/asd}} [2021, April 27]. National Institute of
			Standards and Technology, Gaithersburg, MD}} (\bibinfo {year}
	{2020})\BibitemShut {NoStop}%
	\bibitem [{\citenamefont {Znotins}\ \emph {et~al.}(2021)\citenamefont
		{Znotins}, \citenamefont {Grussie}, \citenamefont {Wolf}, \citenamefont
		{Urbain},\ and\ \citenamefont {Kreckel}}]{znotins_approach_2021}%
	\BibitemOpen
	\bibfield  {author} {\bibinfo {author} {\bibfnamefont {A.}~\bibnamefont
			{Znotins}}, \bibinfo {author} {\bibfnamefont {F.}~\bibnamefont {Grussie}},
		\bibinfo {author} {\bibfnamefont {A.}~\bibnamefont {Wolf}}, \bibinfo {author}
		{\bibfnamefont {X.}~\bibnamefont {Urbain}},\ and\ \bibinfo {author}
		{\bibfnamefont {H.}~\bibnamefont {Kreckel}},\ }\bibfield  {title} {\bibinfo
		{title} {An approach for multi-color action spectroscopy of highly excited
			states of {H$_3^+$}},\ }\href {https://doi.org/10.1016/j.jms.2021.111476}
	{\bibfield  {journal} {\bibinfo  {journal} {J. Molec. Spectrosc.}\ }\textbf
		{\bibinfo {volume} {378}},\ \bibinfo {pages} {111476} (\bibinfo {year}
		{2021})}\BibitemShut {NoStop}%
	\bibitem [{\citenamefont {Roth}\ \emph {et~al.}(2006)\citenamefont {Roth},
		\citenamefont {Koelemeij}, \citenamefont {Daerr},\ and\ \citenamefont
		{Schiller}}]{roth_rovibrational_2006}%
	\BibitemOpen
	\bibfield  {author} {\bibinfo {author} {\bibfnamefont {B.}~\bibnamefont
			{Roth}}, \bibinfo {author} {\bibfnamefont {J.~C.~J.}\ \bibnamefont
			{Koelemeij}}, \bibinfo {author} {\bibfnamefont {H.}~\bibnamefont {Daerr}},\
		and\ \bibinfo {author} {\bibfnamefont {S.}~\bibnamefont {Schiller}},\
	}\bibfield  {title} {\bibinfo {title} {Rovibrational spectroscopy of trapped
			molecular hydrogen ions at millikelvin temperatures},\ }\href
	{https://doi.org/10.1103/PhysRevA.74.040501} {\bibfield  {journal} {\bibinfo
			{journal} {Phys. Rev. A}\ }\textbf {\bibinfo {volume} {74}},\ \bibinfo
		{pages} {040501} (\bibinfo {year} {2006})}\BibitemShut {NoStop}%
\end{thebibliography}
\end{document}


%
	{\raggedright%
		\begin{minipage}[b]{\textwidth} {\bfseries{\small{Online Supplemental
						Material}\par}\vspace{8mm} \bfseries{
					{Laser-probing the rotational cooling of molecular ions by electron
						collisions
					}%
				}\par}
		\end{minipage}\\[2mm]
		\begin{minipage}[b]{\textwidth} \raggedright \small
			\'Abel~K\'alosi$^{1,2}$,
			Manfred~Grieser$^{1}$,
			Robert~von~Hahn$^{1}$,
			Ulrich~Hechtfischer$^{1,\dagger}$,
			Claude~Krantz$^{1,\ddagger}$,
			Holger~Kreckel$^{1}$,
			Damian~M\"ull$^{1}$,
			Daniel~Paul$^{1}$,
			Daniel~W.~Savin$^{2}$,
			Patrick Wilhelm$^{1}$,
			Andreas~Wolf$^{1}$,
			Old\v{r}ich~Novotn\'y$^{1}$
		\end{minipage}\\[3mm]%
		\begin{minipage}{\textwidth} \raggedright\itshape \footnotesize%
			$^1$Max-Planck-Institut f\"ur Kernphysik, 69117 Heidelberg,
			Germany\\
			$^2$Columbia Astrophysics Laboratory, Columbia University, New York, NY 10027, USA\\
			$^\dagger$\erc{Now at: ASML Nederland B.V., Veldhoven, The Netherlands}\\
			$^\ddagger$Present address: GSI Helmholtz Centre for Heavy Ion Research, 64291 Darmstadt
		\end{minipage}
		\par}
	\vspace{5mm}
	\hrule
	\vspace{10mm}
	\thispagestyle{empty}
	\def\theequation{S\arabic{equation}}
	\def\thefigure{S\arabic{figure}}
	\def\thetable{S\arabic{table}}
	
\newcommand{\tnu}{\tilde{\nu}}
\newcommand{\cmm}{cm$^{-1}$}
\newcommand{\citen}[1]{\citeauthor{#1}\ \cite{#1}}


\noindent
In this Supplemental Material we present the background of our measurements on
inelastic electron collisions with CH$^+$ and the modeling calculations we use
for analyzing the results.  We summarize the basic parameters of CH$^+$ and its
radiative rotational cooling.  Details are given for the experimental procedure
of the laser measurements to determine rotational level populations.  We then
describe the parameters of our merged-beams electron--ion interaction setup and
discuss how the measurements are affected by a metastable electronic state of
CH$^+$ and by the rotational-level dependent dissociative recombination of this
ion.  We also summarize calculations on the inelastic electron collision cross
sections and the resulting rotational transition rates for our study.

\vspace{1cm}
{\large\bfseries\noindent List of Supplemental Material\par}
\vspace{0.5cm}


\startcontents{}    
\printcontents{}{1}[2]{}

\newpage

\section{CH$^+$ energy levels and radiative cooling}

%
%
\begin{table}[b]
  \caption{
    \label{tab:molpar} 
    Experimental values for the molecular parameters in the vibrational ground
    states of the three lowest electronic levels of CH$^+$.  The
    $X\,^1\Sigma^+$, $v=0$ dissociation energy to C$^+(^2P_{1/2})$ + H is
    32\,946.7(11) cm$^{-1}$ \cite{hechtfischer_photodissociation_2002}.  $B_0$
    and $D_0$ here are the spectroscopic constants in the usual rotational term
    definition, see Eq.\ (6.47) in Ref.\ \cite{bernath_spectra_2005}; indices 0
    refer to vibrational quantum number $v=0$ and $e$ to the minimum of the
    electronic potential; $\omega_e$ is the vibrational constant; $r_e$, $r_0$
    are internuclear distances; column $T_e$ lists the electronic term value
    unless noted otherwise.  ($D_0$ denotes the dissociation energy in the main
    paper.)}  \vspace{2mm} \centering
  \begin{ruledtabular}
  \begin{tabular}{lllllllll}
\multicolumn{1}{c}{Level} & 
\multicolumn{1}{c}{$T_e$} & 
\multicolumn{1}{c}{$B_0$} & 
\multicolumn{1}{c}{$D_0$} & 
\multicolumn{1}{c}{$\omega_e$} & 
\multicolumn{1}{c}{$B_e$}  & 
\multicolumn{1}{c}{$r_e$} & 
\multicolumn{1}{c}{$r_0$%
\footnote{Scaled by using $r_0=(B_e/B_0)^{1/2}r_e$; rounded value}} \\
  &   \multicolumn{1}{c}{(cm$^{-1}$)} & 
\multicolumn{1}{c}{(cm$^{-1}$)} & 
\multicolumn{1}{c}{($10^{-3}$\,cm$^{-1}$)} & 
\multicolumn{1}{c}{(cm$^{-1}$)} & 
\multicolumn{1}{c}{(cm$^{-1}$)}  & 
\multicolumn{1}{c}{(\AA)} & 
\multicolumn{1}{c}{(\AA)} \\
\hline
$X\,^1\Sigma^+$, $v=0$\footnote{Data from Ref.~\cite{hakalla_new_2006}} & 
~~~~~~~0 & 13.93070(19) & 1.3761(21) & 2857.561(22) & 
14.177461(75) & 1.1308843(30) & 1.1409 \\
~$a\,^3\Pi$, $v=0$\footnote{Data from Ref.~\cite{hechtfischer_photodissociation_2007} 
unless stated otherwise in this row.} &
~~9530(50)\footnote{Band origin of the $a\,^3\Pi$-$X\,^1\Sigma^+$, $v=(0$-$0)$ band 
(\cite{hechtfischer_photodissociation_2007}, Table IX)} & 
13.7770(34) & 1.4760(70) & 
~2631(30)\footnote{Ref.~\cite{kusunoki_triplet_1980} with uncertainty estimated
from remark included there} & 
14.08(2) & 1.1348(8) & 1.147 \\
$A\,^1\Pi$, $v=0$\footnotemark[2] & 
24118.726(14) & 11.42351(18) & 
1.9800(19) & 1864.402(22) & 11.88677(72) & 1.235053(37) & 1.2598
\end{tabular} 
\end{ruledtabular}
\end{table}
%
%

The main properties of the molecular ground state and the first two bound,
electronically excited levels of CH$^+$ are summarized in Table
\ref{tab:molpar}.  This uses the experimental values \cite{hakalla_new_2006}
from the spectroscopy of the allowed $A\,^1\Pi$--$X\,^1\Sigma^+$ transitions.
Moreover, the parameters deduced from experimental studies
\cite{kusunoki_triplet_1980,hechtfischer_photodissociation_2007} of the
metastable $a\,^3\Pi$ level are included.  For the rotational levels of the
$X\,^1\Sigma^+$ ground state (rotational quantum number $J$), we use the recent
rotational constants from terahertz spectroscopy \cite{domenech_first_2018} with
relative deviations from Table \ref{tab:molpar} of $<$10$^{-4}$.  This yields
the energy levels given in Table \ref{tab:levels}.

For calculating the radiative transition probabilities between rotational levels
$J$ in the $X\,^1\Sigma^+$, $v=0$ ground state we use the steps described earlier
\cite{meyer_radiative_2017}.  The Einstein coefficient for the spontaneous decay
of level $J$ is obtained as \cite{bernath_spectra_2005}
\begin{equation}
  A_J \equiv A_{J\to J-1}=(16\pi^3/3\epsilon_0h)\tnu_{J-1}^3\mu_0^2J/(2J+1)
  \label{eq:einstein}
\end{equation}
where $\tnu_J=(E_{J+1}-E_{J})/hc$ with the level energies $E_J$ and $\mu_0$ is
the permanent electric dipole moment of the $X\,^1\Sigma^+$, $v=0$ CH$^+$
molecule.  We use the most recent theoretical value \cite{cheng_dipole_2007}
$\mu_0=0.6623\,ea_0 = 1.6834$ D (1 $ea_0$ = 2.541748 D), calculated at an
internuclear separation of 2.1530 $a_0 = 1.393$ \AA.  This separation agrees
reasonably with the average internuclear distance $r_0$ for $X\,^1\Sigma^+$,
$v=0$ of Table \ref{tab:molpar}.  In these equations, $c$ is the velocity of
light, $\epsilon_0$ the vacuum permeability, $h$ Planck's constant and $e$ the
elementary charge.

%
%
\begin{table}[b]
  \caption{
    \label{tab:levels} 
    Rotational energy levels and Einstein coefficients for 
    spontaneous decay of CH$^+$ for $X\,^1\Sigma^+$, $v=0$. }
  \vspace{2mm}
  \centering
\begin{minipage}{10cm}
\begin{ruledtabular}
 \begin{tabular}
 {@{~~~}ccccc}
 \multicolumn{1}{l}{~~$J$} & 
 \multicolumn{2}{c}{Energy} &
 \multicolumn{1}{c}{$A_J$} &
 \multicolumn{1}{c}{$1/A_J$} \\
 & \multicolumn{1}{c}{(cm$^{-1}$)} &
 \multicolumn{1}{c}{(eV)} &
 \multicolumn{1}{c}{(s$^{-1}$)} &
 \multicolumn{1}{c}{(s)} 
\\
\hline
 0 &  0 & 0 & 0 & \\
 1 & 27.86 & 0.003454 & 0.00640 & 156.1 \\
 2 & 83.54 & 0.010357  & 0.0614  & 16.29 \\
 3 & 167.0 & 0.02070 & 0.2213 & 4.52 \\
 4 & 278.1 & 0.03448 & 0.524 & 1.846 \\
 5 & 416.7 & 0.05166 & 1.076 & 0.929 \\
\end{tabular}
\end{ruledtabular}
\end{minipage}
\end{table}
%
%

The property of the ambient radiation field relevant for absorption and total
emission on a rotational transition with wavenumber $\tnu_J$ is the occupation
number $n(\tnu_J)$ of the vacuum modes at $\tnu_J$, which for a thermal
radiation field at a temperature $T_r$ is
\begin{equation}
  n(\tnu_{J}) \equiv n_{\text{th}}(\tnu_J,T_r)=[\exp(hc\tnu_J/k_BT_r)-1]^{-1}
  \label{eq:bose}
\end{equation}
with the Boltzmann constant $k_B$. The decay rate by spontaneous and stimulated
transitions from a level $J$ into the next lower one is then given as
\begin{equation}
  k^{\rm em}_{J\to J-1}=A_J[1+n(\tnu_{J-1})]
  \label{eq:rad_cooling}
\end{equation}
while the excitation rate by radiative absorption from $J$ into the next higher
level is
\begin{equation}
  k^{\rm abs}_{J\to J+1}=A_{J+1}n(\tnu_{J})(2J+3)/(2J+1).
  \label{eq:rad_excitation}
\end{equation}
We use a superposition of two thermal components, one at $T_r=300$ K at a small
relative fraction $\epsilon$ and a main component at an effective cold radiation
temperature $T_{\rm low}$, ($T_r=T_{\rm low}$), where $T_{\rm low}$ is fitted to
the radiative cooling data.  This yields the effective occupation number
\begin{equation}
  n(\tnu_J)=\epsilon n_{\text{th}}(\tnu_J,\text{300 K}) +(1-\epsilon)
  n_{\text{th}}(\tnu_J,T_{\text{low}}).
  \label{eq:csr-eff-occ}
\end{equation}

When the CH$^+$ ions begin to be stored in the CSR after their production in the
ion source, a significant fraction of them is found to populate the
metastable $a\,^3\Pi$ state.  Most levels (except the $a\,^3\Pi, v=0, J=0$, $f$
symmetry level, which is expected to be much longer lived
\cite{hechtfischer_photodissociation_2007} and is further discussed in
Supplemental Sec.\ \ref{sec:metastableState}) decay with an average lifetime
$\tau_m$ that was previously measured \cite{amitay_dissociative_1996} to be close to 7 s.
As long storage times up to 600 s are the main concern of this work, we begin
modeling the $J$ level populations in the $X\,^1\Sigma^+$, $v=0$ ground state at
a time of 21 s ($> 2\;\tau_m$), starting with the measured populations of
the lowest rotational levels $J=0\ldots3$ in $X\,^1\Sigma^+$, $v=0$.  After this
time, we neglect all population in higher excited levels of the $a\,^3\Pi$ metastable state that could decay into
the selected low-$J$ levels based on the experimentally observed disappearance of the fraction of the dissociative recombination (DR) signal which we can uniquely attribute to the metastable $a\,^3\Pi$ state by storage-time dependent molecular fragment imaging (further discussed in Supplemental Sec.~\ref{sec:metastableState}).

\section{CH$^+$ storage and laser probing of rotational energy levels}

\begin{figure}[b]
  \centering
  \includegraphics[width=0.78\textwidth]{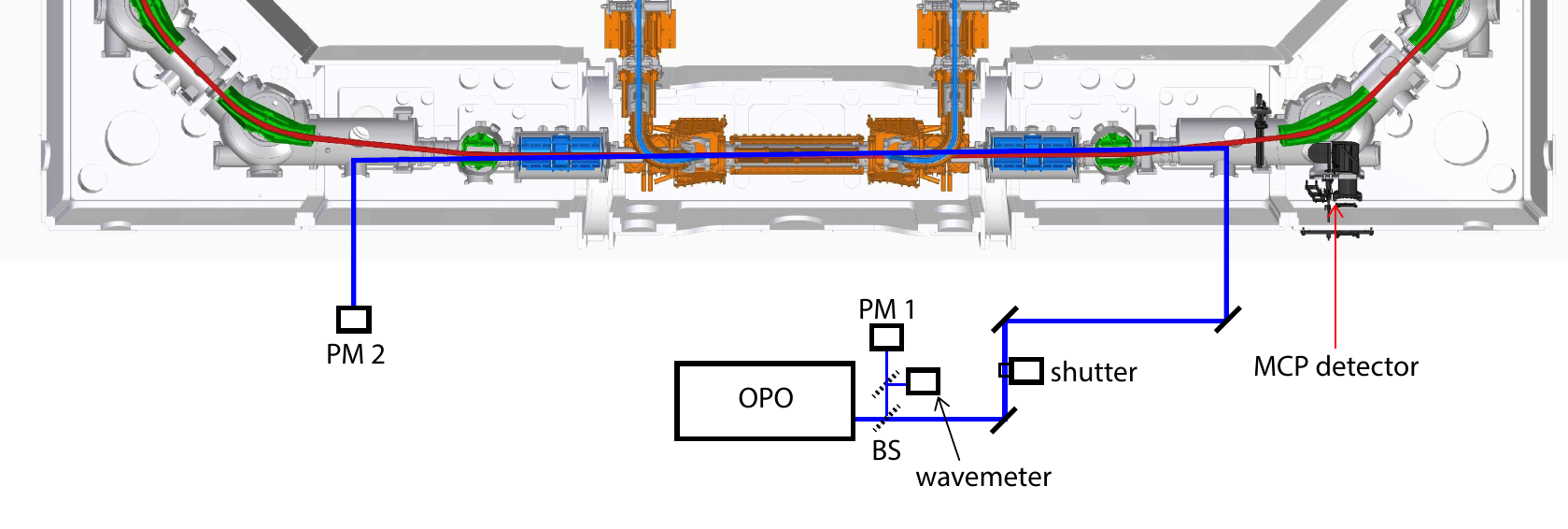}%
  \caption{View of the experimental section of the CSR with the electron beam
    setup in the center, the MCP detector, and the laser beam path.  Red line:
    ion beam; bright blue curve: electron beam; darker blue line: laser beam.
    PM1, PM2: power monitors.  BS: beam splitter.}
 \label{fig:setup}
\end{figure}

The method for laser probing of rotational levels is similar to that used in
previous work \cite{oconnor_photodissociation_2016} at the CSR.  Pulses from a
tunable optical parametric oscillator (OPO) laser (Ekspla NT342b, repetition
rate 20 Hz) are directed into the CSR.  Conditions different in the present
experiment are discussed in the following.  In the present setup, the laser
counterpropagates relative to the ion beam (see Fig.\ \ref{fig:setup}) with a
grazing angle of $\sim$1.5$^\circ$.  The laser wavelength (near 306 nm) is
continuously monitored by a wavemeter (HighFinesse/\AA{}ngstrom Laser Spectrum
Analyzer LSA UV-I, HighFinesse GmbH, Tübingen, Germany).  Hence, the applied
laser wavenumber $\tnu_L$ is specified with $\sim$0.1 \cmm\ stability over the
duration of the experiment and a calibration uncertainty of $\lesssim$1 \cmm.
In addition to $\tnu_L$ the wavemeter yields the spectral width of the applied
laser light.  For modeling the photodissociation spectrum, we use a Gaussian
distribution of $\tnu_L$ with a full width at half maximum (FWHM) of 6.1(3)
\cmm\ [$\sigma_{\tnu_L}=2.6(1)$ \cmm].  The laser pulse energy is measured pulse
by pulse (see power monitors in Fig. \ref{fig:setup}) and variations of the
pulse energy $\epsilon_L$ transmitted through the CSR (typically 0.1--0.2 mJ)
are measured with $\lesssim$5\% uncertainty for normalizing the
photodissociation signal to the irradiated photon flux.

The CH$^+$ ions are stored at a beam energy near 279.5 keV with run-by-run
variations monitored on an $\sim$10$^{-4}$ (0.03 keV) level by ion revolution
frequency measurements.  The absolute energy has a relative uncertainty of
$\sim 3\times 10^{-3}$, corresponding to 0.8 keV.  The beam velocity is
$\beta c$ (with $c$ the velocity of light) where $\beta\approx0.00679$, with its
relative variations being monitored with $\sim0.5 \times 10^{-4}$ uncertainty
for the purpose of Doppler correction of the laser wave number.  The typical
number $N_i$ of ions stored in the CSR is 10$^7$.  The overlap length of the ion
beam with the grazing laser beam is $l_L \approx 0.4$ m, corresponding to
$\sim$1\% of the CSR circumference.  This yields an ion number on the order of
$10^5$ in the laser overlap region.  For recombination rate measurements $N_i$
was reduced by a factor of $\sim$10$^1$ to avoid saturating the neutral product
detector.

By the Doppler correction, the wave number in the ion
rest frame ($\tnu$) is obtained from the laser wave number $\tnu_L$ as
\begin{equation}
\tnu = \tnu_L\;\sqrt{1-\beta^2}/(1-\beta\cos\theta)
\end{equation}
The Doppler correction $\tnu-\tnu_L \approx 224$ \cmm\ can be monitored for each
run with an uncertainty of $\sim 0.5\times 10^{-4}$ (0.01 \cmm), while the
uncertainty of the absolute size of the Doppler correction is
$\sim 1.5\times 10^{-3}$ (0.35 \cmm).  Owing to the low beam velocity of the
present experiment, this uncertainty is about a factor of 3 lower than in the
CH$^+$ photodissociation experiment at the TSR ($\beta\approx0.034$)
\cite{hechtfischer_photodissociation_2002}, where the Doppler correction
constitutes the dominant component of the uncertainty (1.1 \cmm) in the CH$^+$
dissociation energy measured by that experiment.

\begin{figure}[t]
  \centering
  \includegraphics[width=0.6\textwidth]{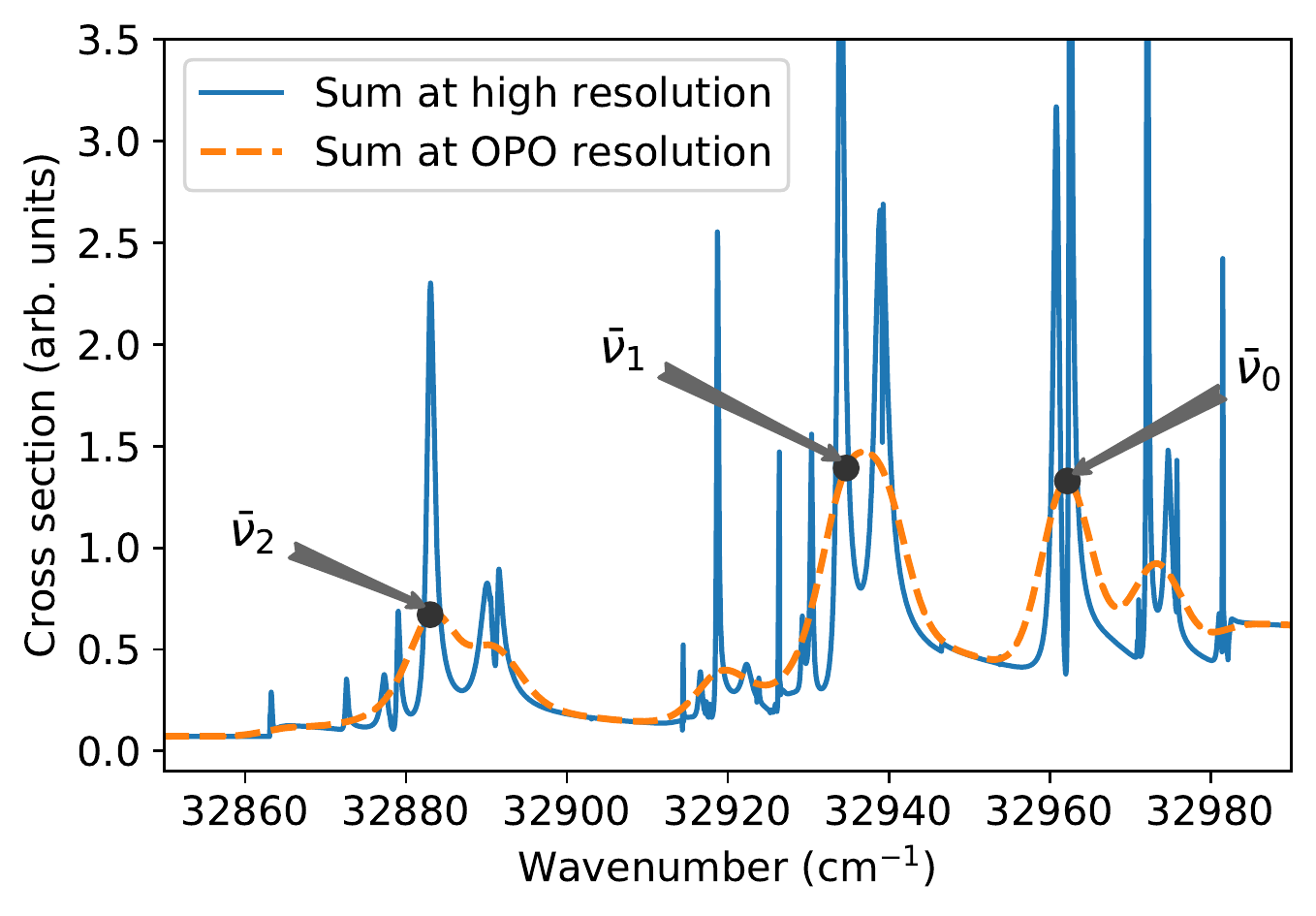}%
  \caption{Calculated near-threshold photodissociation spectrum of CH$^+$ for a
    rotational temperature of 77~K (mainly $J=0\ldots2$) including natural
    resonance broadening only (blue line) and at the present resolution of 6.1
    \cmm\ (FWHM) (yellow dashed line). The probing wave numbers
    $\tnu_{0\ldots2}$ are marked.}
 \label{fig:spectrum}
\end{figure}

The microchannel plate (MCP) detector (see Fig.\ \ref{fig:setup}) detects
neutral fragments from the interaction section of the CSR.  The fragments
induced by laser pulses reach the detector with a time spread of
$l_L/\beta c \approx 0.2$ $\mu$s, spaced by 50 ms corresponding to the pulse
repetition frequency.  In the breaks between pulses, a continuous neutralization
rate of the circulating CH$^+$ ions of the order of 10$^3$ s$^{-1}$ is observed,
stemming from the CH$^+$ collisions with the residual gas.  Laser-induced
neutral count rates $R_i$ were recorded in time gates of 0.6 $\mu$s duration
synchronized with the laser pulses at wavenumbers $\tnu_i$ ($i=0\ldots2$) and
$\tnu_{b'}$ [see Fig.\ 1(b) of the main paper].  The value $\tnu_{b'}=32\,750$
\cmm\ was added to the probing wavenumbers in order to monitor the full
background rate $R_{b'}$, which includes any laser-induced events not related to
the $J\leq3$ levels of $X\,^1\Sigma^+$, $v=0$ CH$^+$.  Moreover, a summed rate
$R_b$ for the continuous background from any neutrals formed in collisions of
CH$^+$ ions with residual gas molecules was obtained from a 90-$\mu$s gate
ending 12 $\mu$s before and another 40-$\mu$s gate starting 2.4 $\mu$s after
each laser pulse.  Together with the laser pulse energy $\epsilon_L$, the
normalized photodissociation signal is found as
\begin{equation}
  S_i=S(\tnu_i) = (R_i-R_{b'})/\epsilon_LR_b.
  \label{eq:signal}
\end{equation}
To yield the photodissociation signals, pulses at a given laser wavenumber were
typically repeated during $\sim$\,4~s before setting the next value of $\tnu_i$
($i=0\ldots2$) or $\tnu_{b'}$ in the measurement cycle.  Doppler corrected
values are given for all probing wave numbers.

\begin{figure}[t]
  \centering \raisebox{4.5cm}{(a)}%
  \includegraphics[width=0.4\textwidth]{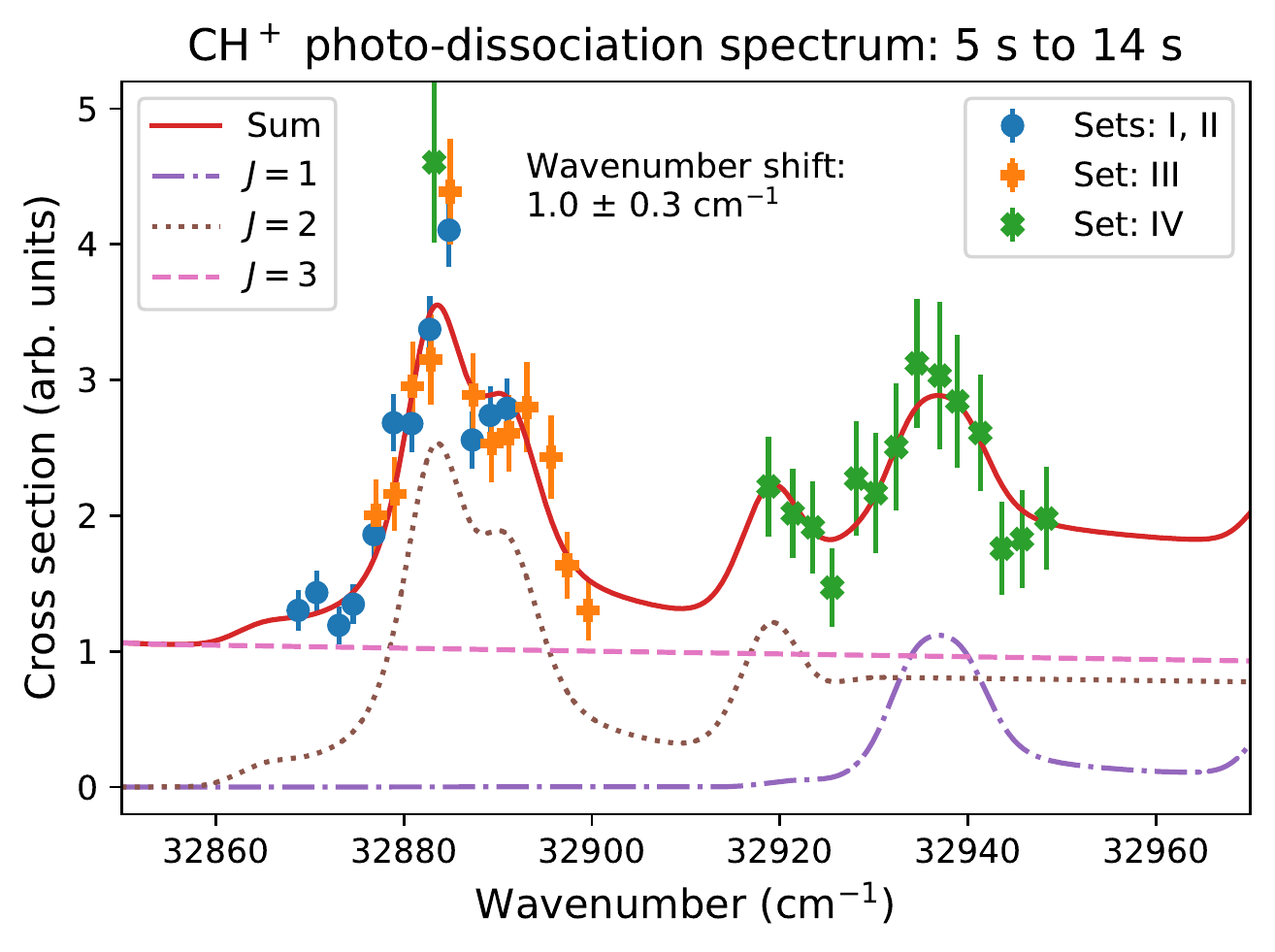}~~~~
  \raisebox{4.5cm}{(b)}%
  \includegraphics[width=0.4\textwidth]{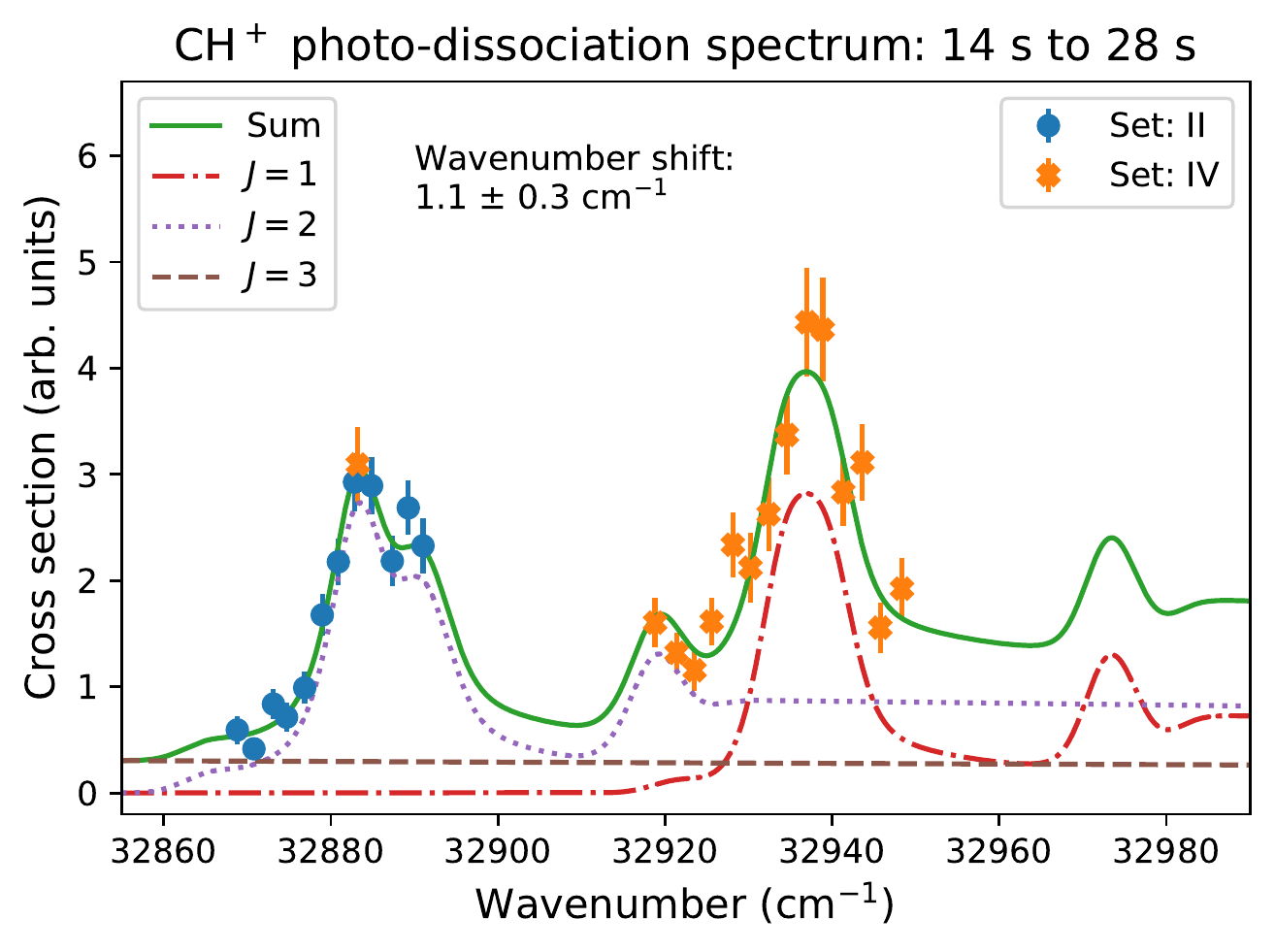}\\
  \raisebox{4.5cm}{(c)}%
  \includegraphics[width=0.4\textwidth]{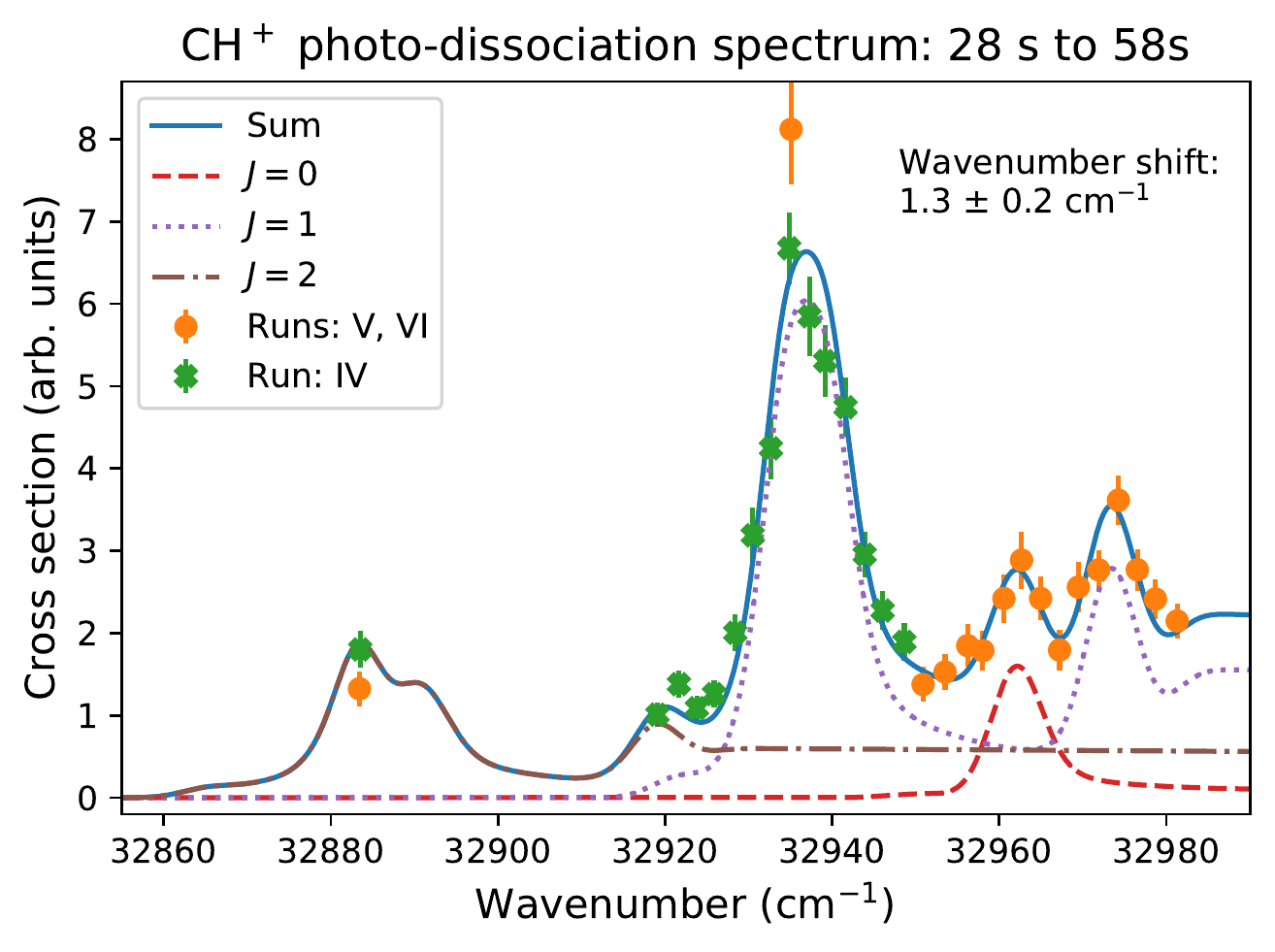}~~~~
  \raisebox{4.5cm}{(d)}%
  \includegraphics[width=0.4\textwidth]{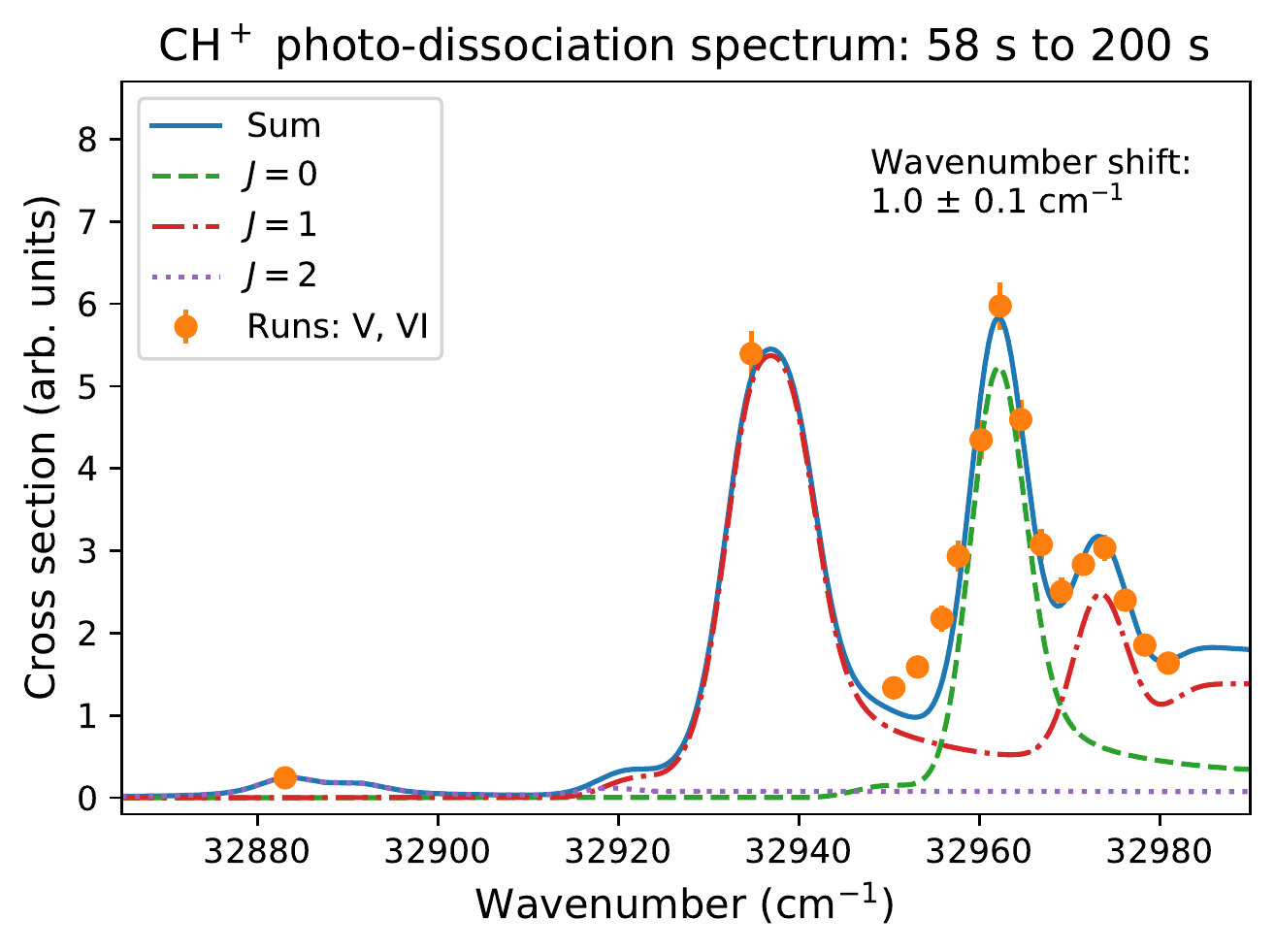}
  \caption{Preparative laser scans of the CH$^+$ photodissociation signal for
    the storage time ranges given in the headings, showing sequence labels of
    the experimental runs and the individually fitted values of the wave number
    shift $\delta\tnu$ discussed in the text.  Individual $J$ components of the
    fitted convoluted spectra are distinguished by line styles according to the
    legends.}
  \label{fig:laserscans}
\end{figure}

The rotational probing makes use of the resonant structure of the near-threshold
CH$^+$ photodissociation, in which rotationally resolved predissociation
resonances are observed and are found to be in excellent agreement with
theoretical predictions.  The underlying theory uses experimentally adjusted
fine-structure resolved molecular potential curves and is based on
\citen{hechtfischer_photodissociation_2002} and \citen{barinovs_ch_2004}, with
details discussed in the supplement of \citen{oconnor_photodissociation_2016}.
Using the calculated cross sections $\sigma_J(\tnu)$ for separate initial
CH$^+$($J$) ions, a rotationally averaged spectrum
\begin{equation}
  \bar{\sigma}(\tnu)=\sum_{J=0}^{J_{\rm max}} p_J \sigma_J(\tnu)
\end{equation}
($J_{\rm max}=5$) is shown in Fig.\ \ref{fig:spectrum} after convolution over a
Gaussian broadened wavenumber distribution, illustrating the present
experimental resolution.  The rotational populations $p_J$ are set for a thermal
distribution with a rotational temperature at the upper end of the range of
interest here, yielding dominant resonant contributions for $J=0...2$, together
with small, smooth contributions for (mainly) $J=3$.  In contrast to previous
work \cite{oconnor_photodissociation_2016} at CSR, the data are not acquired at
full two-dimensional (time and wavenumber) resolution, but at the set $\tnu_i$
of probing wave numbers only.

For rotational probing, the measured signals $S_i=S(\tnu_i)$, with $S$ from Eq.\
(\ref{eq:signal}), are expressed as
\begin{equation}
  S_i=C\sum_{J=0}^{J_{\rm max}} Q_{Ji}\; p_J
\end{equation}
with $J_{\rm max}=2$ and the probing amplitudes
$Q_{Ji}=\sigma_J(\tnu_i-\delta\tnu)$.  Detailed scans of the photodissociation
signal (Fig. \ref{fig:laserscans}) were performed for calibrating the probing
amplitudes and in particular the shift $\delta\tnu$, which accounts for the
residual uncertainty of the absolute wave number scale and the dissociation
energy $D_e$ of CH$^+$.  In fact, these scans already observe the temporal
change of the rotational populations $p_J$.  Considering the radiative lifetimes
of the $J$ levels in the $X^1\Sigma^+, v=0$ ground state (Table
\ref{tab:levels}), significant contributions are expected to include
$J=0\ldots3$ for $t\gtrsim5$ s and $J=0\ldots2$ for $t\gtrsim20$ s.  The spectra
observed in the indicated time windows were fitted to theoretical cross sections
\begin{equation}
  S(\tnu)=\sum_{J=0}^{J_{\rm max}} c_J \sigma_J(\tnu-\delta\tnu)
\end{equation}
with $J_{\rm max}=3$ in Fig. \ref{fig:laserscans}(a), (b) and $J_{\rm max}=2$ in
Fig. \ref{fig:laserscans}(c), (d).

The ensemble of fit results shows that, relating to the $\tnu$ scale of the
applied wavemeter after the Doppler correction, the probing amplitudes can be
obtained from the calculated, Gaussian convoluted cross sections when shifting
the wavenumber scale of these results up by $\delta\tnu=1.0(1)$ \cmm.  The
fitted relative amplitudes of the $J$-specific contributions in the time windows
of Fig. \ref{fig:laserscans}(b) and (c) then allow us to derive the rotational
populations within the $X^1\Sigma^+$ electronic state.

The purpose of the measurements is to track the rotational populations in the
$X^1\Sigma^+$ electronic state once that its feeding from the metastable
$a\,^3\Pi$ state, with a measured average lifetime $7(1)$\,s, can be neglected.
We consider that we can neglect this feeding for storage times $t > 21$\,s.
From the results of Fig. \ref{fig:laserscans}(b) and (c) (with only a minor
effect of the $a\,^3\Pi$ in the earlier one of these time windows), we estimate
the starting populations in the $J=0\ldots3$ levels of the $X^1\Sigma^+, v=0$
ground state as given in Table \ref{tab:rotpop}.  Effects of $a\,^3\Pi$ decay in
the earlier one of these time windows is included in the given uncertainties.
Populations in $J\geq4$ rotational levels with lifetimes of $<$2 s are
neglected.
 
%
%
\begin{table}[t]
  \caption{
    \label{tab:rotpop} Experimentally derived relative populations $\tilde{p}_J$
    at $t=21$ s, the start of the radiative relaxation model in the
    $X\,^1\Sigma^+$, $v=0$ level.  The uncertainties resulting from the fitting
    procedure are given for the last digit as upper and lower indices at a
    confidence level corresponding to a one-sigma uncertainty in the symmetric
    case.}
  \vspace{2mm}
  \centering
\begin{minipage}{2.5cm}
\begin{ruledtabular}
 \begin{tabular}{@{~~}cc}
 $J$ & $\tilde{p}_J$\\
\hline
 0 & 0.03$^{-3}_{+2}$ \\
 1 & 0.39$^{-5}_{+5}$ \\
 2 & 0.43$^{-4}_{+7}$ \\
 3 & 0.15$^{-7}_{+4}$ 
\end{tabular}
\end{ruledtabular}
\end{minipage}
\end{table}

%
%


\section{Merged-beams electron collisions}

%
%

The electron beam is merged with the circulating ion beam (see Fig.\
\ref{fig:setup}) with the help of a bent magnetic guiding field (20 mT in the
bending regions and 10 mT in the overlap region with the ion beam).  After being
transported from the electron source at 20 eV, the electrons are decelerated in
0.85 m long drift tube to an energy of 11.79 eV, at which the beam velocity is
matched to that of the stored ion beam (279.4 keV kinetic energy).  The velocity
distribution in the co-moving frame of the electron beam is anisotropic with
respect to the beam direction with a transverse temperature $T_\perp$ and a
longitudinal temperature $T_\|$.  The electrons are emitted from a
laser-illuminated photocathode operated near room temperature.  It was shown
\cite{orlov_cold_2003} that the transverse temperatures of the emitted electrons
then approach the bulk temperature (i.e., thermal energy $\gtrsim26$~meV/$k_B$).
The magnetic guiding field strength at the cathode exceeds the field strength in
the interaction region by a factor of 20.  Correspondingly, through the inverse
magnetic-bottle effect, the transverse temperature $T_\perp$ in the interaction
region is reduced compared to the emission temperature by approximately this
factor \cite{pastuszka_electron_1996} (i.e., $T_\perp\gtrsim1.3$~meV).  From
initial operating experience at the CSR electron cooler, $k_BT_\perp$ is
estimated to be somewhat larger than the theoretical limit (1.3 meV) and to lie
between 1.5 meV and 3 meV.  In Supplemental Sec.\ \ref{sec:trtemp} we discuss
how we determined the best-fit value of $k_BT_\perp$ for this work.

The current profile of the electron beam was measured and is approximated by a
cylindrical beam with a diameter of $2\rho = 10.2(5)$ mm\label{rhoref} in the
interaction region.

\begin{figure}[t]
  \centering
  \includegraphics[width=0.6\textwidth]{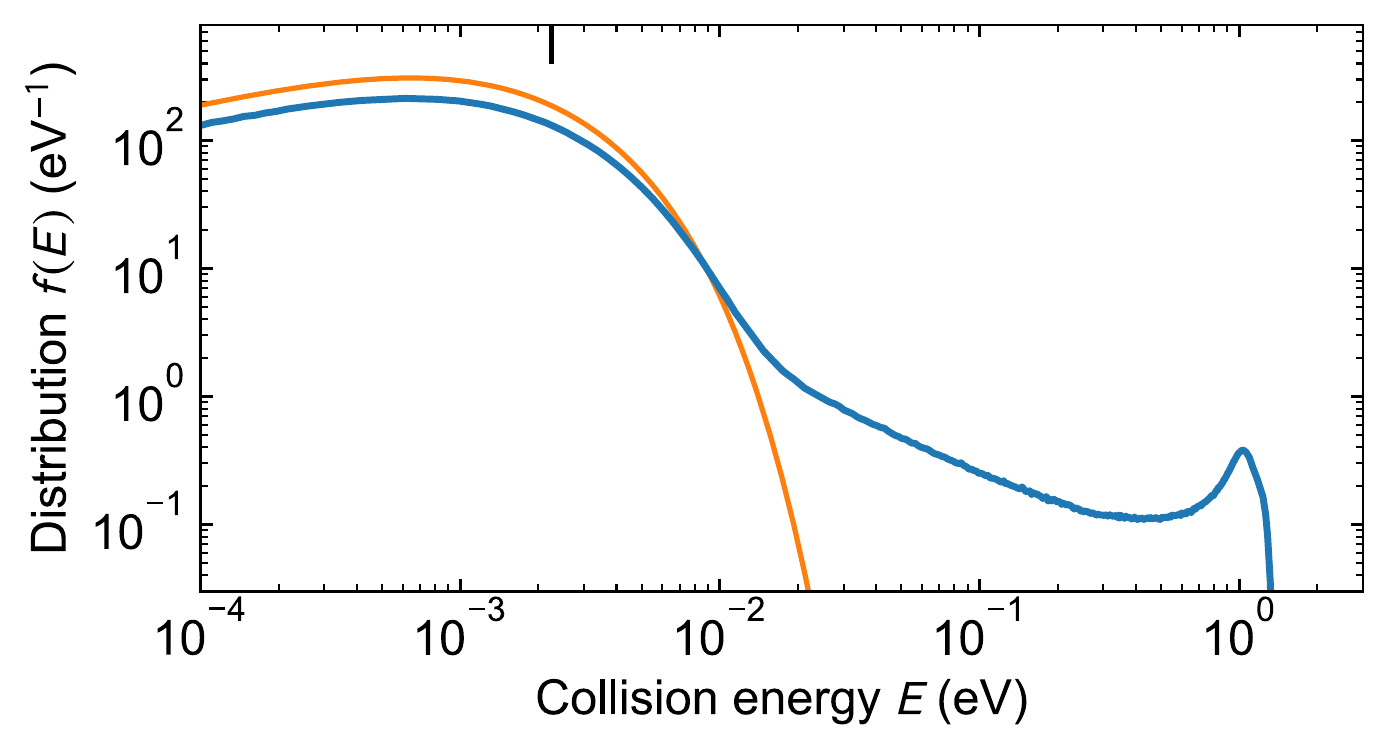}
  \caption{Collision energy distributions (normalized) in the merged electron
    and ion beams at matched averaged beam velocities.  Blue: Distribution from
    a Monte-Carlo simulation with $k_BT_\perp = 2.25$ meV (vertical mark at the
    top), including the experimental beam and drift-tube geometry, and the
    acceleration voltage fluctuations.  Orange: Flattened Maxwellian energy
    distribution $f(E)$ from Eq.\ (\ref{eq:flat}) of the same $T_\perp$,
    determined such that $af(E)$ with $k_BT_\|= 0.59$ meV and $a=0.693$ fits the
    simulated data at $E<k_BT_\perp$. }
  \label{fig:edist}
\end{figure}

While the relative velocity between electrons and ions is minimized by adjusting
the precise electron velocity such that the electron cooling effect
\cite{budker_electron_1978} on the ion beam is maximized, the velocity spread
due to the beam temperatures leads to finite collision energies.  Moreover, in
the merging regions, electrons and ions interact at increased relative
velocities.  Using the geometry of the merging regions and the deceleration
fields around the drift tube, the relative velocities are modeled by Monte-Carlo
integration over the complete ion beam length of $l_0=1.136(14)$ m where both,
electrons and ions, are present.  This model also takes into account the
electron beam temperatures (assuming Maxwellian velocity distributions) and the
experimental ripple and noise on the voltage difference between the cathode and
the interaction drift tube.  Based on measurements for the situation during the
CH$^+$ beam time, root mean square (r.m.s.)\ voltage fluctuations of 0.1 V are
inserted in the model calculations.  The resulting collision energy distribution
$f(E)$ is shown in Fig.\ \ref{fig:edist} for $k_BT_\perp = 2.25$ meV.  The
low-energy part ($E<k_BT_\perp$) can be very well fitted by the scaled energy
distribution, $af(E)$, resulting from a bi-modal Maxwellian electron velocity
distribution, with the normalized energy distribution reading
\begin{equation}
  f(E)=\frac{1}{\zeta k_BT_\perp}\,e^{-E/k_BT_\perp}\,{\rm
    erf}\left(\zeta\sqrt{\frac{E}{k_BT_\|}}\right)
  \;,\;\;\zeta=\sqrt{1-\frac{T_\|}{T_\perp}}~. \label{eq:flat}
\end{equation}
The parameters of this fit are $a$ and $T_\|$.  The case shown in Fig.\
\ref{fig:edist} leads to a fitted $k_BT_\|= 0.59$~meV, with this value being
dominated by the effect of the electron beam acceleration voltage fluctuations.
Moreover, the fraction of the modeled distribution fitted by the normalized
function of Eq.\ (\ref{eq:flat}) is $a=0.693$.  Hence, $\sim$70\% of the total
ion--electron overlap length, corresponding to $\hat{l}_0=al_0=0.79(1)$~m,
contributes with a narrow energy distribution ($E\lesssim 2k_BT_\perp$), while
for $\sim$30\% of $l_0$ the effective collision energy distribution is much
wider with up to $\sim$1 eV of collision energy.

For the laser probing studies of rotationally inelastic collisions, reported
here, the electron current was 18.5 $\mu$A.  Together with the measured
effective electron beam radius and the electron velocity in the central part of
the interaction region, this yields an electron density of
$n_e = 7.0(6)\times10^5$ cm$^{-3}$\label{edens}, where the uncertainty is
dominated by the knowledge of the electron current profile.

Dissociative recombination (DR) measurements used the same MCP detector (see
Fig.\ \ref{fig:setup}) as that applied for the photodissociation measurements.
In DR measurements, this detector continuously counts the neutral products from
the electron--ion interaction region and also analyzes their kinetic energy
release (KER) by measuring the transverse distance between the pairs of neutrals
formed by DR.

For any electron-induced process (such as DR or inelastic collisions) the energy
dependent cross section $\sigma(E)$ leads to a merged-beams rate coefficient
\begin{equation}
  \alpha^{\rm mb}=\sqrt{\frac{2}{m}}\int_0^\infty f(E)\sqrt{E}\sigma(E)dE
  \label{eq:alphasigma}
\end{equation}
where $m$ is the electron mass.

Based on the experience that by optimizing the electron cooling the ion beam is
approximately centered within the electron beam, we assume concentric beams for
our modeling of the experimental situation.  The ion beam profile in the
interaction region is measured by imaging the center-of-mass of coincident
two-fragment DR events on the imaging detector and derived from these data by
taking into account the projection of the neutral fragment trajectories from the
interaction region to the detector.  Here, the ion beam size and its divergence
are assumed to be related \cite{krantz_transverse_2021} by the known focusing
properties of the storage ring \cite{von_hahn_cryogenic_2016}.  The determined
r.m.s.\ ion beam sizes $\sigma_x$ in the horizontal direction (bending plane of
the storage ring) and the vertical size $\sigma_y$, reached after after electron
cooling times of $t=10$~s and 20~s, are listed in Table~\ref{tab:ionsize}.
Assuming a Gaussian beam profile, we describe the normalized transversal density
distribution by
\begin{equation}
  f(x,y)dxdy=\frac{1}{2\pi \sigma_x\sigma_y}\;
  e^{-x^2/2\sigma_x^2}\;e^{-y^2/2\sigma_y^2}dxdy.
\end{equation}
Conversion to polar coordinates $r,\phi$ yields
\begin{equation}
  f(r,\phi)d r d\phi=\frac{\sqrt{a_xa_y}}{\pi}\;
  e^{-r^2(a^+ + a^-\cos2\phi)}rdrd\phi
  \label{eq:trdenspolar}
\end{equation}
with $a_i=1/2\sigma_i^2$ for $i = x$ and $y$, and $a^\pm=(a_x \pm a_y)/2$.  The
transverse overlap fraction $\eta_\perp$ for an electron beam radius $\rho$ is
then
\begin{equation}
  \eta_\perp = \sqrt{a_xa_y} \int_0^{\rho^2}e^{-a^+u}~ {I}_0(a^-u)~du,
  \label{eq:etaeq}
\end{equation}
with the angular integration of Eq.\ (\ref{eq:trdenspolar}) being equivalent to
the integral representation of the modified Bessel function $I_0$.  The integral
of Eq.\ (\ref{eq:etaeq}) is easily calculated numerically.  Using the measured
electron beam diameter [$2\rho =10.2(5)$ mm, see p.\ \pageref{rhoref}] leads to
the results for $\eta_\perp$ listed in Table~\ref{tab:ionsize}.  We conclude
that the overlap efficiency appropriate to our 50\% duty cycle measurement
conditions is $\eta_\perp=0.91(4)$.

%
%
\begin{table}[t]
  \caption{
    \label{tab:ionsize} 
    Measured r.m.s.\ ion beam radii $\sigma_x$ and $\sigma_y$, the deduced
    overlap ratio $\eta_\perp$, and the effective r.m.s.\ radii and FWHM
    diameters of a round ion beam with the same overlap ratio.  Estimated
    standard deviation uncertainties are included.}  \vspace{2mm} \centering
\begin{minipage}{9cm}
\begin{ruledtabular}
 \begin{tabular}
 {@{~~~}rcc}
 \multicolumn{1}{c}{~~} & 
 \multicolumn{1}{c}{$t=10$ s} &
 \multicolumn{1}{c}{$t=20$ s} \\
\hline
   $\sigma_x$ (mm) & 2.98(30) & 2.40(24) \\
   $\sigma_y$ (mm) & 1.70(17) & 1.56(16) \\
   $\eta_\perp$  & 0.88(4) & 0.95(2) \\
   $\hat{\sigma}$ (mm) & 2.45(22) & 2.07(19) \\
   Effective FWHM diam.\ (mm) & 5.77(52) & 4.87(45) \\ 
\end{tabular}
\end{ruledtabular}
\end{minipage}
\end{table}
%
%

For a round beam of r.m.s.\ radius $\hat{\sigma}=1/\sqrt{2\hat{a}}$ the overlap
fraction would be $\hat{\eta}_\perp=1-e^{-\hat{a}\rho^2}$.  The $\hat{\sigma}$
that yields $\hat{\eta}_\perp=\eta_\perp$ is
\begin{equation}
  \hat{\sigma}=\rho/\sqrt{2\ln[1/(1-\eta_\perp)]}.
\end{equation}
The radii $\hat{\sigma}$ and FWHM diameters for an equivalent round ion beam are
included in Table~\ref{tab:ionsize}.  They lead to the $<$\,6~mm effective ion
beam diameter quoted in the main paper.

For ions circulating in the storage ring, the rate of the collisionally induced
process is then
\begin{equation}
  R=\eta_\perp (l_0/C_0)n_e\alpha^{\rm mb}\;.
  \label{eq:rates}
\end{equation}
All results were obtained with the given $n_e$ (see p.\ \pageref{edens}).  The
factor linking all collisional rates $R$ to $\alpha^{\rm mb}$ is
$\bar{n}_e=\eta_\perp(l_0/C_0)n_e=2.06(21)\times10^{4}$ cm$^{-3}$, representing
the effective average of the electron density over the storage ring
circumference.

\section{Theoretical results on CH$^+$ rotationally inelastic collisions}

The cross section for rotationally inelastic collisions of electrons with
positive molecular ions can be calculated
\cite{boikova_rotational_1968,neufeld_electron-impact_1989} in compact form
using the Coulomb--Born (CB) approximation, which assumes the free-electron wave
functions for a $1/r$ Coulomb potential of the ion and multipole interaction
with the molecular charge distribution to be valid at all distances from the
molecular center of mass.  For free--free transitions of the incident electron
leading to molecular excitation, the cross section for electric dipole
interaction in first-order perturbation theory is obtained for rotational
excitation $J\rightarrow J+1$ as
\begin{equation}
  \sigma_{J,J+1}(E) =
  \frac{4\pi^2a_0^2}{3\sqrt{3}} \frac{E_0}{\Delta E}
  \left(\frac{\mu_0}{ea_0}\right)^2 \frac{J+1}{2J+1}
  \frac{\Delta E}{E}g_{ff}(w,\epsilon),
  \label{eq:sigma_exc}
\end{equation}
where $E\geq\Delta E$ is the incident electron energy and $\Delta E$ the
excitation energy, $\mu_0$ the molecular electric dipole moment (in the
vibrational ground state for the present case). $E_0$ and $a_0$ are the atomic
unit of energy and the Bohr radius, respectively.
\begin{figure}[t]
 \centering
  \includegraphics[width=0.55\textwidth]{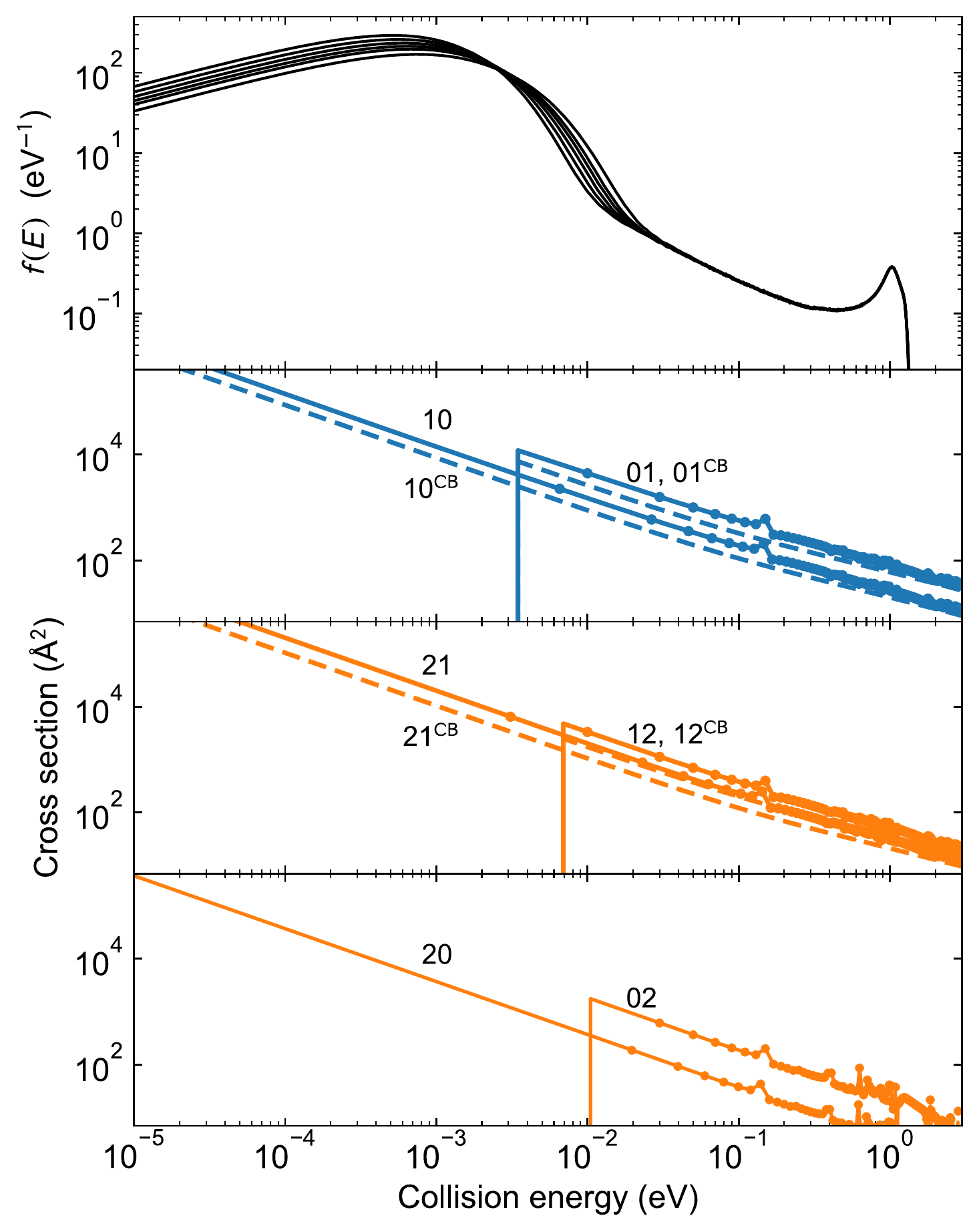}
  \caption{Upper panel: Modeled merged-beams velocity distributions (normalized)
    for matched average beam velocities and transverse electron temperatures of
    $k_BT_\perp$ varying from 1.5 meV to 3 meV (in steps of 0.25 meV from 1.5 to
    2.5 meV).  Lower panels: Inelastic electron cross sections
    $\sigma_{J,J'}(E)$ for $J,J'\leq2$.  Dots and full lines with labels $JJ'$:
    results by \citen{hamilton_electron-impact_2016} (dots) with interpolation
    and extrapolation as described in the text. Dashed lines and labels
    $JJ'^{\rm CB}$: Coulomb-Born approximation calculated with Eqs.\
    (\ref{eq:sigma_exc}) and (\ref{eq:det_bal}).}
 \label{fig:cross_sec}
\end{figure}
The angular factors apply to rotational levels in a $^1\Sigma^+$ electronic
state.  The Gaunt factor $g_{ff}(w,\epsilon)$ represents the integral over the
continuum Coulomb wave functions with the parameters $w=2\Delta E/E_0$ and
$\epsilon=2(E-\Delta E)/E_0$.  We interpolate $g_{ff}(w,\epsilon)$ from the
recent high-accuracy tabulation by \citen{van_hoof_accurate_2014}.  The Gaunt
factors near threshold ($\epsilon\ll w$) are close to $1$ and significantly vary
only for $\epsilon\gg w$.  Hence, Eq.\ (\ref{eq:sigma_exc}) implies that
$\sigma_{J,J+1}$ abruptly rises to $\sigma_0^{J,J+1}$ at $E=\Delta E$ and, at
higher $E$, decreases approximately proportional to $1/E$.

The CB expression Eq.\ (\ref{eq:sigma_exc}) is also suitable for estimating
vibrational excitation.  The related excitation cross section for $v\to v+1$ in
the harmonic approximation, after summing over rotational final states, is
obtained \cite{boikova_rotational_1968} by replacing
\begin{equation}
  \left(\frac{\mu_0}{ea_0}\right)^2 \frac{J+1}{2J+1} \rightarrow  
  \left(\left.\frac{\sigma_0}{ea_0}\frac{\partial \mu_0}{\partial R}\right\vert_{r_e}\right)^2
\end{equation}
with the dipole moment derivative at the equilibrium distance $r_e$ (where
$\sigma_0=\sqrt{\hbar/4\pi c\,\omega_em_r}$, with the reduced mass $m_r$, is the
root-mean-square spatial extension of the vibrational ground-state probability
distribution), using also the appropriate excitation energy $\Delta E$.  The
increasing $\Delta E$ and the mostly smaller size of the dipole moment
expression in general lead to a significantly smaller excitation cross section
for vibration than for rotation.

The significant deviations expected from the assumptions of the CB approximation
in the inner molecular region were considered for rotational excitation in
recent $R$-matrix calculations for a number of small diatomic molecules,
including CH$^+$ \cite{hamilton_electron-impact_2016}.  Excitation cross section
data underlying the rate coefficients presented in Ref.\
\cite{hamilton_electron-impact_2016} were made available to us by the authors.
They are given on an electron energy grid from 0.01 eV in steps of 0.02 eV, thus
starting somewhat above the $J=0$ to $1$ threshold (0.00346 eV).  Using the
cross-section data on this rather coarse energy grid, we determine intermediate
values by linear interpolation on a double-logarithmic scale (thus requiring a
constant power law between any two points).  Moreover, from the lowest grid
point above the threshold we extrapolate down in $E$ using the power law of the
next-higher interpolation interval.

Inelastic electron-impact cross sections for CH$^+$ based on Eq.\
(\ref{eq:sigma_exc}) and on the calculations by
\citen{hamilton_electron-impact_2016} are shown in Fig.\ \ref{fig:cross_sec}.
In addition to excitation cross sections $\sigma_{J,J+1}(E)$, the de-excitation
cross sections $\sigma_{J+1,J}(E)$ are also shown, which for a given pair of
levels can be obtained using the principle of detailed balance,
\begin{equation}
  g(J+1)\,E\,\sigma_{J+1,J}(E)=g(J)\,(E+\Delta E)\,\sigma_{J,J+1}(E+\Delta E)
  \label{eq:det_bal}
\end{equation}
where $g(J)=2J+1$ are the statistical weights of the bound molecular levels.
The $R$-matrix results in Fig.\ \ref{fig:cross_sec} show very similar trends as
the compact CB results, with the difference essentially consisting in the value
of the excitation cross section at threshold,
$\sigma_{J,J+1}(\Delta E)=\sigma_0^{J,J+1}$.  Moreover, the CB results follow
the dipole selection rule of $|\Delta J| =1$, while the $R$-matrix calculations
show significant contributions also for $|\Delta J| >1$.  In the present context
we consider $J \leq3$ including the $R$-matrix results with $|\Delta J| \leq3$.

The experimental collision energy distributions $f(E)$ shown in the top frame of
Fig.\ \ref{fig:cross_sec} suggest that, while de-excitation will be dominant,
also excitation will occur and sensitively depend on the transverse electron
temperature.

\begin{figure}[t]
 \centering
  \includegraphics[width=0.55\textwidth]{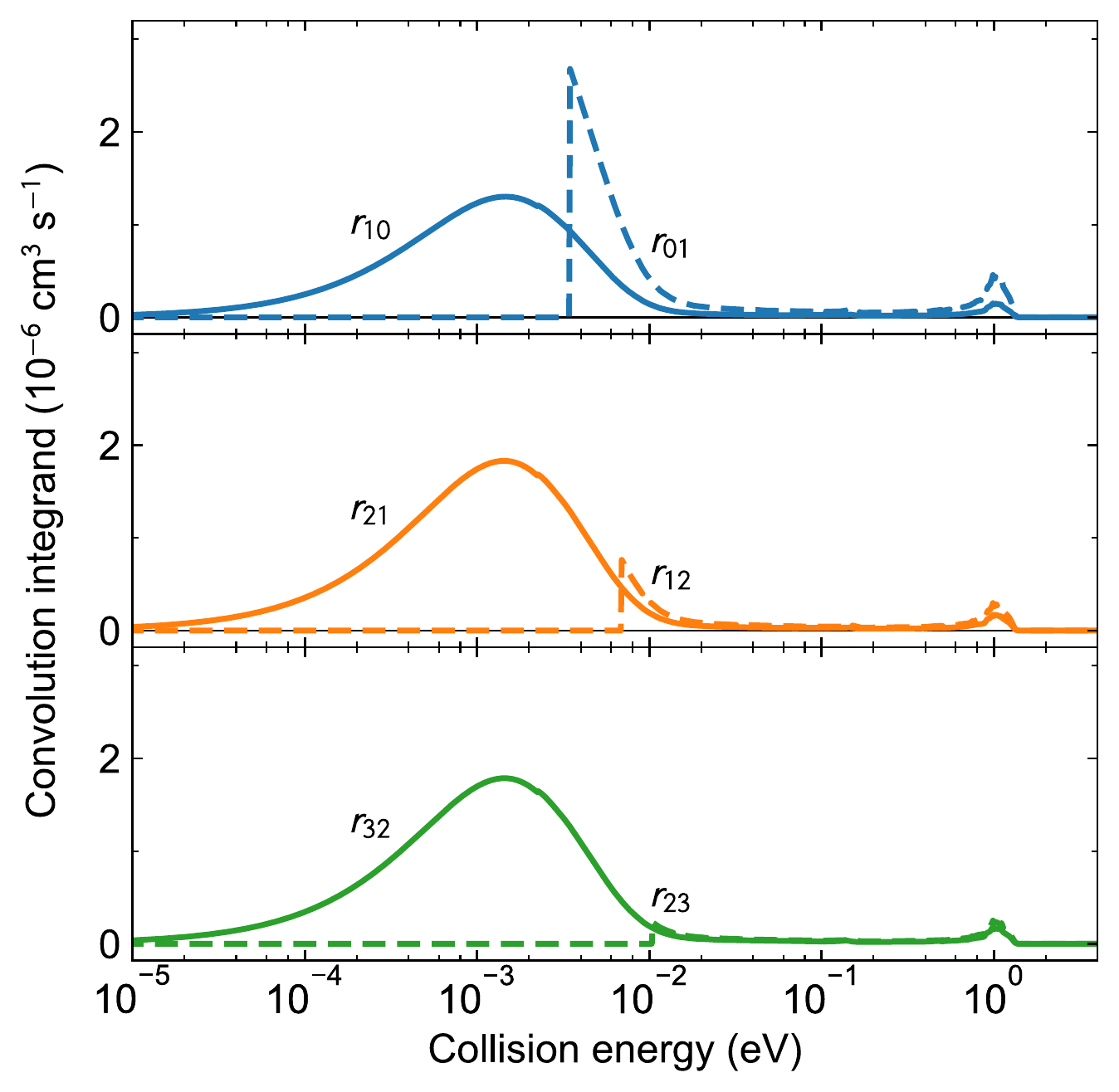}
  \caption{Logarithmic convolution integrand $r_{J,J'}$ to base $a=10$ from Eq.\
    (\ref{eq:conv_int}) for the rate coefficient calculation of Eq.\
    (\ref{eq:rate_coeff}) using the product of the modeled electron energy
    distribution ($k_BT_\perp=2.25$ meV) and the cross sections of Ref.\
    \cite{hamilton_electron-impact_2016} for cooling (full lines) and excitation
    (broken lines).}
 \label{fig:conv_int}
\end{figure}

The rates $R_{J,J'}$ of inelastic transitions $J\to J'$ are obtained from the
merged-beams rate coefficients $\alpha^{\rm mb}_{J,J'}$ according to Eq.\
(\ref{eq:alphasigma}), using the energy distribution $f(E)$ for matched average
electron and ion beam velocities and the cross sections $\sigma_{J,J'}(E)$.  To
illustrate the relevance of the various electron energies, we consider equally
spaced bins on a logarithmic energy scale such that, for $x=\log_a E$ to a base
$a$,
\begin{equation}
  \alpha^{\rm mb}_{J,J'}=\int_{-\infty}^\infty r_{J,J'}(x)\,dx\;.
  \label{eq:alp_integral}
\end{equation}  
With $x\ln a=\ln E$ and $dE = E(\ln a)dx$ we obtain the integrand
\begin{equation}
  r_{J,J'}(x)=\ln a \; \sqrt{\frac{2}{m}}\,E^{3/2}f(E)\sigma(E)\;.
  \label{eq:conv_int}
\end{equation}
The energy dependent contributions to the logarithmic integral Eq.\
(\ref{eq:alp_integral}) can be visualized directly through the graphical area
under the functions $r_{J,J'}$ when plotted with the logarithmic energy scale.
They are presented in Fig.\ \ref{fig:conv_int} for the dipole-allowed excitation
and de-excitation channels with the lowest $J$.

\begin{figure}[t]
  \centering%
  \includegraphics[width=0.55\textwidth]{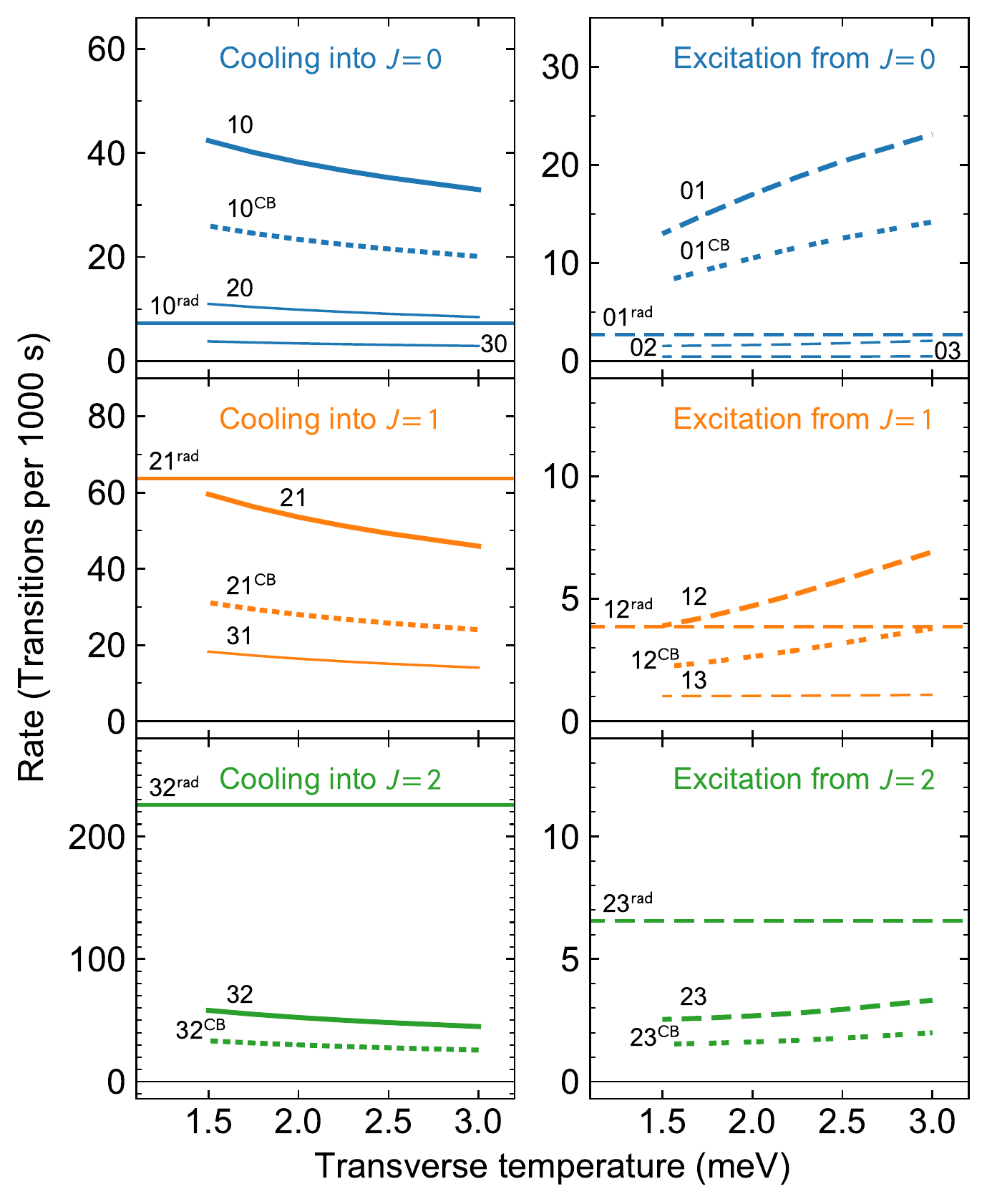}%
  \caption{Modeled rates $R_{J,J'}$ (units of $10^{-3}$ s$^{-1}$) for cooling
    into and excitation from low-$J$ levels as indicated, using the experimental
    parameters with Eqs.\ (\ref{eq:rates}) and (\ref{eq:rate_coeff}).  The
    dependence on the transverse electron temperature $k_BT_\perp$ is shown.
    Results for the cross section of Ref.\ \cite{hamilton_electron-impact_2016}
    (full and long-dashed curves, labels $JJ'$) and for the CB cross sections
    (short-dashed curves, labels $JJ'^{\rm CB}$).  The collisional rates are
    calculated as $R_{J,J'}=\bar{n}_e\alpha^{\rm mb}_{J,J'}$ setting the
    ring-averaged electron density to $\bar{n}_e=2.06\times10^{4}$ cm$^{-3}$.
    Also included are the radiative cooling and excitation rates from Eqs.\
    (\ref{eq:rad_cooling}) and (\ref{eq:rad_excitation}) (horizontal full and
    dashed lines; labels $JJ'^{\rm rad}$).  Note the scale change for the
    cooling rates as $J$ increases.}
 \label{fig:rates_plot}
\end{figure}

The numerical integration makes use of the exponential decrease of $r_{J,J'}(x)$
at high energy and can be performed conveniently from the sum over bins on a
fine grid in $x$.  At low energy ($x\to-\infty$) both, $f(E)$ and
$\sigma_{J,J'}(E)$, remain finite for the cooling channels ($J'<J$).  Here we
extend the explicit calculation of $r_{J,J'}$ down to a limit
$\epsilon\ll k_BT_\perp, k_BT_{||}$ so that for $E<\epsilon$ the energy
distribution and the cross sections are well approximated by
\begin{equation}
  f(E)= \sqrt{\frac{E}{\epsilon}}\;f(\epsilon)\;\;\;{\rm and}\;\;\;\
  \sigma_{J,J'}(E)=\frac{\epsilon}{E}\;\sigma_{J,J'}(\epsilon)\;.
\end{equation} 
With $x_0=\log_a \epsilon$ this implies for $x<x_0$
\begin{equation}
  r_{J,J'}(x)=\sqrt{\frac{2}{m}}\,\epsilon^{3/2}
  f(\epsilon)\sigma_{J,J'}(\epsilon)\;\frac{E}{\epsilon}=r_{J,J'}(x_0)\,
  e^{(x-x_0)\ln a}
\end{equation}
such that Eq.\ (\ref{eq:alp_integral}) can be replaced by
\begin{equation}
  \alpha^{\rm mb}_{J,J'}=\frac{r_{J,J'}(x_0)}{\ln a}+\int_{x_0}^\infty r_{J,J'}(x)dx\;.
\label{eq:rate_coeff}
\end{equation}
The chosen limit $\epsilon$ corresponds to the lowest value on the logarithmic
scale shown in Fig.\ \ref{fig:conv_int}.  

The inelastic rates $R_{J,J'}$ for our experimental conditions are obtained
according to Eq.\ (\ref{eq:rates}) by multiplication of $\alpha^{\rm mb}_{J,J'}$
with the fixed ring-averaged electron density, for which we here assume the
value of $\bar{n}_e=2.06 \times10^{4}$ cm$^{-3}$.  The results are shown in
Fig.\ \ref{fig:rates_plot} for the estimated experimental range of the
transverse electron temperature $T_\perp$ discussed above.

For comparison, Fig.\ \ref{fig:rates_plot} includes also the radiative
transition rates for rotational cooling and excitation from Eqs.\
(\ref{eq:rad_cooling}) and (\ref{eq:rad_excitation}).  It can be seen that the
additional effect of inelastic electron collisions is expected to substantially
exceed the radiative rates for the transitions between the rotational ground
state and the $J=1$ excited state.  Thus, the time constant (reciprocal rate) is
141 s for radiative cooling $J=1\to J'=0$ [at the fitted CSR effective
occupation number, Eq.\ (\ref{eq:csr-eff-occ})] and is expected to become $\sim$
25 s through inelastic electron collisions.  Between the levels $J=2$ and 1,
electronic and radiative rates are of similar size for both cooling and
excitation, while for the level pair $J=3$ and 2 the electronic rates are small
compared to radiational cooling and excitation.  Note that the collisional
cooling rates keep their magnitude for higher $J$, while the radiative cooling
rates rapidly increase with $J$.  For the comparison with the experimental
results, we consider that rotational levels $J\geq3$ will have relaxed
radiatively after $>10$ s of storage and neglect their population at later
times.  The rotational population model is hence restricted to the three levels
$J\leq2$.

\section{Time dependent modeling\label{sec:timedep} }

The relative rotational populations in the stored beam are defined as
$p_J=N_J/N$ where $N=\sum_JN_J=N_i$ is the total number of stored ions.  We
consider total rates $a_{JJ'}$ leading to transitions $J\to J'$, which conserve
the $N$.  The master equation for the relative populations is given by
\begin{equation}
  \dot{p}_J = \sum_{J'}a_{J'J}\,p_{J'} - p_J \sum_{J'}a_{JJ'}
  \label{eq:master}
\end{equation}
where
\begin{equation}
  a_{JJ'} =  \delta_{J',J-1}\,k^{\rm em}_{J\to J-1} +  
  \delta_{J',J+1}\,k^{\rm abs}_{J\to J+1}
  + R_{JJ'}.
  \label{eq:ratedef}
\end{equation}
with the radiative rates from Eqs.\ (\ref{eq:rad_cooling}) and
(\ref{eq:rad_excitation}) and the inelastic collision rates $R_{JJ'}$ as
presented in Fig.\ \ref{fig:rates_plot}.  ($\delta_{i,j}$ denotes the Kronecker
delta.)

The equilibrium populations $p_{J,{\rm eq}}$ follow from setting $\dot{p}_J=0$
in Eq.\ (\ref{eq:master}).  Considering radiation only ($R_{JJ'}=0$), Eqs.\
(\ref{eq:rad_cooling}), (\ref{eq:rad_excitation}) and (\ref{eq:ratedef}) lead
for the equilibrium population ratio of two subsequent rotational levels to
\begin{equation}
  (p_{J+1}/p_J)_{{\rm eq}} = \frac{2J+3}{2J+1}\; \frac{n(\tnu_{J})}{1+n(\tnu_{J})}.
\end{equation}
Assuming a fully thermal radiation field, i.e.,
$n(\tnu_{J})=n_{\text{th}}(\tnu_J,T_r^{\rm eff})$ according to Eq.\
(\ref{eq:bose}) with an effective temperature $T_r^{\rm eff}$, results for $J=0$
in
\begin{equation}
  (p_{1}/p_0)_{{\rm eq}} = 3 e^{-{hc\tnu_0/k_BT_r^{\rm eff}}}
  ~~~\Rightarrow~~~
  T_r^{\rm eff}= -\frac{hc\tnu_0}{k_B}
  \frac{1}{\ln [(1/3)(p_{1}/p_0)_{{\rm eq}}]}.
\end{equation}
This is used to derive $T_r^{\rm eff}$ in the main paper.

Changes of the relative rotational populations can also be caused by the DR in
merged-beams collisions at matched electron and ion velocities, according to the
level-dependent loss rates $r_J=-\dot{N}_J/N$.  As these collisions lead to a
change in the total ion number $N$, the effect on $p_J$ is described by the
non-linear equation
\begin{equation}
  \left(\dot{p}_J\right)_{\rm DR} = \left(\sum_{J'}r_{J'}\,p_{J'} - r_J  \right) p_J
  \label{eq:dr_effect}
\end{equation}
which follows from $\dot{p}_J=(d/dt)(N_J/N)$, considering the time derivatives
of both $N_J$ and $N=\sum_JN_J$.  We explicitly consider Eq.\
(\ref{eq:dr_effect}) in the model used for Fig.\ 3 of the main paper.  For the
purpose of a rough estimate, consider the rate of population change in a
two-level system ($J=0$ and 1) with $p_0=p_1=\frac{1}{2}$.  Eq.\
(\ref{eq:dr_effect}) then leads to
\begin{equation}
  \begin{aligned}
    \left(\dot{p}_0\right)_{\rm DR} &=
    \textstyle\frac{1}{4}(r_1-r_0) = \textstyle\frac{1}{2}(r_1-r_0)p_0\\
    \left(\dot{p}_1\right)_{\rm DR} & = \textstyle\frac{1}{4}(r_0-r_1) =
    \textstyle\frac{1}{2}(r_0-r_1)p_1\;.
  \end{aligned}
  \label{eq:dr_effect_approx}
\end{equation}
The result of this estimate is given in Supplemental Sec.\ \ref{sec:rotdep}.

\section{Constraining the transverse temperature\label{sec:trtemp}} 

\begin{figure}[t]
 \centering
  \includegraphics[width=5cm]{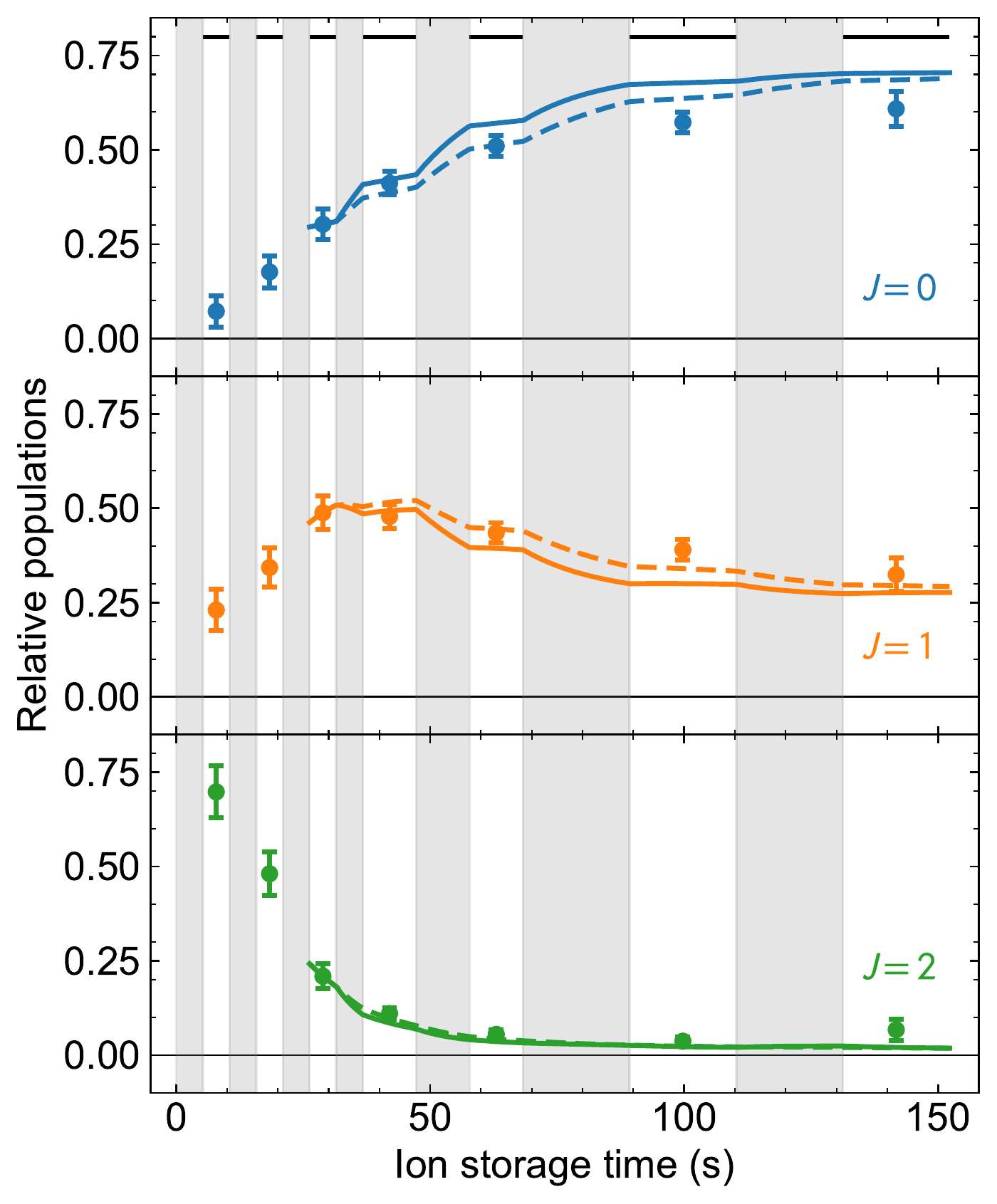}
  \includegraphics[width=5cm]{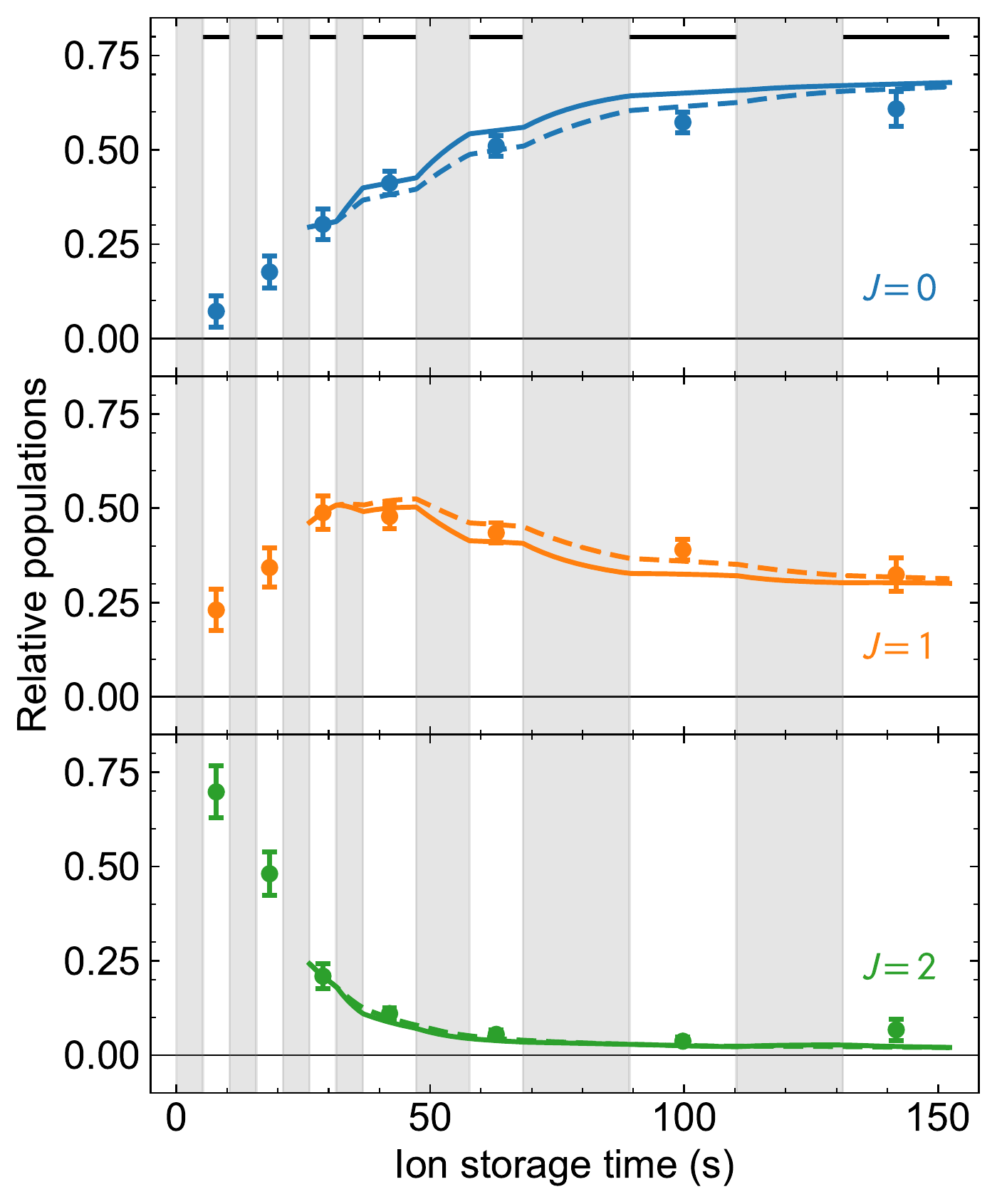}
  \includegraphics[width=5cm]{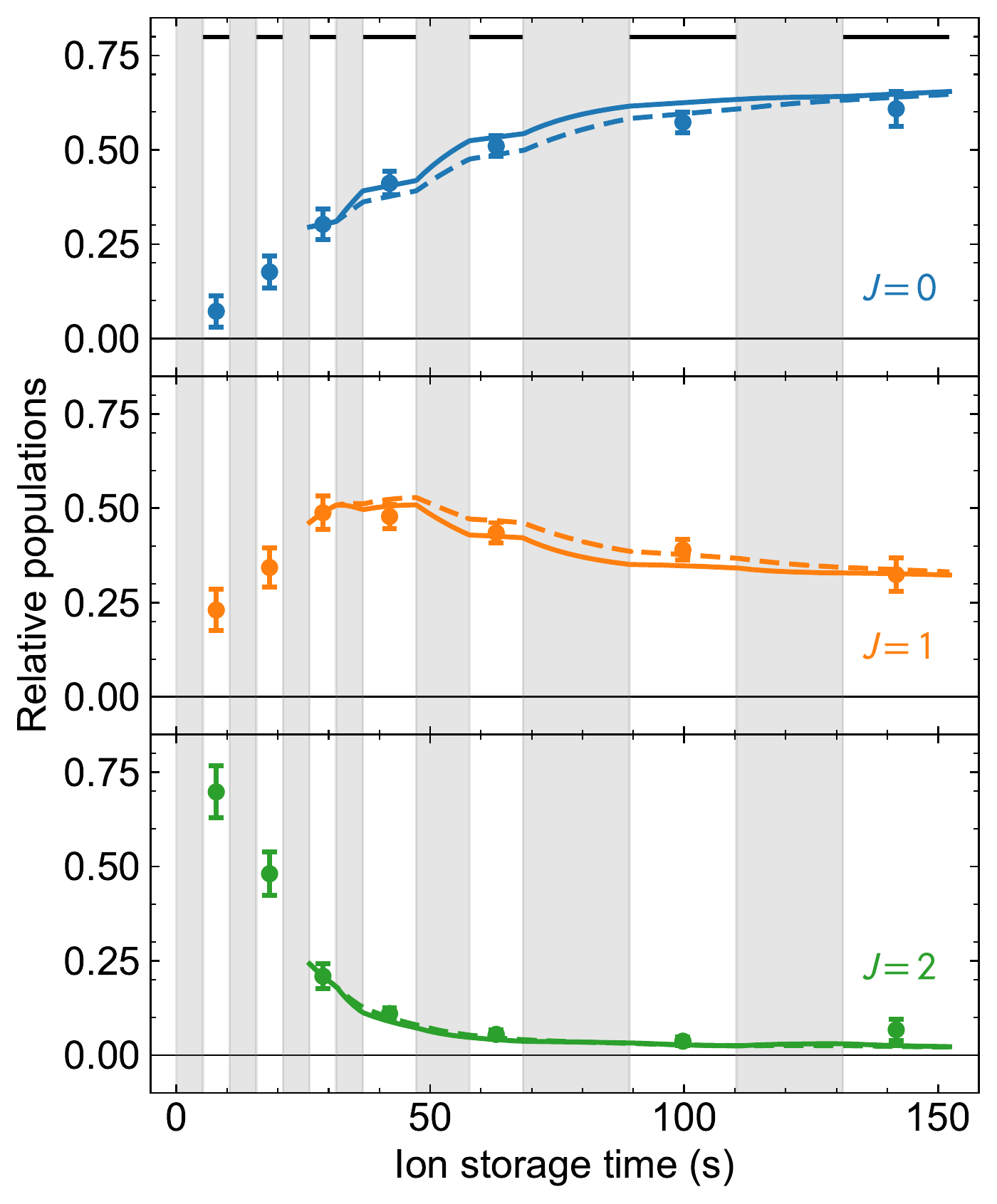}\\[2cm]
  \includegraphics[width=5cm]{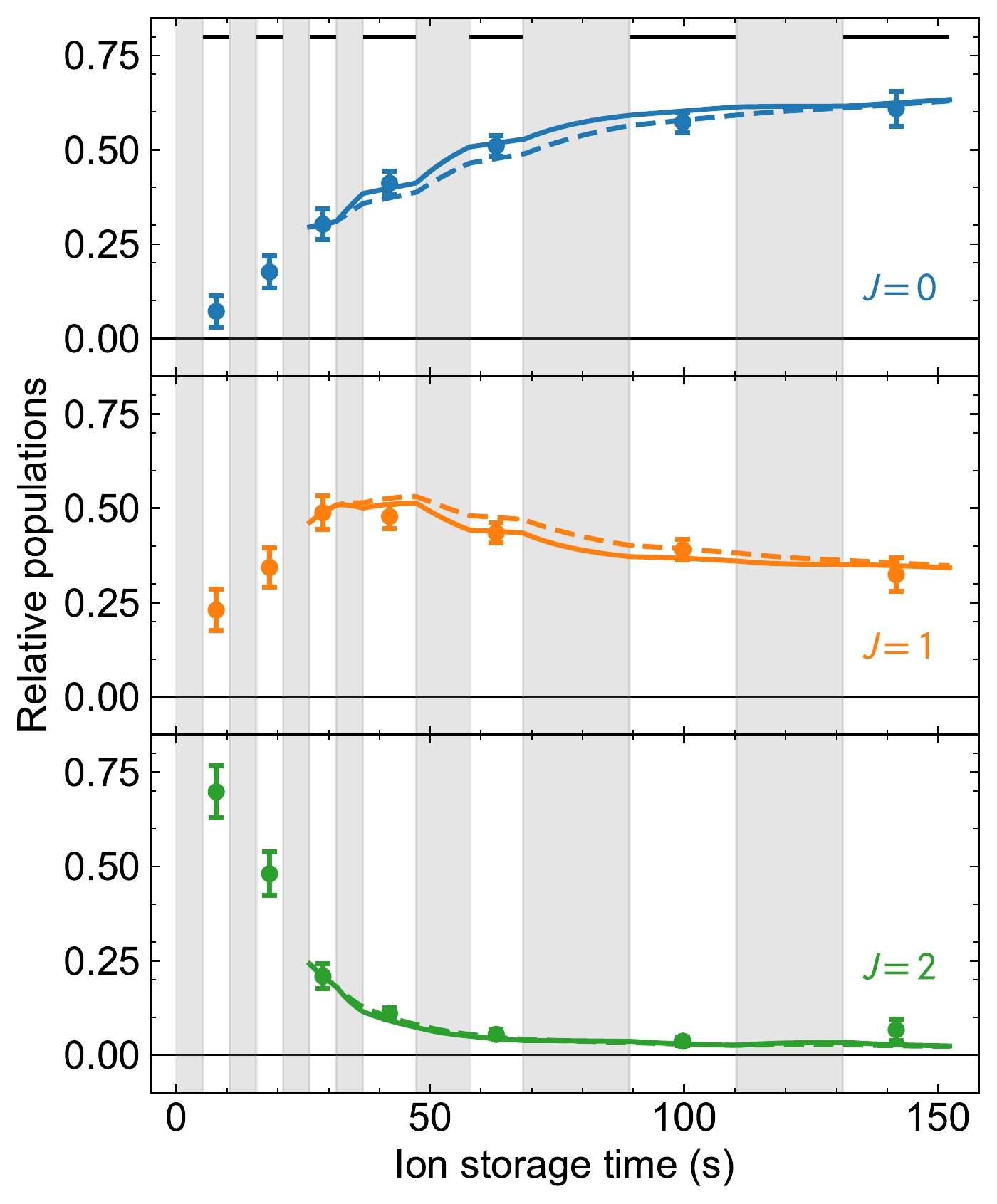}
  \includegraphics[width=5cm]{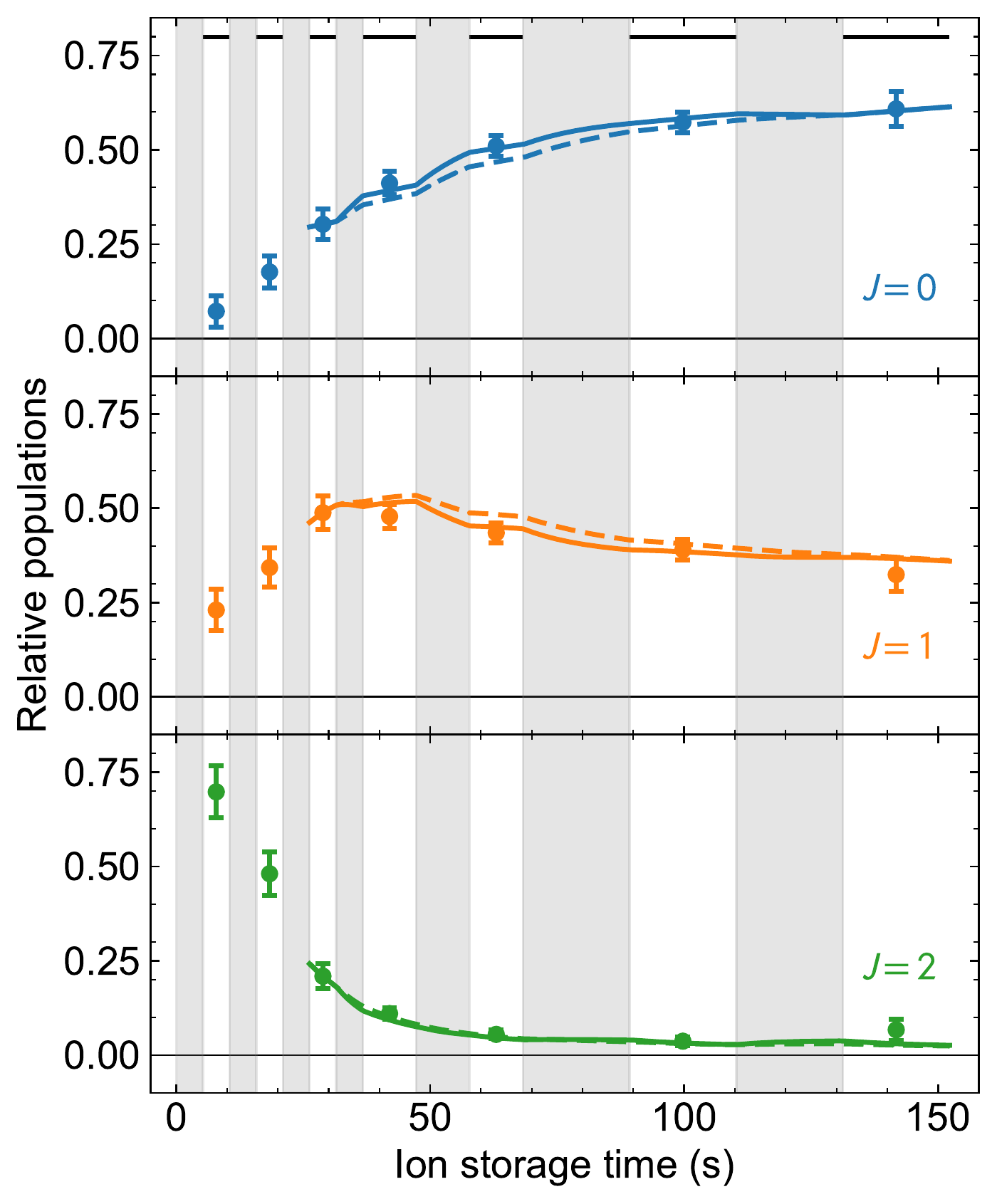}
  \includegraphics[width=5cm]{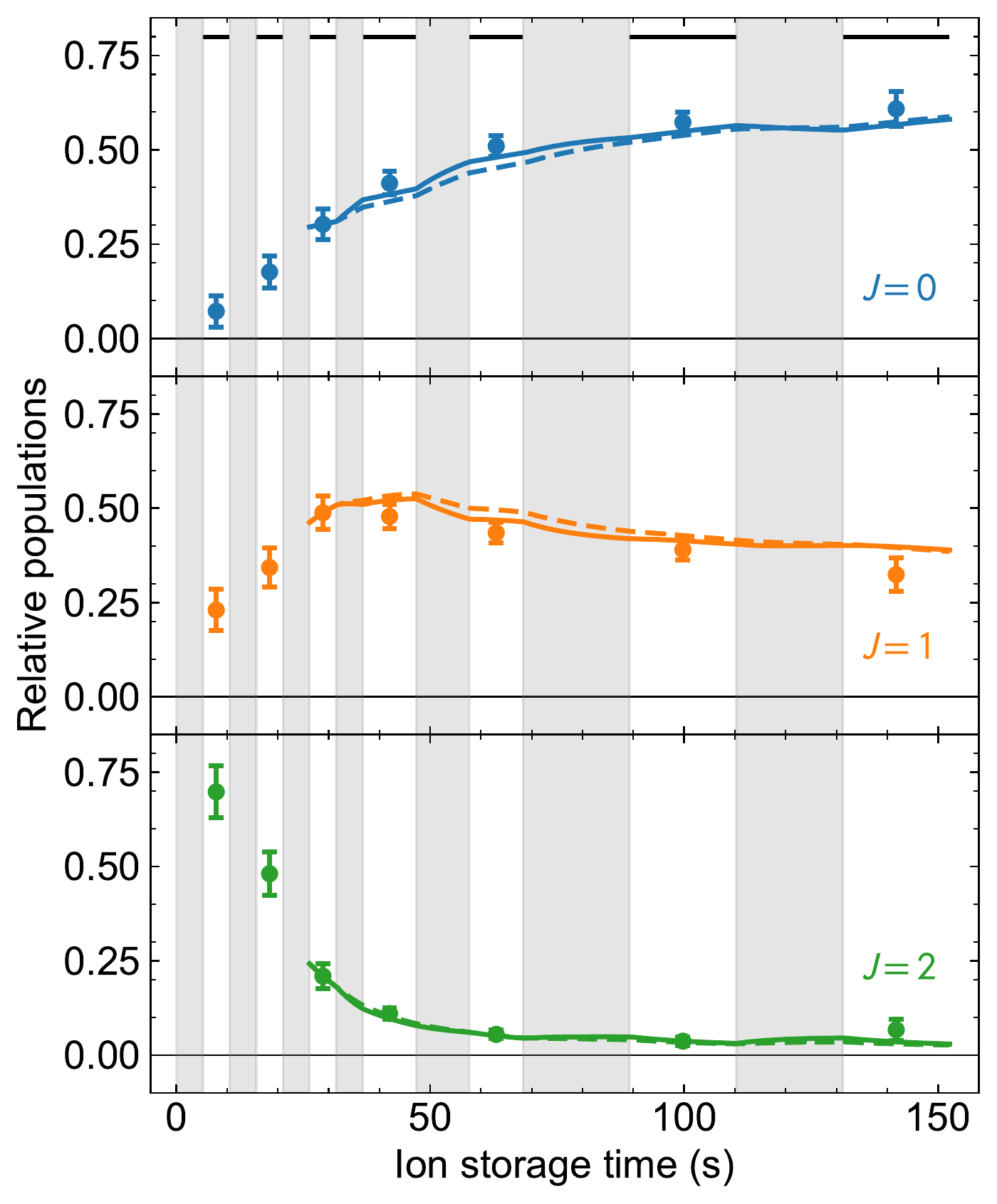}\\[2cm]
  \caption{Modeled relative rotational populations $p_J(t)$ for the complete
    rate-equation model as described in the main paper for the following
    transverse temperatures: upper row, left to right, $k_BT_\perp=1.5,1.75,2.0$
    meV; lower row, left to right, $k_BT_\perp=2.25,2.5,3.0$ meV.  As in the
    main paper, the rate-model results from the CB approximation and the
    $R$-matrix calculations \cite{hamilton_electron-impact_2016} are shown by
    dashed and full curves, respectively.}
  \label{fig:pop_temp}
\end{figure}

The uncertainty of $T_\perp$ translates into an uncertainty of the modeled
inelastic rates.  Since we measure the rotational populations for up to 150 s,
while the expected total cooling time constants ($1/R$) amount to at most 25 s
(see Fig.\ \ref{fig:rates_plot}), the longest storage time should already come
close to the collisional equilibrium populations.  The rotational levels most
sensitive to this equilibrium are $J=0$ and 1.  We have calculated
time-dependent models of $p_J$ for a set of discrete values of $k_BT_\perp$ in
the estimated experimental range (see Fig.\ \ref{fig:pop_temp}) and compare
these results to the populations measured by laser probing.  The best agreement
of the modeled $p_{0,1}$ with the measurement in the final probing interval
($t=141$ s) is obtained for the model with $k_BT_\perp=2.25$ meV.  From this we
conclude the experimentally constrained value of $k_BT_\perp=2.25(25)$ meV used
in the main paper.

\section{Metastable electronically excited CH$^+$ state
\label{sec:metastableState}}

The excitation of the metastable $a\,^3\Pi$ electronic state in the stored ion
beam is revealed by the KER of the neutral DR products observed at the MCP.  The
$1\sigma^22\sigma^23\sigma^2~X^1\Sigma^+$ ground state of CH$^+$ lies $\sim$7.18
eV above the lowest neutral level C($2s^22p^2~^3P$) + H($1s$).  In DR at
near-zero electron collision energy (matched beam velocities) the mainly
populated final state is C($2s^22p^2~^1D$) + H($1s$) with a KER close to 5.9 eV.
Higher excited singlet and triplet terms, starting at 7.48 eV with
C($2s^22p3s~^3P$), cannot be reached from the CH$^+$ ground state.  Conversely,
the lowest triplet state $1\sigma^22\sigma^23\sigma1\pi~a\,^3\Pi$ of CH$^+$ lies
$\sim$1.18 eV above CH$^+(X^1\Sigma^+)$.  Hence, DR of low-energy electrons with
the metastable CH$^+$ ions can reach such higher excited terms.  The
C($2s2p^3~^3D$) term (7.95 eV above the C ground state) is found here (as earlier by~\citen{amitay_dissociative_1996}) to be by far the dominating channel for
low-energy DR of CH$^+(a\,^3\Pi)$.  The KER, whose value identifies the final as
well as the initial levels of the DR process, is near 0.4 eV, much lower than
that for the CH$^+$ ground state.  DR events with higher KER, and thus lower
lying terms of the C atom as final channels, were not detected from metastable
CH$^+$ ions.

\begin{figure}[t]
	\centering%
	\includegraphics[width=0.65\textwidth]{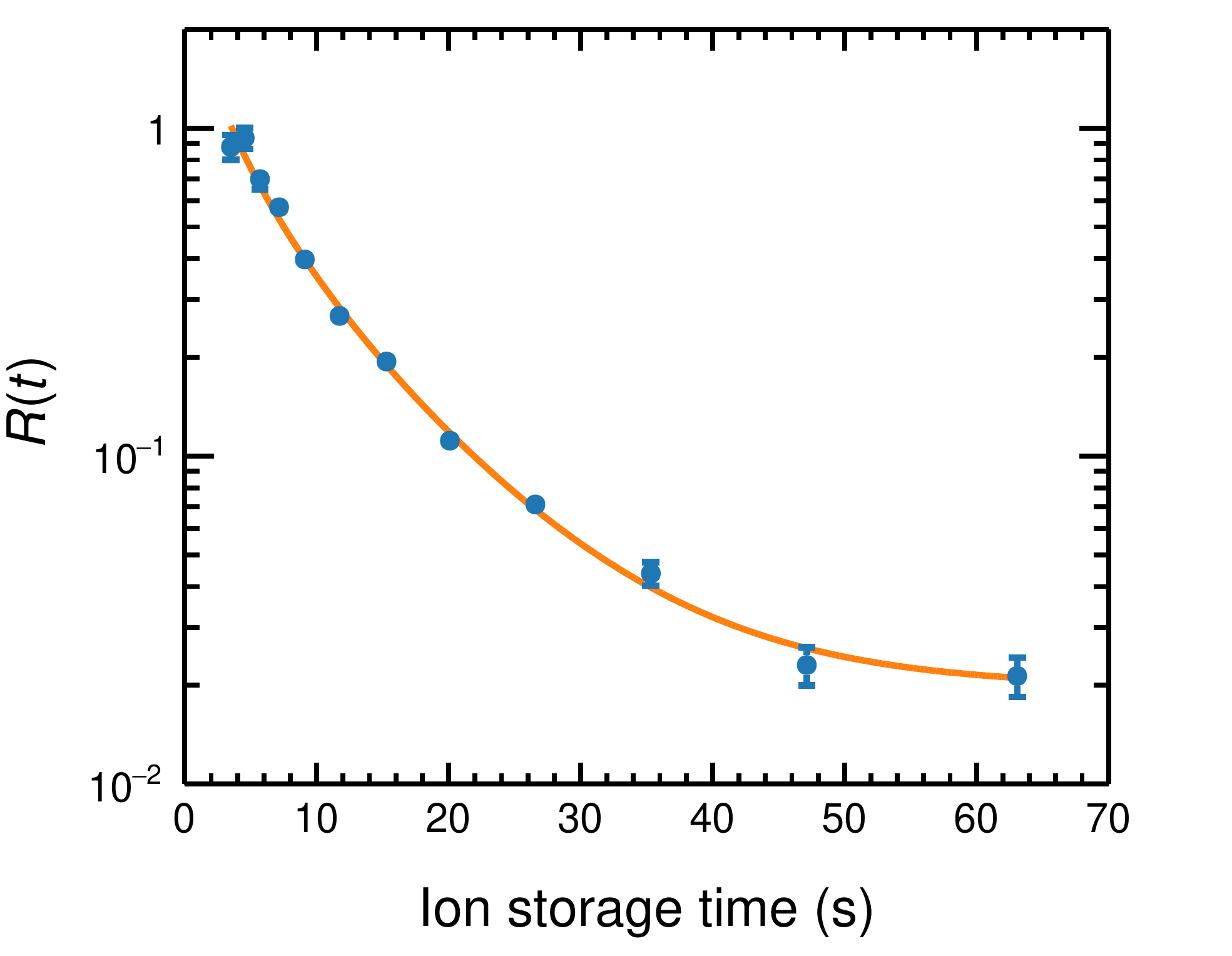}%
	\caption{Metastable state characterization from DR-product imaging. The observed relative contribution of low-KER DR signal $R(t)$ as a function of storage time is given by the symbols with $1\sigma$ statistical errorbars. Full line: fit of the data using Eq.~\ref{eq:metastable_ratio} and the metastable population model of Eq.~\ref{eq:metastable_pop}.}
	\label{fig:metastable_R}
\end{figure}

We follow the earlier procedure \cite{amitay_dissociative_1996} and extract the
	relative DR signals (at matched beam velocities) from the $X^1\Sigma^+$ ($S_\text{g}$) and $a\,^3\Pi$ ($S_\text{m}$) states by analyzing the
	magnitude of the contribution with low KER, assigned to the metastable state, to the observed DR-product imaging
	distribution as a function of storage time. Assuming a constant ratio of the effective (i.e., $J$-averaged)
	DR cross sections for the metastable $\sigma_\text{m}$ and the ground states $\sigma_\text{g}$, the ratio of these DR signals can be expressed in terms of the relative metastable population $\hat{p}_m(t)$ as
	\begin{equation}
		R\left(t\right) = \frac{S_\text{m}}{S_\text{g}} = \frac{\sigma_\text{m}}{\sigma_\text{g}}\frac{\hat{p}_m(t)}{1-\hat{p}_m(t)}.
		\label{eq:metastable_ratio}
	\end{equation}
	The storage times for the present data extend up to 100 s while they were limited to 25 s in the previous measurements \cite{amitay_dissociative_1996}. Starting at the times of $\sim 3$~s after injection, we can account for the $R(t)$ data (see Fig.\ \ref{fig:metastable_R}) by assuming a single-exponential model for $\hat{p}_m(t)$
	\begin{equation}
		\hat{p}_{m}(t)=\hat{p}_{m,t_0}\,e^{-(t-t_0)/\tau_m}+\hat{p}_{m,\infty}
		\label{eq:metastable_pop}
	\end{equation}
	with an exponential decay [$\tau_m=(10.1\pm 1.0)$~s and $\hat{p}_{m,t_0}=(0.10\pm 0.03)$ for $t_0=\SI{21}{s}$] and constant offset [$\hat{p}_{m,\infty}=(0.025\pm 0.007)$]. Around the start of the observation, this fitted $\hat{p}_m(t)$ amounts to more than $0.4$. This causes the relation of $R(t)$ to $\hat{p}_m(t)$ to be significantly non-linear, such that we can independently fit both ${\sigma_\text{m}}/{\sigma_\text{g}} = 0.8\pm0.3$ and the $\hat{p}_m(t)$ model parameters.
	 A probable candidate as the source of the long-lived ($\gg100$ s) $a\,^3\Pi$ contribution signal can be attributed to the $a\,^3\Pi(v=0,J=0,f~{\rm symmetry})$
	sublevel.  The existence of this longer lived metastable rotational level was
	pointed out earlier \cite{hechtfischer_photodissociation_2007}, predicting that
	it can radiatively decay only via much slower {\em magnetic} dipole transitions
	(mean lifetimes $\gg\SI{100}{s}$).

We find a lifetime $\tau_m$ somewhat longer than the previous value \cite{amitay_dissociative_1996} of $(7\pm 1)$ s, which we explain by the extended time range of the measurement, our different cooling conditions for the rotational populations, and the
modified fit model including a longlived contribution. We consider the fit results using Eq.~\ref{eq:metastable_pop} as the best estimate of the metastable population and apply it for the analysis of the DR rates in Supplement Sec.~\ref{sec:rotdep}. Regarding possible feeding of $J$-levels in the ground state by metastable decays, this is equivalent to the fractional metastable populations $\hat{p}_{m}$ $< 0.1$ for $t > \SI{25}{\second}$ and  $< 0.05$ for $t > \SI{40}{\second}$ as stated in the main paper. Considering faster earlier decay components in a more complex model of $\hat{p}_{m}(t)$ would imply smaller non-linear effects in the function ${\hat{p}_m(t)}/({1-\hat{p}_m(t)})$, which after fitting the parameters would lead to even lower values for $\hat{p}_{m,t_0}$ and to larger values of ${\sigma_\text{m}}/{\sigma_\text{g}}$. As for the assumption of a constant ratio of ${\sigma_\text{m}}/{\sigma_\text{g}}$ in the observed time window, we examined the underlying mechanism that drives DR for the metastable state. For the DR from ground-state CH$^{+}$, we observe a weak rotational dependence among the relevant low-$J$ levels (see Supplement Sec.~\ref{sec:rotdep}, Table~\ref{tab:rates_DR}). Theoretically, such a weak dependence is predicted \cite{mezei_atom_2019} for a strong neutral dissociating resonance for low-energy electronic continuum states with the CH$^{+}$ core, which manifests itself in an anti-crossing pattern within the energetic structure of the $^2\Pi$ Rydberg states of CH (see Fig. 1 of Ref.~\cite{vazquez_rydbergch_2007}). A similar efficient direct DR route likely also exists for the $a\,^3\Pi$ metastable state of CH$^{+}$, considering the energetic structure in the Rydberg states of $^2\Sigma^{-}$ and $^2\Delta$ symmetries of CH (see Fig. 3 of Ref.~\cite{vazquez_rydbergch_2007}). Given the non-resonant nature of such a direct DR pathway, only a weak rotational dependence for the DR rate coefficient \cite{drbook} is likely expected for metastable ($a\,^3\Pi$) CH$^{+}$.

\section{Rotational dependence of dissociative recombination 
  \label{sec:rotdep}}

\begin{figure}[t]
  \centering%
  \includegraphics[width=0.65\textwidth]{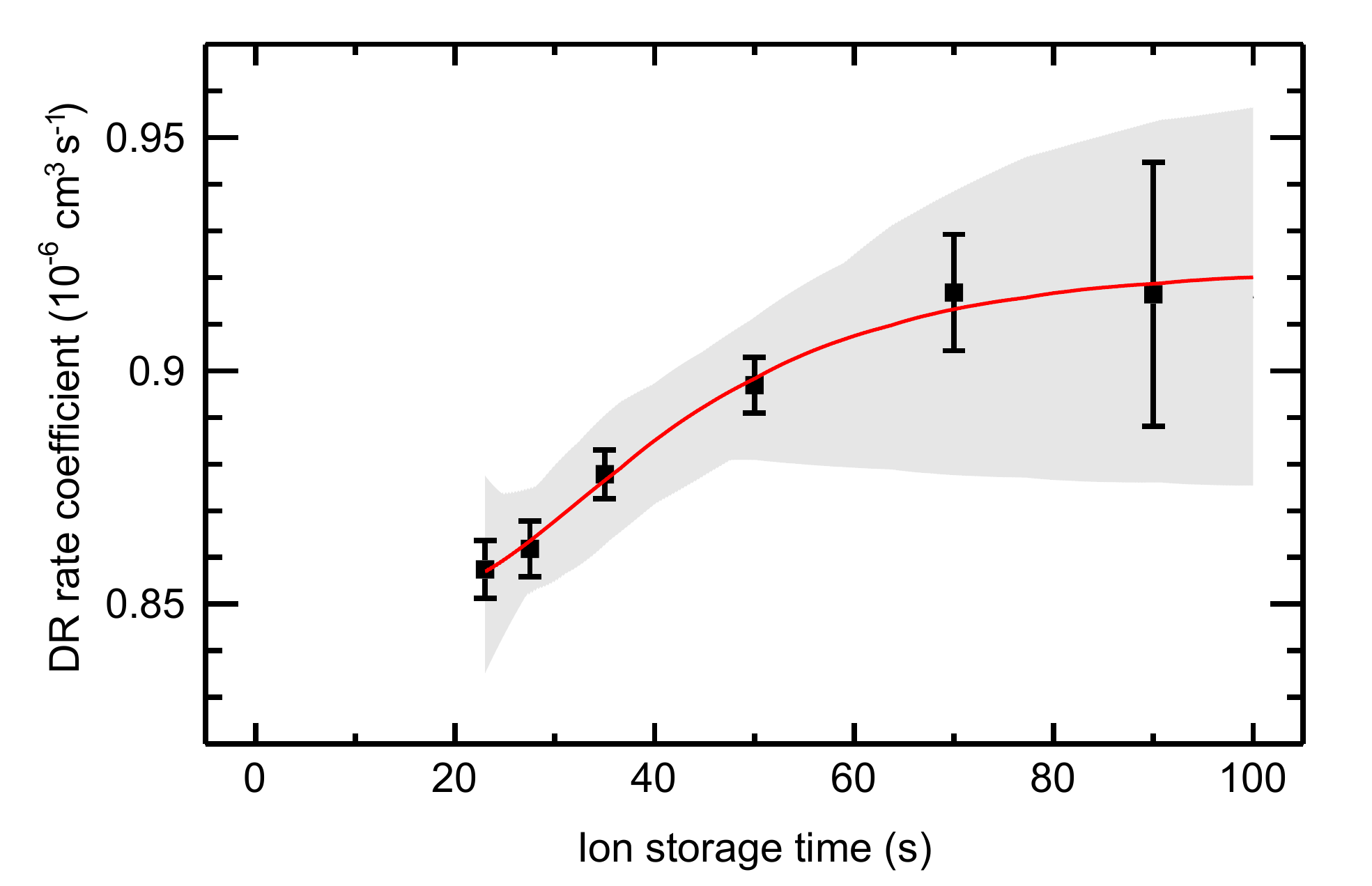}%
  \caption{Total DR rate coefficient storage time dependence.  Black symbols:
    measured values and $1\sigma$ statistical uncertainty; red line: time
    dependence from the model of Eq.\ (\ref{eq:DR_model}) using the best
    estimator values for $\alpha_J$ and $\alpha_m$ from the MCMC analysis; gray
    area: space covered by all parameter combinations sampled in the MCMC
    analysis.}
  \label{fig:rates_DR}
\end{figure}

\begin{figure}[t]
  \centering%
  \includegraphics[width=0.6\textwidth]{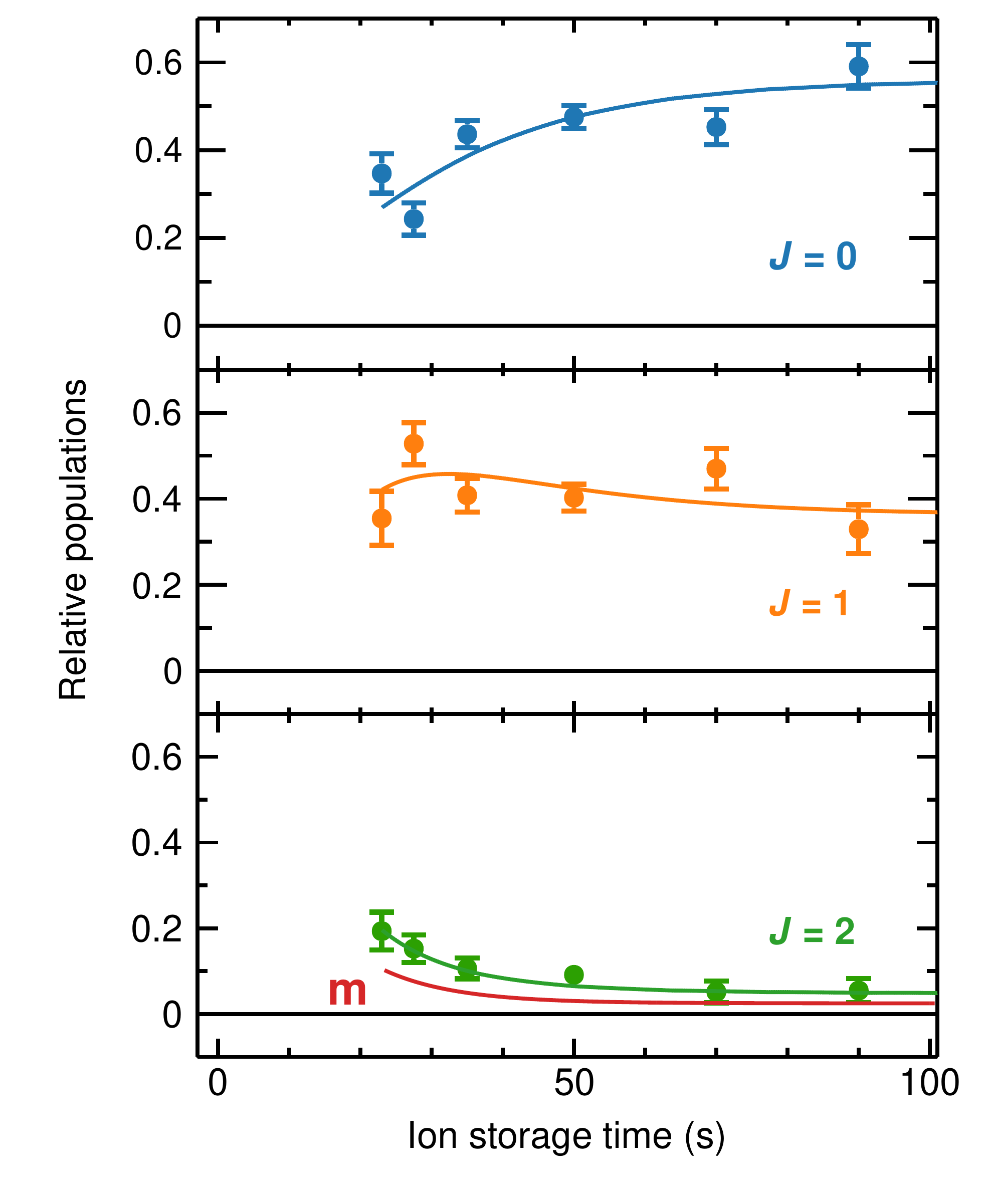}%
  \caption{Relative populations $\hat{p}_J(t)$, $J=0\ldots2$, determined from
    laser probing measurements performed with the same control cycle for the
    electron beam intensity and energy as in the DR rate measurements, but with
    increased stored ion number.  Smooth lines were added to the panels, which
    were drawn to best represent the observed storage time evolution.  The
    metastable $a\,^3\Pi$ state population $\hat{p}_{m}(t)$ (Eq.\
    \ref{eq:metastable_pop}) is displayed in the bottom panel.}
  \label{fig:populations_DR}
\end{figure}

The laser probing was also combined with studies of the DR rate.  Laser probing
and DR measurements could in general not be performed simultaneously, as laser
probing required high stored ion numbers (typically $10^7$ ions) to observe
sufficient photodissociation signal with the pulsed laser, while the continuous
count rates from DR had to be limited by lower stored ion numbers (typically
$10^6$ ions) to avoid saturation of the detector.  However, a control cycle was
developed where the electron velocity was repeatedly set (in a wobbling scheme
with typical dwell times of 30 ms) to values detuned from the ion beam velocity.
Velocity-detuned time windows were used to measure the background in the
detector count rate not induced by electron collisions (subtracted for
determining the DR merged-beams rate coefficient) and for laser probing to find
the time dependent relative rotational populations.  This way, laser-probing and
DR measurements were performed for the same control cycle such that the effect
of changing, laser-probed rotational populations on the DR rate coefficient
could be analyzed.  To distinguish the relative populations from those discussed
in the main paper (measured while only matched electron and ion beam velocities
were used) they are denoted as $\hat{p}_J$ in the following.  The given DR rate
coefficients always refer to those measured in the wobbling phases when the
average electron and ion beam velocities were matched to each other.

As shown in Fig.\ \ref{fig:rates_DR}, the DR rate at matched average beam
velocities (corresponding to the present laser probing of inelastic collisions)
showed a slight time dependence that is attributed to the variation of the
relative populations $\hat{p}_J(t)$ of the three lowest levels $J=0\ldots2$.  We
assume for these levels DR rate coefficients $\alpha_J$ [corresponding to the
quantities $\alpha_{\rm DR}^{\rm mb}(J)$ used in the main paper;
$\alpha_J=\alpha_{\rm DR}^{\rm mb}(J)$] as well as $\alpha_{m}$ for the $a\,^3\Pi$
state (see Sec.\ \ref{sec:metastableState} of this Supplemental Information).
With this we model the total DR rate by
\begin{equation}
  \alpha(t)=\sum_{J=0}^{2}\alpha_J \hat{p}_J(t) + \alpha_{m}\hat{p}_{m}(t)
  \label{eq:DR_model}
\end{equation}
where the populations $\hat{p}_J(t)$ are normalized such that their sum amounts
to $1-\hat{p}_{m}(t)$ [see Eq.\ (\ref{eq:metastable_pop})].

The laser-probed rotational populations are visualized in Fig.\
\ref{fig:populations_DR}.  Within the statistical scatter, the data for $J=0$
and $J=2$ show clear trends in $\hat{p}_J(t)$ that should lead to changes in
$\alpha(t)$ depending on the $J$-dependence of the DR rate coefficients.  We
represent the trends in $\hat{p}_J(t)$ by smooth curves with normalized relative
populations, which is included in Fig.\ \ref{fig:populations_DR}.

\begin{table}[t]
  \caption{ Level-specific DR rate coefficients
    $\alpha_J=\alpha_{\rm DR}^{\rm mb}(J)$ ($J=0\ldots2$) and $\alpha_m$ of
    CH$^{+}$ in the merged beams at matched average beam velocities as
    determined from the MCMC analysis.  The limits given are the estimated
    $1\sigma$ uncertainties of the fitted results except for $\alpha_2$ and
    $\alpha_m$ where both limits are sharp corresponding to the positive
    parameter values required in the MCMC sampling. }
  \label{tab:rates_DR}
  \vspace{2mm}
  \centering
  \begin{minipage}{7cm}
    \begin{ruledtabular}
      \begin{tabular}{@{~~}cc}
        \raisebox{-3mm}{Quantity} &  Value \\[-2mm]
                  &\raisebox{0mm}[0mm][3mm]{($10^{-6}$ cm$^3$\,s$^{-1}$)}\\
        \hline
        $\alpha_0$ & $1.053\pm0.062$ \\
        $\alpha_1$ & $0.741\pm0.083$ \\
        $\alpha_2$ & $0.8\pm0.8$ \\
        $\alpha_m$ & $1.4\pm1.4$ \\
        $\bar{\alpha}_{01}=  (\alpha_0 + \alpha_1)/2 $ & $0.897\pm0.015$ \\
        $\alpha_0 - \alpha_1 $ & $0.31\pm0.15$ \\
      \end{tabular}
    \end{ruledtabular}
  \end{minipage}
\end{table}

To estimate the parameters $\alpha_J$ and $\alpha_m$ in the model of Eq.\
(\ref{eq:DR_model}) we use the smoothed trends of $\hat{p}_J$.  A linear
least-squares minimization procedure cannot be applied in our case since its
results violate the physical constraints $\alpha_J>0$ and $\alpha_m>0$.  Thus,
we use the more robust Markov-chain Monte-Carlo (MCMC) analysis method
\cite{mcmc}.  The latter is based on the Bayesian inference principle and
samples the marginal-likelihood function of the model over the relevant
parameter space with Markov chains, i.e., in a more efficient way than sampling
uniformly.  As a result, the MCMC analysis provides the probability
distributions for all parameters, i.e., $\alpha_J$ and $\alpha_m$, and their
mutual correlations.

We use the MCMC package {\sc emcee} \cite{Foreman_Mackey_2013} and constrain the
valid parameter space to positive values for all rate coefficients.  As a prior
we choose for all parameters a broad uniform distribution over the interval
$[0,1]\times 10^{-5}$ cm$^3$\,s$^{-1}$.  After $\sim$\,60 iterations the MCMC
routine converges and a further $10^5$ iterations are used to generate the
aforementioned probability distributions for the individual rate coefficients
$\alpha_J$ and $\alpha_m$ and specific linear combinations.  From those, the
most probable values and the uncertainties of the parameters are derived as mean
values and standard deviations of the distributions for $\alpha_{0}$ and
$\alpha_{1}$ (with sharp uncertainties given from the extent of the
distributions for $\alpha_{2}$ and $\alpha_{m}$), as summarized in Table
\ref{tab:rates_DR}.  The results document that the average $\bar{\alpha}_{01}$
of the $J=0$ and $1$ rate coefficients is found with a smaller relative
uncertainty than the difference $\alpha_0 - \alpha_1$.  To calculate the effect
of the $J$-specific DR rates on the populations according to Eq.\
(\ref{eq:dr_effect}), we use $r_j=\bar{n}_e\alpha_{\rm DR}^{\rm mb}(J)$ with the
best estimator values given for $\alpha_J=\alpha_{\rm DR}^{\rm mb}(J)$ in Table
\ref{tab:rates_DR}.  The results of this time dependent model are shown in Fig.\
3 of the main paper.

We remark that the experimental result for $\alpha_0 - \alpha_1 $ (Table
\ref{tab:rates_DR}), the ring-averaged electron density, and the estimate of
effects caused by $J$-specific DR at the end of Supplemental Sec.\
\ref{sec:timedep}, Eq.\ (\ref{eq:dr_effect_approx}), lead to
$\left(\dot{p}_0/p_0\right)_{\rm DR} \approx -\left(\dot{p}_1/p_1\right)_{\rm
  DR} \approx -(3.2\pm1.5)\times10^{-3}$ s$^{-1}$.  This estimate can be
directly compared to the inelastic and radiative rates shown in Fig.\
\ref{fig:rates_plot}.  The effect of the $J$-specific DR rates counteracts the
increase of $p_0$ due to the radiative cooling.  Its magnitude corresponds to
only about half the $J=1\to0$ cooling rate $k^{\rm em}_{1\to0}$ included in this
figure (label $10^{\rm rad}$).  However, the DR-induced $J$-changing rates are
about an order of magnitude lower than those caused by the inelastic collisions.

\bibliography{chlaser}